%% file: MuonDISfaser.tex
\documentclass[11pt,a4paper]{article}
\pdfoutput=1

\usepackage[colorlinks=true, linkcolor=black!50!blue, urlcolor=blue, citecolor=blue, anchorcolor=blue]{hyperref}
\usepackage[font=small,labelfont=bf,margin=0mm,labelsep=period,tableposition=top]{caption}
\usepackage[a4paper,top=1.5cm,bottom=2cm,left=1.5cm,right=1.5cm,bindingoffset=0mm]{geometry}

\usepackage{placeins,cite}
\usepackage{graphicx}
\usepackage{soul}
\usepackage{subcaption}
\usepackage{float}
\usepackage{afterpage}
\usepackage{epsfig}
\usepackage{amssymb}
\usepackage{amsmath}
\usepackage{bm}
\usepackage{multirow}
\usepackage{url}
\usepackage{xcolor}
\usepackage{url}
\usepackage{booktabs,multirow}
\usepackage{textcomp}
\usepackage{comment}
\graphicspath{{../}{./figures/}}

\usepackage{makecell}

\makeatletter
\def\thickhline{%
             \noalign{\ifnum0 =`}\fi\hrule \@height \thickarrayrulewidth \futurelet
             \reserved@a\@xthickhline}
\def\@xthickhline{\ifx\reserved@a\thickhline
                \vskip\doublerulesep
                \vskip -\thickarrayrulewidth
                \fi
                \ifnum0 =`{\fi}}
\makeatother
\newlength{\thickarrayrulewidth}
\usepackage{tikz}
\usetikzlibrary{positioning}
\usetikzlibrary{shapes.geometric, arrows}
\usetikzlibrary{arrows.meta}
\usepackage{varwidth}
\usepackage{xcolor}
\definecolor{mtplotlib1}{HTML}{1f77b4}
\definecolor{mtplotlib2}{HTML}{ff7f0e}
\definecolor{mtplotlib3}{HTML}{2ca02c}
\definecolor{mtplotlib4}{HTML}{d62728}

\tikzset{%
  >={Latex[width=2mm,length=2mm]},
            base/.style = {rectangle, rounded corners, draw=black,
                           minimum width=4cm, minimum height=1cm,
                           text centered}, 
            mystyle/.style={rectangle, rounded corners, draw=black,
            minimum width=12cm, minimum height=1cm,
            text centered}, 
    col0/.style = {base, fill=white!30},
    col1/.style = {base, fill=mtplotlib1!30},
    col11/.style = {mystyle, fill=mtplotlib1!30},
    col2/.style = {base, fill=mtplotlib2!30},
    col3/.style = {base, fill=mtplotlib3!30},
    col4/.style = {base, minimum width=2.5cm, fill=mtplotlib4!15,}
}

\RequirePackage[normalem]{ulem} 

\newcommand{\be}{\begin{equation}}
\newcommand{\ee}{\end{equation}}
\newcommand{\bea}{\begin{eqnarray}}
\newcommand{\eea}{\end{eqnarray}}
\newcommand{\bi}{\begin{itemize}}
\newcommand{\ei}{\end{itemize}}
\newcommand{\ben}{\begin{enumerate}}
\newcommand{\een}{\end{enumerate}}

\newcommand{\lc}{\left[}
\newcommand{\rc}{\right]}
\newcommand{\lp}{\left(}
\newcommand{\rp}{\right)}

\def\frac#1#2{{{#1}\over {#2}}}
\def\gsim{\mathrel{\rlap{\lower4pt\hbox{\hskip1pt$\sim$}}
    \raise1pt\hbox{$>$}}}       
\def\lsim{\mathrel{\rlap{\lower4pt\hbox{\hskip1pt$\sim$}}
    \raise1pt\hbox{$<$}}}

\newcommand{\tr}{\mathrm{tr}}

\newcommand{\draft}[1]{}
\def\beq{\begin{equation}}
\def\eeq{\end{equation}}

\numberwithin{equation}{section}
\numberwithin{figure}{section}
\numberwithin{table}{section}

\usepackage{tabularx}
\newcolumntype{C}[1]{>{\centering\arraybackslash}p{#1}}

\definecolor{darkblue}{rgb}{0.0,0,0.5}
\definecolor{darkgreen}{rgb}{0.0,0.3,0.0}
\definecolor{redish}{rgb}{0.675,0,0.2}
\definecolor{red}{rgb}{0.8,0,0}
\definecolor{green}{rgb}{0,0.6,0}
\definecolor{bluish}{rgb}{0.2,0.2,0.675}
\definecolor{mygrey}{rgb}{0.6,0.6,0.6}

\usepackage{tikz} 
\usetikzlibrary{shapes,arrows,positioning,automata,backgrounds,calc,er,patterns,arrows.meta}
\usepackage{tikz-feynman}
\tikzfeynmanset{compat=1.0.0}
\usepackage{varwidth}
\usepackage{xcolor}
\definecolor{mtplotlib1}{HTML}{1f77b4}
\definecolor{mtplotlib2}{HTML}{ff7f0e}
\definecolor{mtplotlib3}{HTML}{2ca02c}
\definecolor{mtplotlib4}{HTML}{d62728}

\tikzset{%
  >={Latex[width=2mm,length=2mm]},
            base/.style = {rectangle, rounded corners, draw=black,
                           minimum width=4cm, minimum height=1cm,
                           text centered}, 
            mystyle/.style={rectangle, rounded corners, draw=black,
            minimum width=12cm, minimum height=1cm,
            text centered}, 
    col0/.style = {base, fill=white!30},
    col1/.style = {base, fill=mtplotlib1!30},
    col11/.style = {mystyle, fill=mtplotlib1!30},
    col2/.style = {base, fill=mtplotlib2!30},
    col3/.style = {base, fill=mtplotlib3!30},
    col4/.style = {base, minimum width=2.5cm, fill=mtplotlib4!15,}
}

\usepackage{tabularx}
\usepackage{subcaption}
\newcolumntype{C}[1]{>{\centering\arraybackslash}p{#1}}

\definecolor{lightblue}{rgb}{0.0,0.5,1.0}

\begin{document}
\newgeometry{top=1.5cm,bottom=1.5cm,left=1.5cm,right=1.5cm,bindingoffset=0mm}

%

\begin{center}
  {\Large \bf Deep-Inelastic Scattering at TeV Energies with LHC  Muons}\\
  \vspace{1.1cm}
  {\small
  Reinaldo Francener$^{1,2,3}$, Victor P. Gon\c{c}alves$^{4}$, 
 Felix Kling$^{5,6}$, Peter Krack$^{2,3}$ 
  and Juan Rojo$^{2,3}$
  }\\
  
\vspace{0.7cm}

{\it \small
  ~$^1$Instituto de Física Gleb Wataghin - Universidade Estadual de Campinas (UNICAMP), \\[0.1cm]
13083-859, Campinas, SP, Brazil\\[0.1cm]
  ~$^2$Nikhef Theory Group, Science Park 105, 1098 XG Amsterdam, The Netherlands\\[0.1cm]
  ~$^3$Department of Physics and Astronomy, Vrije Universiteit, NL-1081 HV Amsterdam\\[0.1cm] 
  ~$^4$Institute of Physics and Mathematics, Federal University of Pelotas (UFPel),\\[0.1cm]
Postal Code 354, 96010-900, Pelotas, RS, Brazil \\[0.1cm]
   ~$^5$Deutsches Elektronen-Synchrotron DESY, Notkestr. 85, 22607
Hamburg, Germany\\[0.1cm]
   ~$^6$Department of Physics and Astronomy, University of California, Irvine, CA 92697-4575, USA\\[0.1cm] 
 }

\vspace{1.0cm}

{\bf \large Abstract}

\end{center}

The LHC far-forward experiments FASER and SND@LHC have pioneered the detection of TeV-energy neutrinos produced in hard-scattering proton-proton collisions at the LHC. 
In addition to neutrinos, an intense flux of TeV-energy muons reaches these detectors, representing a dominant background for both neutrino studies and beyond the Standard Model searches. 
Here we demonstrate that this forward muon flux enables a comprehensive neutral-current deep-inelastic scattering (DIS) program at FASER with a strong kinematical overlap with the Electron Ion Collider.
For the Run 3 luminosity of $\mathcal{L}_{\rm pp}=250$~fb$^{-1}$, more than $10^5$ inclusive muon DIS events, of which up to $10^4$ from charm production, are expected at FASER$\nu$.
As a representative application, we demonstrate the sensitivity of muon DIS at FASER$\nu$ to probe the (intrinsic) charm content of the proton at large-$x$.
We also provide predictions for event yields of muon DIS for future FASER runs and for the proposed Forward Physics Facility. 

\clearpage

\tableofcontents

\input{sec-introduction.tex}

\input{sec-settings}

\input{sec-inclusiveDIS}
\input{sec-charmDIS}
\input{sec-summary}

\bibliographystyle{utphys}
\bibliography{MuonDISfaser}

\end{document}

%% file: sec-introduction.tex
\section{Introduction}

Charged-lepton deep-inelastic scattering (DIS) has represented since decades a cornerstone in our investigations of Quantum Chromodynamics (QCD) and of the internal structure of the nucleon.
Following the groundbreaking electron-proton DIS experiments at SLAC~\cite{Breidenbach:1969kd, Miller:1971qb}, which conclusively demonstrated the existence of point-like quarks in the proton, several charged-lepton DIS experiments have scrutinized proton structure and provided novel insights on the nature of the strong interaction.
These include BCMDS~\cite{BCDMS:1989qop, BCDMS:1989ggw}, EMC~\cite{Kullander:1990se}, NMC~\cite{NewMuon:1996fwh}, COMPASS~\cite{COMPASS:2007rjf} at CERN, HERA~\cite{Klein:2008di} at DESY, and the Jefferson Lab experiments~\cite{CLAS:2003umf}.
In the near future, a next-generation charged-lepton DIS experiment, the Electron-Ion Collider (EIC)~\cite{AbdulKhalek:2021gbh}, should record first collisions in the early 2030s and enable options of polarised collisions and collisions involving different nuclear species.

Measurements of structure functions and related observables, such as multi-differential distributions, in charged-lepton DIS provide the central ingredients for a wealth of global perturbative QCD analyses of the proton structure, from unpolarised and polarised parton distribution functions (PDFs)~\cite{NNPDF:2021njg, Hou:2019efy, Bailey:2020ooq,Cruz-Martinez:2025ahf,Borsa:2024mss} to nuclear PDFs~\cite{AbdulKhalek:2022fyi, Eskola:2021nhw, Duwentaster:2022kpv}, transverse-momentum dependent (TMD) PDFs~\cite{Angeles-Martinez:2015sea}, and higher-dimensionality non-perturbative objects such as Generalised PDFs~\cite{Kumericki:2016ehc}.  
In turn, the precise knowledge of these non-perturbative QCD objects is of paramount importance to perform predictions at high-energy colliders such as the LHC~\cite{Gao:2017yyd, Campbell:2022qmc, LHCHiggsCrossSectionWorkingGroup:2016ypw} as well as for astroparticle physics experiments such as IceCube~\cite{IceCube:2013low} and KM3NET~\cite{KM3Net:2016zxf}. 
Furthermore, charged-lepton DIS has enabled remarkable discoveries in QCD from the unexpected small contribution of valence quarks to the total proton spin~\cite{EuropeanMuon:1987isl} and the suppression of nuclear structure functions with respect to the free nucleon baseline~\cite{EuropeanMuon:1988lbf} to the presence of intrinsic charm quarks in the proton~\cite{Brodsky:1980pb,Ball:2022qks,NNPDF:2023tyk} and the onset of BFKL dynamics in HERA data~\cite{xFitterDevelopersTeam:2018hym, Ball:2017otu}, to name a few representative examples.

The recent observation of neutrinos produced in LHC collisions by the FASER~\cite{FASER:2022hcn, FASER:2023zcr} and SND@LHC~\cite{SNDLHC:2022ihg, SNDLHC:2023pun} far-forward experiments, followed by the subsequent (differential) measurements of the neutrino-nucleon cross-section at TeV energies by FASER~\cite{FASER:2024hoe, FASER:2024ref}, have opened a new area of research in particle physics, the so-called ``collider neutrino era''. 
These experiments benefit from the unprecedentedly intense neutrino fluxes generated by LHC collisions in the forward region~\cite{Kling:2021gos, Buonocore:2023kna, FASER:2024ykc} and enable a broad portfolio of physics studies in QCD, neutrino physics, and astroparticle physics, see~\cite{Ariga:2025qup, Kling:2025zon} and references therein. 
In particular, it has been demonstrated in~\cite{Cruz-Martinez:2023sdv} that neutrino-nucleus scattering at FASER and its possible upgrades such as the Forward Physics Facility (FPF)~\cite{Anchordoqui:2021ghd, Feng:2022inv, Adhikary:2024nlv, FPFWorkingGroups:2025rsc} realise an effective ``neutrino-ion collider'', with a kinematic coverage comparable to the EIC but with charged-current scattering, providing constraints on proton structure which significantly reduce PDF uncertainties~\cite{PDF4LHCWorkingGroup:2022cjn} for Higgs and electroweak cross-sections and disentangling possible BSM signals from QCD effects in high-$p_T$ tails in LHC collisions~\cite{Greljo:2021kvv, Hammou:2024xuj}.

In addition to neutrinos and possible BSM particles, such as feebly interacting particles (FIPs) and long-lived particles (LLPs)~\cite{Feng:2017uoz, FASER:2018eoc, FASER:2023tle, FASER:2024bbl}, an intense flux of muons produced in the primary pp collision and downstream secondary interactions reaches the FASER and SND@LHC far-forward detectors. 
This muon flux represents the main background for many SM and BSM measurements, and therefore most of the experimental efforts have focused on its characterization, reduction and subtraction~\cite{Sabate-Gilarte:2023aeg}.
Measurements of the muon fluxes have been performed by both collaborations~\cite{FASER:2018bac, FASER:2020gpr, FASER:2021mtu, SNDLHC:2023mib} and are used to estimate muon-induced backgrounds in neutrino and BSM analyses.

Beyond their usual role as backgrounds, the availability of a high-energy (TeV scale), high-intensity flux of forward muons offers also novel physics opportunities, as highlighted in studies on muon-philic particle searches at FASER~\cite{Ariga:2023fjg, Batell:2024cdl, MammenAbraham:2025gai}.
In this work we explore the potential of TeV-scale muon DIS at FASER for QCD studies, in particular to scrutinise the charm content of the proton. 
We demonstrate that, even when restricted to the Run-3 luminosities, FASER$\nu$ will record more than $10^5$ inclusive muon DIS events of which up to $10^4$ will be charm-production events, a sufficiently high yield to provide competitive constraints.
To highlight this potential, we propose the measurement of charm production and its asymmetry in muon DIS at large Bjorken-$x$, a process which provides unparalleled sensitivity on the charm quark content of the proton~\cite{Brodsky:1980pb, Brodsky:2015fna, NNPDF:2023tyk}.
Our analysis reveals that Run 3 statistics enable a measurement of charm production that could disentangle between intrinsic and extrinsic calculations of the charm PDF, providing closure on the decades-long controversy concerning whether the EMC $F_2^c$ measurements in the early 1980s~\cite{Aubert:1982km} did or did not reveal intrinsic charm.
For completeness, we also provide predictions for event yields in inclusive and charm production for muon DIS in future FASER runs and for the proposed FASER$\nu$2 detector at the FPF.

The outline of this paper is as follows.
First in Sect.~\ref{sec:settings} we describe the settings of the simulation, including the calculation of the muon flux, the modelling of muon DIS and the associated final states, and the selection and acceptance cuts that are applied.
In Sect.~\ref{sec:inclusiveDIS} we apply our simulation pipeline to predictions for inclusive muon DIS at FASER, both for the current detector and for its upgrades in the HL-LHC era.
This pipeline is then applied to the case of charm production and its asymmetry in muon DIS in Sect.~\ref{sec:charmDIS}, where we demonstrate the sensitivity of these measurements to the (intrinsic) charm content of the proton.
Finally, we conclude in Sect.~\ref{sec:summary} and outline possible future directions.

%% file: sec-settings.tex
\section{Analysis settings}
\label{sec:settings}
\label{subsec:muonflux}
 \label{sec:muonDIS}

Here we describe the settings of the simulation, including the calculation of the muon flux, the modelling of muon DIS process and the associated final state, and the applied selection and acceptance cuts.

\paragraph{The muon flux.}
The muon flux reaching the FASER$\nu$ far-forward detector is modelled according to {\sc\small FLUKA} simulations~\cite{Battistoni:2015epi, Sabate-Gilarte:2023aeg}.
While part of this flux is constituted by high-energy muons produced directly in the primary pp collision at the ATLAS interaction point, there is also a sizeable secondary component arising from particle interactions with the LHC infrastructure. 
After production, the muons are propagated through the LHC infrastructure and magnetic fields, as well as through about 100~m of rock before reaching the FASER location.  
These simulations for the muon flux reaching FASER can be accurately validated independently of any scattering cross-section by measuring charged tracks in the spectrometer of the electronic detector. 
In this respect, and as opposed to the neutrino flux case, the muon flux reaching FASER has effectively negligible modelling uncertainties.

When evaluating the muon flux reaching FASER$\nu$, we consider two geometries: a rectangular area of $25\;{\rm cm}\times30\;\mathrm{cm}$ corresponding to the full size of FASER$\nu$, and a circular area with radius $r=9\;\mathrm{cm}$ corresponding to the overlap with the FASER spectrometer and which should be used in measurements which involve the latter.
In both cases, the axis of the FASER$\nu$ detector is taken to be located at $(x,y)=(1~{\rm cm},{\rm -3.3~{\rm cm}})$ in the transverse plane with respect to the nominal line of sight (LoS), matching the position of the detector at the beginning of Run 3~\cite{FASER:2024hoe}. 
Our calculation assumes an integrated luminosity of $\mathcal{L}_{\mathrm{pp}} = 250~\mathrm{fb}^{-1}$ which is the total accumulated luminosity expected by FASER$\nu$ by the end of Run 3.
While the muon flux actually reaching FASER$\nu$ changes every data-taking year, the associated differences can be neglected in this study and do not alter our qualitative findings. 
The length of the FASER$\nu$ tungsten detector is set to be $L_T = 50\;\mathrm{cm}$. 
This is not the full longitudinal size, since the measurement of the momentum of the incoming and outgoing muon via multiple Coulomb scattering requires the propagation through at least two interaction length of material up and downstream of the location of the muon interaction. Therefore, the following estimates are conservative, since we are not considering the full size of FASER$\nu$.

By consistency with the treatment of the muon fluxes in terms of a ``neutrino PDF'' in our previous studies~\cite{vanBeekveld:2024ziz, Cruz-Martinez:2023sdv}, $f_{\nu}(x_\nu)$ with $x_{\nu}=2E_\nu/\sqrt{s_{\rm pp}}$, here we normalise the incoming neutrino flux as follows
\be
\label{eq:muon_fluxes}
f_\mu(x_\mu) \equiv  n_T L_T \frac{dN_{\mu}}{dx_\mu}(x_\mu) 
\, ,\qquad  x_\mu \equiv \frac{2E_\mu}{\sqrt{s_{\mathrm{pp}}}}\, ,
\ee
with $x_{\mu}=1$ corresponding to the maximal energy kinematically allowed for a muon given the energy available in the primary pp collision, and $n_T$ is the tungsten target nucleon density.
The muon fluxes Eq.~(\ref{eq:muon_fluxes}) are interpolated separately for positively and negatively charged muons and stored in a {\sc\small LHAPDF} grid~\cite{Buckley:2014ana}, enabling its seamless integration within the {\sc\small POWHEG} event generation framework~\cite{Banfi:2023mhz, FerrarioRavasio:2024kem, vanBeekveld:2024ziz} used here.

\begin{figure}[t]
    \centering
\includegraphics[width=0.49\linewidth]{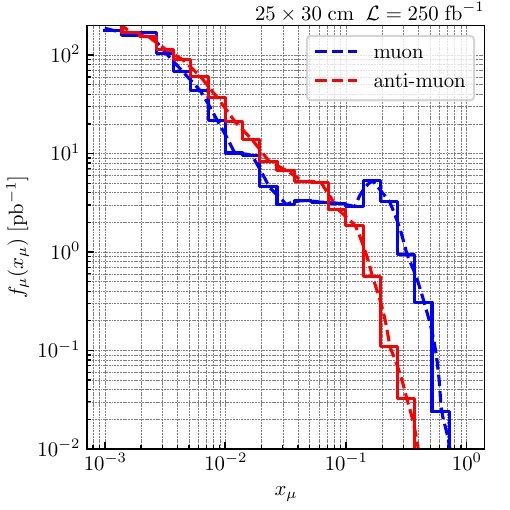}
\includegraphics[width=0.49\linewidth]{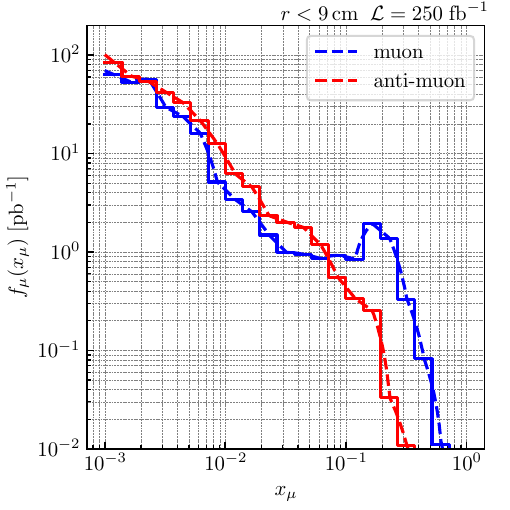}
    \caption{The (anti)muon fluxes for the rectangular geometry of FASER$\nu$ (left) and the circular geometry of the FASER spectrometer (right panel).
    An integrated luminosity of $\mathcal{L}_{\rm pp}=250$ fb$^{-1}$ is assumed.
    The dashed line indicates the interpolation provided by {\sc\small LHAPDF} and which is used as input to the {\sc\small POWHEG} simulations of muon DIS presented in this work.
    These fluxes are normalised according to the definition of Eq.~(\ref{eq:muon_fluxes}).
    }
    \label{fig:muonflux}
\end{figure}

Fig.~\ref{fig:muonflux} displays the muon and antimuon fluxes for rectangular and circular geometries and for an integrated luminosity of $\mathcal{L}_{\rm pp}=250$ fb$^{-1}$.
The histogram corresponds to the original {\sc\small FLUKA} calculation and the dashed continuous curve to the {\sc\small LHAPDF} interpolation. 
We note the large asymmetry muons and antimuons, induced by the LHC magnets, with the former dominating the large $x_\mu$ region.
As we show in Sect.~\ref{sec:inclusiveDIS}, muon scattering at FASER$\nu$ is dominated by the $Q^2\ll m_Z^2$ region and hence NC DIS mediated by a virtual photon does not distinguish between muons and antimuon: only knowledge of their sum is needed for event rate predictions, and no physical observable will be sensitive to a muon/antimuon asymmetry.

\paragraph{Muon DIS at FASER$\nu$.}
The muon fluxes reaching FASER$\nu$ displayed in Fig.~\ref{fig:muonflux} are interfaced to the POWHEG deep-inelastic scattering generator~\cite{Banfi:2023mhz, FerrarioRavasio:2024kem, vanBeekveld:2024ziz} to provide fully differential cross-sections at NLO accuracy in the QCD coupling.
Parton-level cross-sections are then interfaced to {\sc\small Pythia8}  \cite{Bierlich:2022pfr} for the showering and hadronisation, with the Monash 2013 tune~\cite{Skands:2014pea} used for the latter.
Event selection is carried out at the final-state level (after hadronisation).
Specifically, we simulate the inclusive process
\be
\mu^\pm + W \to \mu^\pm + X_h \, ,
\ee
which contains as a subset charm production events
\be
\mu^\pm + W \to \mu^\pm + c\,(\bar{c}) +\widetilde{X}_h \, ,
\ee
with $W$ indicating the tungsten target, out of which FASER$\nu$ is composed, and $X_h,\widetilde{X}_h$ stand for hadronic final states.
We assume that FASER$\nu$ will measure the incoming muon energy $E_\mu$, the outgoing muon energy $E'_\mu$, and the muon scattering angle $\theta_{\mu}$.
As well known, measurements of $\lp E_\mu,E'_\mu,\theta_{\mu}\rp$ fully specify the DIS kinematics, with complementary kinematic constraints provided by measurements of the energy $E_h$ and invariant mass $W$ of the hadronic final state.

To select muon DIS events, we restrict the outcome of the {\sc\small POWHEG+Pythia8} simulations to satisfy the following conditions and acceptance cuts:

\begin{itemize}
\item The energy of the final-state muon should satisfy $E'_\mu > 100 \;\mathrm{GeV}$.
The distribution of muon DIS events at FASER$\nu$ peaks at energies around $E_\mu\sim1$ TeV, and therefore one can verify that this cut in $E'_\mu$ has a limited impact on the total rates. 

\item We impose a cut in the number of charged tracks in the final state $n_{\mathrm{tr}}$.
Specifically, we consider that a hadronic charged track is recorded if the final-state has a charged particle with a tri-momentum of at least $|\vec{p}_{\rm tr}|=1\;\mathrm{GeV}$. 
We consider three cases, namely $n_{\rm tr}\ge 3,4,$ or $5$.
A cut $n_{\rm tr}\ge 5$ is required by the current interaction vertex finding algorithm, with improvements to lower this threshold being considered. 
Since muon DIS events at FASER$\nu$ are dominated by low-$Q^2$ scattering, the transfer of energy to the hadronic final state is moderate and a too stringent cut in $n_{\rm tr}$ sizeable reduces the selected sample, mainly in large-$x$ region, where we have the largest contribution of intrinsic charm. 
Here we present our baseline results with the requirement that $n_{\rm tr}\ge 3$.
This cut $n_{\rm tr}$ is highly correlated to the cut in $E_h$ applied in~\cite{vanBeekveld:2024ziz}.

\item In addition to the inclusive selection cuts, we are also interested to tag events with charm hadrons in the final state in order to probe charm production in muon DIS.
Tagging charm hadrons in an emulsion detector as FASER$\nu$ is facilitated by their characteristic displaced vertex signature.
Here we consider an efficiency $\epsilon_c=70\%$ for charm hadrons tagging~\cite{Aki2025}. 
Charm production events are therefore defined as those event satisfying fiducial cuts with at least one charm hadron tag.

\item No cuts on the scattered muon angle $\theta_\mu$ are applied, since NC DIS events are fully contained by the FASER$\nu$ detector volume. 

\item To ensure that the scattering cross-section can be reliable described by perturbative QCD, we impose $Q \ge 1.65$ GeV and $W\ge 2$ GeV, with $W$ being the invariant mass of the hadronic final state.

\end{itemize}

With these settings, one can evaluate the predicted number of neutral-current muon DIS events at FASER$\nu$ in the fiducial region.
These event yields are presented in Table~\ref{table:Nevents} summing over muon and antimuon scattering.
As in the rest of the paper, we take a baseline integrated luminosity of $\mathcal{L}_{\rm pp}=250$ fb$^{-1}$.
These predictions are based on the muon fluxes of Fig.~\ref{fig:muonflux} and the {\sc\small POWHEG+Pythia8} NLO simulations for the cross-section modelling.
The nuclear PDFs of tungsten are evaluated as 
\be
f_{i/N}(x,Q^2) = (Z\,f_{i/p}(x,Q^2) + (A - Z)\,f_{i/n}(x,Q^2))/A  \, ,\qquad i = q,\,\bar{q}, \,g \, ,
\ee
where $f_{i/N}$ represents the PDF of an average tungsten nucleon and $f_{i/p(n)}$ the PDF of a free proton (neutron), hence neglecting genuine nuclear modifications which are small in the FASER region~\cite{Feng:2022inv}. 
In this work we consider four NNLO  proton PDF sets: NNPDF4.0~\cite{NNPDF:2021njg}, which contains a fitted charm PDF as default, NNPDF4.0 with perturbative charm,
CT18 fitted charm~\cite{Guzzi:2022rca} with the BHPS3 model~\cite{Brodsky:1980pb} for $\Delta\chi^2=30$,  and the baseline CT18 set~\cite{Hou:2019efy} with perturbative charm.
For each PDF set, we provide predictions both for inclusive DIS and for charm production, in the latter case assuming a charm tagging efficiency of $\epsilon_c=70\%$ and where a single charm hadron tag is required. 

\begin{table}[t]
\centering
\renewcommand{\arraystretch}{1.5}
\begin{tabularx}{\textwidth}{Xllcc}
\toprule
\multicolumn{5}{c}{Muon DIS neutral-current event yields at FASER$\nu$ for $\mathcal{L}_{\rm pp}=250$ fb$^{-1}$}\\
\midrule
PDF set     & Charm PDF  & Process			                 & DIS cuts              & DIS + fiducial cuts\\	
\toprule
\multirow{2}{*}{NNPDF4.0} & \multirow{2}{*}{Fitted charm} &$\mu^\pm + W \rightarrow \mu^\pm + X$      & 2.7$\times10^{5}$         & 1.7$\times10^{5}$       \\
&& $\mu^\pm + W \rightarrow \mu^\pm + X + c(\bar{c})$  & 7.2$\times10^{3}$         & 5.8$\times10^{3}$ \\
\midrule
\multirow{2}{*}{NNPDF4.0~PC} & \multirow{2}{*}{Pert. charm} & $\mu^\pm + W \rightarrow \mu^\pm + X$      & 2.6$\times10^{5}$         & 1.8$\times10^{5}$      \\
& &$\mu^\pm + W \rightarrow \mu^\pm + X + c(\bar{c})$  & 7.5$\times10^{3}$         & 6.3$\times10^{3}$\\      
\midrule
\multirow{2}{*}{CT18 FC}     & \multirow{2}{*}{Fitted charm (BHPS3)} & $\mu^\pm + W \rightarrow \mu^\pm + X$      & 2.6$\times10^{5}$         & 1.8$\times10^{5}$        \\
& &$\mu^\pm + W \rightarrow \mu^\pm + X + c(\bar{c})$  & 1.6$\times10^{4}$         & 1.2$\times10^{4}$\\
\midrule
\multirow{2}{*}{CT18}     & \multirow{2}{*}{Pert. charm}& $\mu^\pm + W \rightarrow \mu^\pm + X$      & 2.6$\times10^{5}$         & 1.7$\times10^{5}$        \\
&& $\mu^\pm + W \rightarrow \mu^\pm + X + c(\bar{c})$  & 1.2$\times10^{4}$         & 9.5$\times10^{3}$\\
\bottomrule
\end{tabularx}
\vspace{0.3cm}
\caption{Predicted number of neutral-current muon DIS events at FASER$\nu$ in the fiducial detector region, summing over muon and antimuon scattering, and
assuming an integrated luminosity of $\mathcal{L}_{\rm pp}=250$ fb$^{-1}$.
The forward muon fluxes used in the calculation are shown in Fig.~\ref{fig:muonflux}  and the  cross-section modelling is based on {\sc\small POWHEG+Pythia8} NLO simulations.
We consider four NNLO PDF sets: from top to bottom, NNPDF4.0 (which by default includes a fitted charm PDF), NNPDF4.0 with perturbative charm, CT18 fitted charm with the BHPS3 model (for $\Delta\chi^2=30$), and the baseline CT18 set with perturbative charm.
For each PDF set, we provide predictions both for inclusive DIS and for charm production, in the latter case assuming a charm tagging efficiency of $\epsilon_c=70\%$.
We provide event yields after only DIS cuts are applied, and then after also the fiducial acceptance kinematic cuts applied.
}
\label{table:Nevents}
\end{table}

The results of Table~\ref{table:Nevents} indicate that large samples of muon DIS events, both inclusive and charm production, become available for physics analyses at FASER$\nu$ with the Run-3 luminosity ($\mathcal{L}_{\rm pp}=250$ fb$^{-1}$). 
The predicted inclusive rates are almost independent of the choice of PDF set, and indicate that around $2\times 10^{5}$ events are expected after cuts.
The effect of the fiducial cuts, when added to the baseline DIS cuts, is moderate, and lead to an event yield reduction by around a factor 30\%. 
In the case of charm production, the predicted integrated yields depend on the treatment of the charm PDF by the considered input PDF set, and up to $ 10^{4}$ events after cuts can be expected, a factor 20 reduction as compared to the inclusive rates. 
For charm production, the effects of the fiducial cuts are milder than in the case of inclusive DIS, since these events are on average more energetic and satisfy more easily the $n_{\rm tr}$ conditions. 
We discuss in Sects.~\ref{sec:inclusiveDIS} and~\ref{sec:charmDIS} how these inclusive rates are distributed across the relevant kinematic variables. 

The predictions for inclusive muon DIS event yields in Table~\ref{table:Nevents} can be compared with the number of muons detected in the pilot run for the emulsion detector~\cite{FASER:2021mtu}.
This FASER$\nu$ pilot detector employing emulsion films was exposed to 12.2 fb$^{-1}$, and its active mass was around factor 100 smaller as compared to the FASER$\nu$ volume.
This implies, by dividing the event yields of Table~\ref{table:Nevents} by a factor $2\times 10^3$, that this pilot emulsion detector was expected to detect around 83 muon scattering events, to be compared with the 78 observed charged vertices, most of which coming from muon interactions. 
Therefore, within $\mathcal{O}(1)$ factors, our {\sc\small POWHEG}-based predictions for the number of NC muon DIS events  at FASER$\nu$ are consistent with the findings of the pilot run detector.

%% file: sec-inclusiveDIS.tex
\section{Inclusive muon DIS}
\label{sec:inclusiveDIS}

By leveraging the simulation pipeline presented in Sect.~\ref{sec:settings}, we now take a closer look at inclusive neutral-current muon DIS, namely the process
\be
\mu^\pm + W \rightarrow \mu^\pm + X_h \, ,
\ee
subject to acceptance and selection cuts.
In the FASER$\nu$ kinematics, muon scattering satisfies $Q^2\ll m_Z^2$ and we can neglect $Z$-boson exchange contributions, which in turn means that the cross-section becomes insensitive to the muon charge.
Here here we always present results summing over muons and antimuons.

\begin{figure}[t]
    \centering
\includegraphics[width=0.49\linewidth]{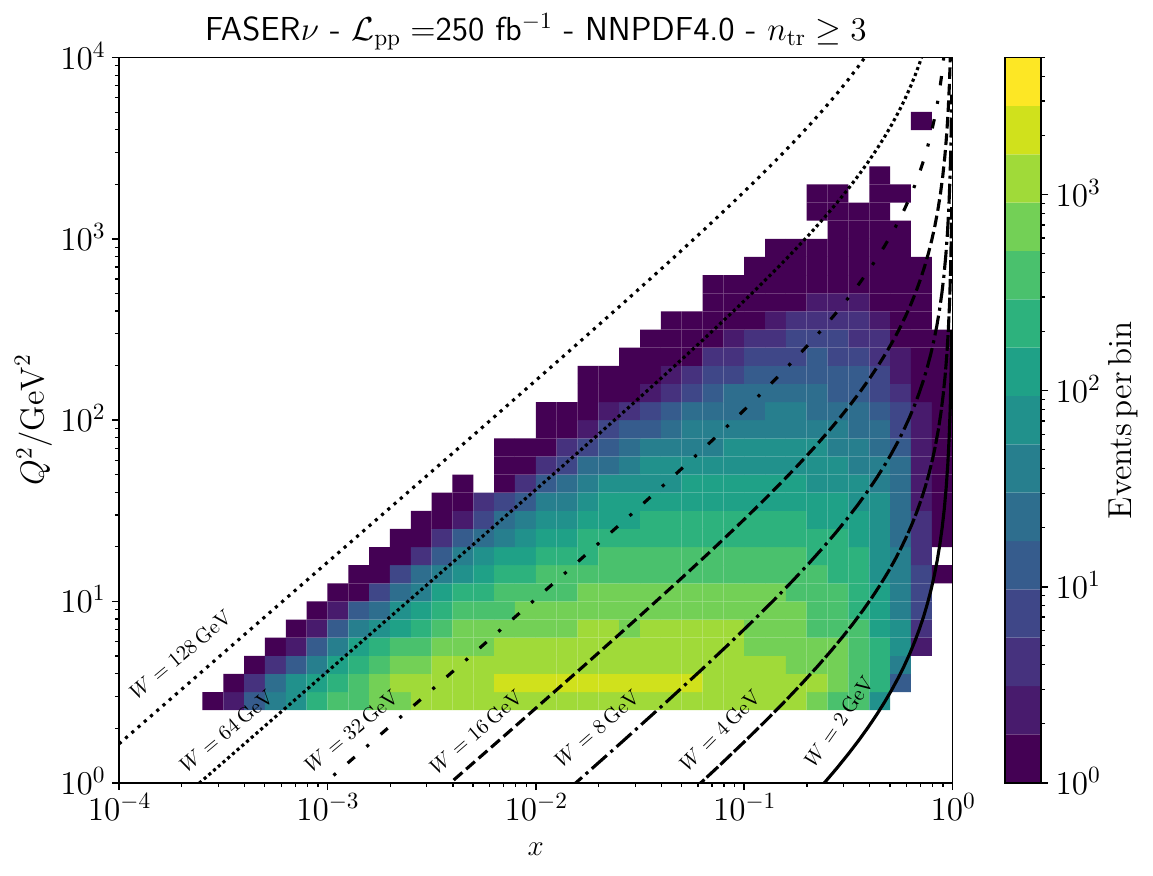}
\includegraphics[width=0.49\linewidth]{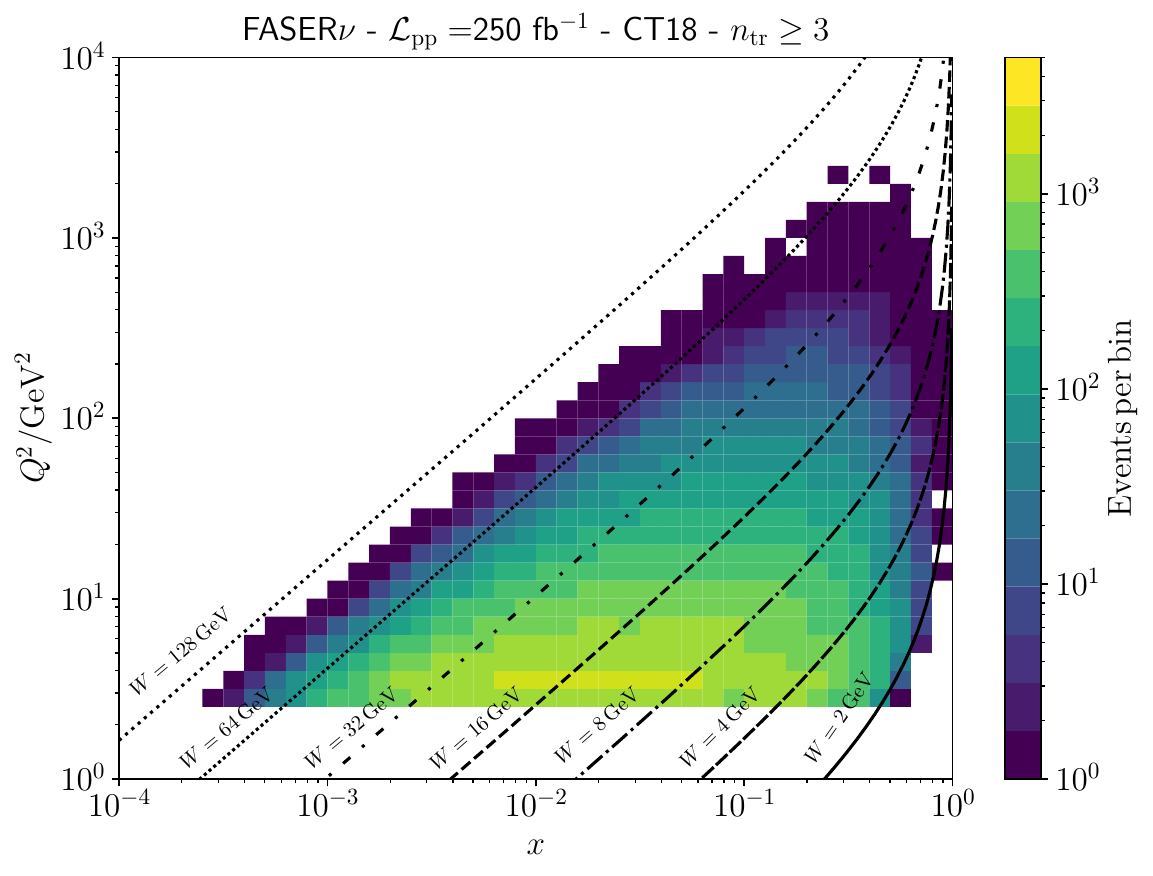}
\caption{The number of muon DIS neutral-current inclusive events predicted for FASER$\nu$ as a function of $Q^{2}$ and $x$ satisfying the acceptance and selection cuts described in Sect.~\ref{sec:muonDIS} for an integrated luminosity of $\mathcal{L}_{\rm pp}=250$ fb$^{-1}$.
We display results for the NNPDF4.0 (left) and CT18 (right) baseline sets, see also Table~\ref{table:Nevents} for the corresponding integrated  yields.
Similar results are obtained for other PDF sets.
The curves indicate fixed values of $W$.
}
\label{fig:DIStotal_Q_x}
\end{figure}

Fig.~\ref{fig:DIStotal_Q_x} displays the number of muon DIS neutral-current inclusive events at FASER$\nu$ satisfying the cuts described in Sect.~\ref{sec:muonDIS} in bins of $Q^{2}$ and $x$ and for an integrated luminosity of $\mathcal{L}_{\rm pp}=250$ fb$^{-1}$.
We display results for the NNPDF4.0 and CT18 baseline sets (hence with fitted and with perturbative charm respectively), see also Table~\ref{table:Nevents} for the associated total events yields.
The curves indicate fixed values of $W$, the invariant mass of the hadronic final state $X_h$.
From Fig.~\ref{fig:DIStotal_Q_x} we see that FASER$\nu$ at Run~3 covers the region $x \gsim 3\times 10^{-4}$ in the perturbative region $Q \ge 1.65$ GeV, and extends all the way until $x\sim 0.8$.
In terms of the coverage in $Q^2$, FASER$\nu$ reaches up to $Q^2_{\rm max}\sim 10^{3}$ GeV$^2$, justifying that $Z$-boson exchange contributions can be neglected.
The overall kinematic coverage of muon DIS at FASER$\nu$ is hence similar to that of neutrino DIS studied in~\cite{Cruz-Martinez:2023sdv}, which in turn mostly overlap with that of the EIC~\cite{AbdulKhalek:2021gbh}.
We conclude that the large energy of the muon flux reaching FASER$\nu$, up to several TeV, enables muon-nucleon collisions which in the CoM reference frame are comparable with those of the EIC.

The kinematic coverage of Fig.~\ref{fig:DIStotal_Q_x} also indicates that the bulk of neutral current scattering events to be detected at FASER$\nu$ are located at small $Q^2$ and large $x$.
The dominance of the low-$Q^2$ region follows from the neutral-current nature of muon DIS, where $\gamma^*$ exchange between the leptonic and hadronic states leads to a $1/Q^4$ suppression of the cross-section due to the massless propagator.
This is at odds with neutrino DIS, where $W$-boson exchange implies that the bulk of event have associated higher $Q^2$ values.
This difference between muon and neutrino DIS at FASER$\nu$ has consequences for event tagging and reconstruction, with the neutrino DIS leading to hadronic final states with on average higher energies than in the muon case. 

Fig.~\ref{fig:DIStotal_Q_x} also displays curves with constant value of $W$, the invariant mass of the hadronic final state
\be
\label{eq:W_def}
W^2 = m_p^2 + \frac{Q^2(1-x)}{x} \, ,
\ee
where recall that we impose a selection cut of $W^2\ge 4$ GeV$^2$.
Here we evaluate $W$ in terms of $Q^2$ and $x$ using Eq.~(\ref{eq:W_def}), rather than directly in terms of adding up the four-momenta of the hadronic final state particles.
We find that the majority of the events are composed by hadronic states of relatively low mass, and that a moderate cut of $W \ge 8 $ GeV would remove the bulk of the events.
This finding indicates that a careful analysis of the low-$Q^2$, large-$x$ region, both from the experimental and theoretical viewpoints, is essential to fully exploit the potential of muon DIS at FASER.
Therefore, this kinematic region, dominated by high-$x$ and low-$Q^2$ (thus low-$W$), is more sensitive to non-perturbative effects such as target mass corrections and higher-twist contributions \cite{Accardi:2011fa,VIRCHAUX1992221}. 
These effects must be properly understood and quantified before the corresponding data can be reliably included in global PDF fits.

\begin{figure}[t]
\centering    \includegraphics[width=0.49\linewidth]{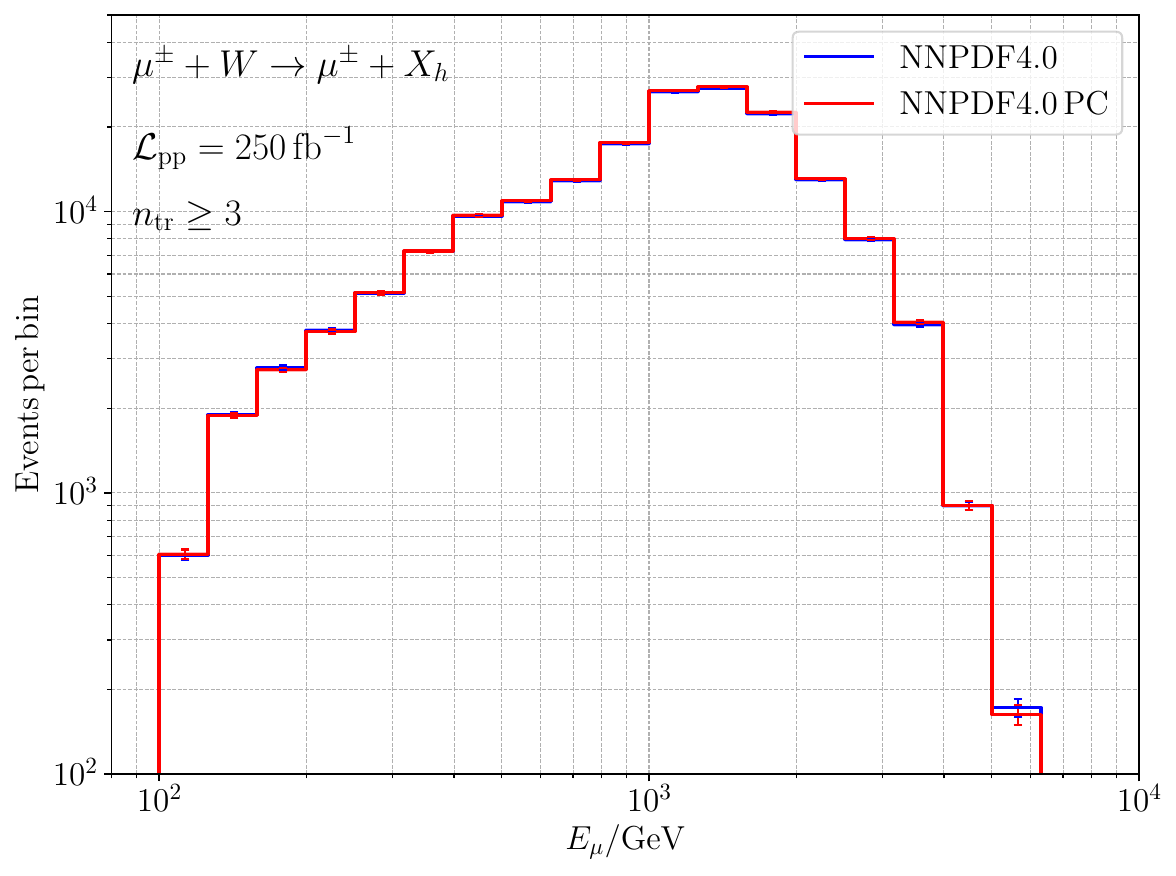}
\includegraphics[width=0.49\linewidth]{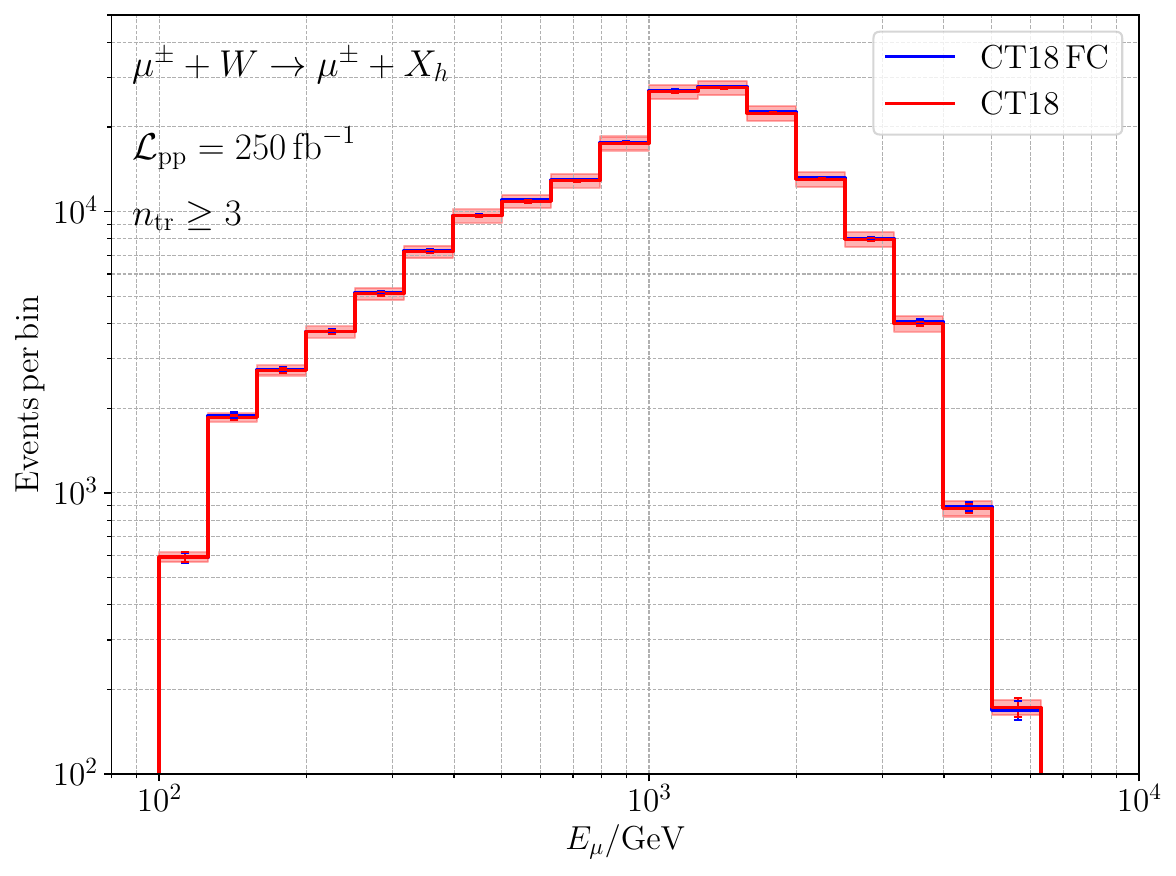} \\
    \caption{The number of inclusive muon DIS events at FASER$\nu$ for $\mathcal{L}_{\rm pp}=250$ fb$^{-1}$ as a function of the incoming muon energy $E_\mu$.
    The left (right) panel shows the predictions based on NNPDF4.0 (CT18) for both the fitted and perturbative charm variants.
    The error bars (bands) in each bin indicates the associated statistical (PDF) uncertainty.
}    
\label{fig:DIStotal_cuts_E_distrib}
\end{figure}

Binning the muon inclusive DIS events shown in Fig.~\ref{fig:DIStotal_Q_x} in terms of the energy of the incoming muon, $E_\mu$, leads to the distributions of Fig.~\ref{fig:DIStotal_cuts_E_distrib}.
The left panel shows the predictions based on NNPDF4.0 (fitted and perturbative charm variants), while the right panel displays the same for the CT18 sets, see Table~\ref{table:Nevents} for the associated total number of muon DIS events in each case.
The error bars in each bin indicates the associated statistical uncertainty in each bin.
The muon energy distribution of Fig.~\ref{fig:DIStotal_cuts_E_distrib} shows that muon DIS events at FASER$\nu$ are dominated by energies $E_\mu$ between 1 and 2 TeV, with a steep fall off of the cross section for higher and lower energies.
For $E_\mu\lsim 100$ GeV the event rates become negligible, while for the high energy tails we find that even for muons with $E_\mu\sim 4$ TeV one expects around $10^3$ inclusive events.
Therefore muon DIS at FASER$\nu$ is dominated by TeV-energy muons, extending by an order of magnitude the coverage of previous muon DIS (lab-based) 
experiments. 
From Fig.~\ref{fig:DIStotal_cuts_E_distrib} we also see that, consistently with Table~\ref{table:Nevents}, the $E_\mu$ distribution is stable upon variations of the input PDF set, and in particular that the treatment of the charm PDF does not modify the event yields. 

\begin{figure}[t]
    \centering
\includegraphics[width=0.49\linewidth]{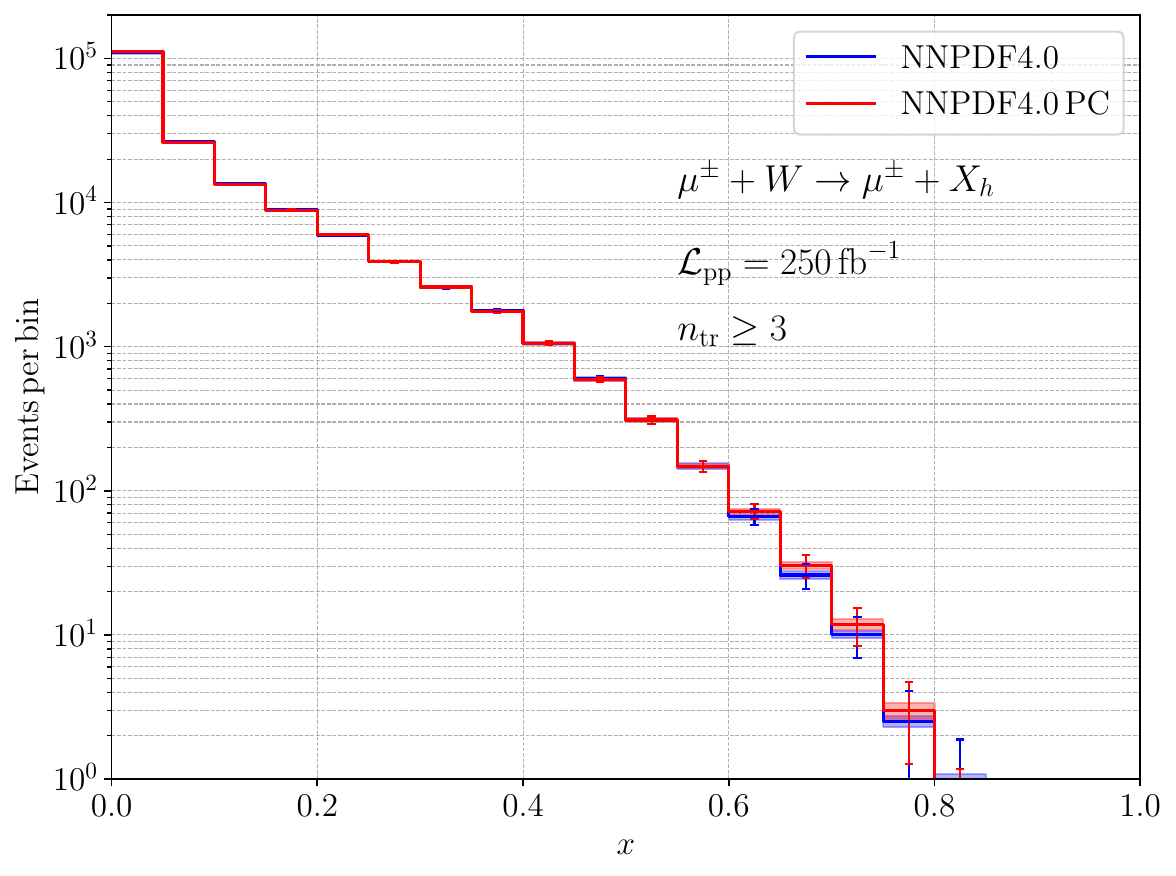}
\includegraphics[width=0.49\linewidth]{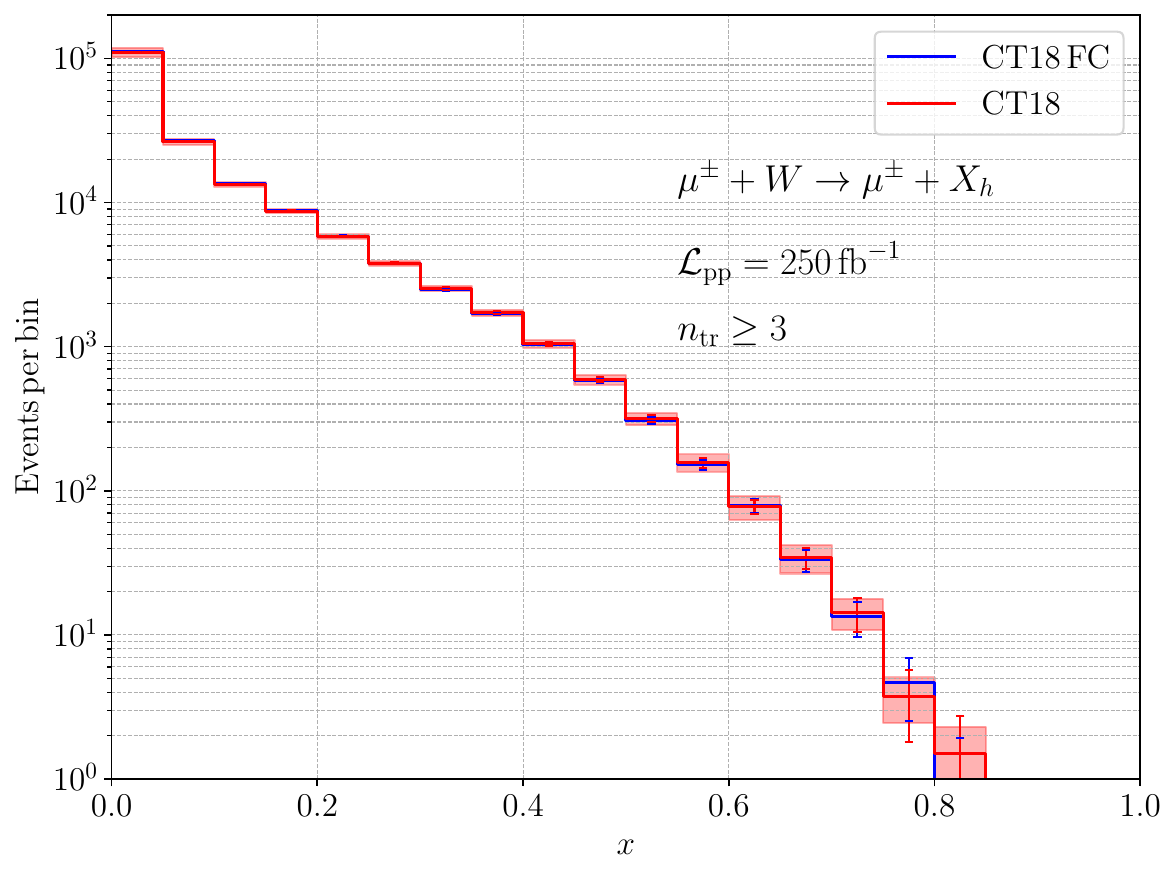} \\
    \caption{Same as Fig.~\ref{fig:DIStotal_cuts_E_distrib} now with events binned in terms of $x$. 
    }
\label{fig:DIStotal_cuts_x}
\end{figure}

The same events as in Fig.~\ref{fig:DIStotal_cuts_E_distrib} now binned in $x$ are displayed in Fig.~\ref{fig:DIStotal_cuts_x}.
This distribution is interesting due to its direct correlation with the large-$x$ quark PDFs.
In the large-$x$ region, the distribution of event yields falls off steeply with $x$, following the associated fall off of the underlying PDFs~\cite{Ball:2016spl}.
Nevertheless, events are expected up to $x\sim 0.8$, a region where PDF uncertainties are large and that is essential for searches of new heavy particles at the LHC~\cite{Ball:2022qtp}.
As in the $E_\mu$ case, for inclusive scattering there are no major differences between the predictions of the four PDF sets considered. 
In Sect.~\ref{sec:charmDIS} we demonstrate that this distribution, when restricted to charm production events, is  sensitive to the charm content of the proton.

\begin{figure}[t]
    \centering
\includegraphics[width=0.49\linewidth]{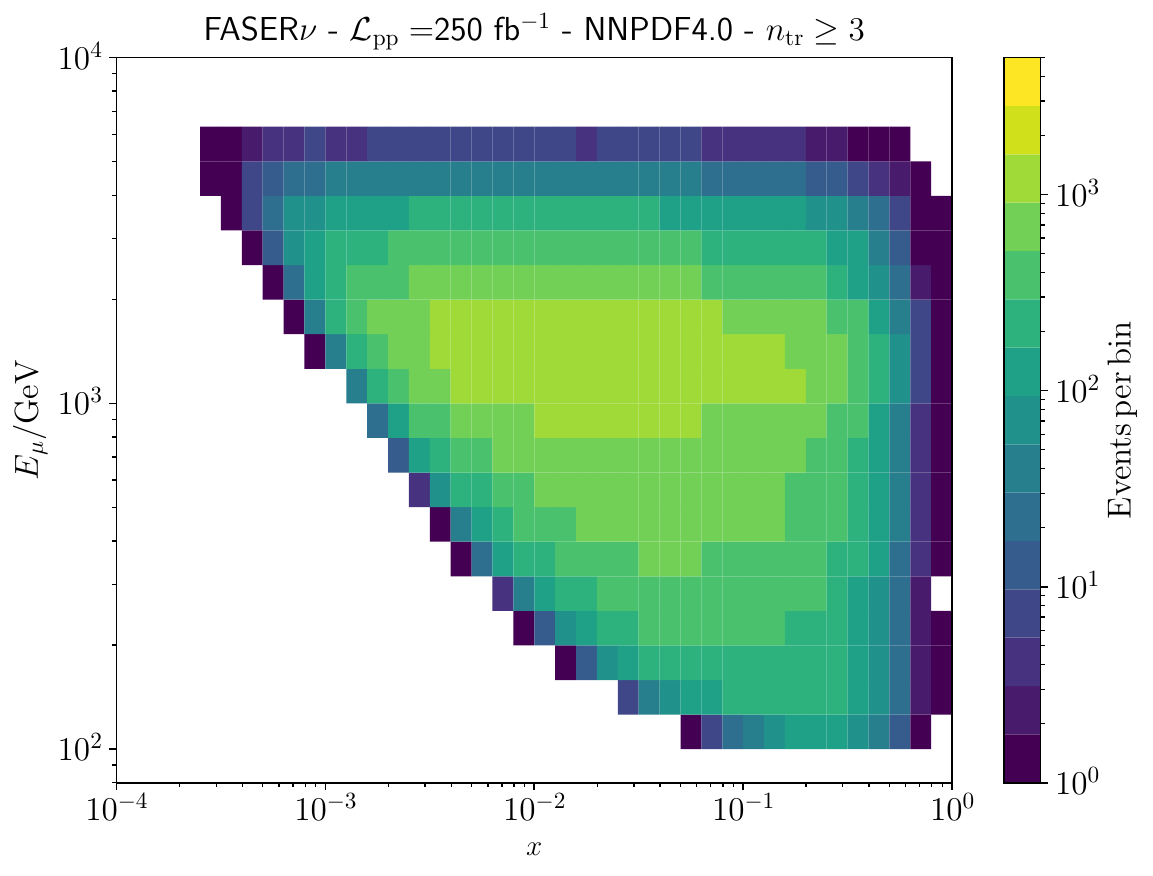}
\includegraphics[width=0.49\linewidth]{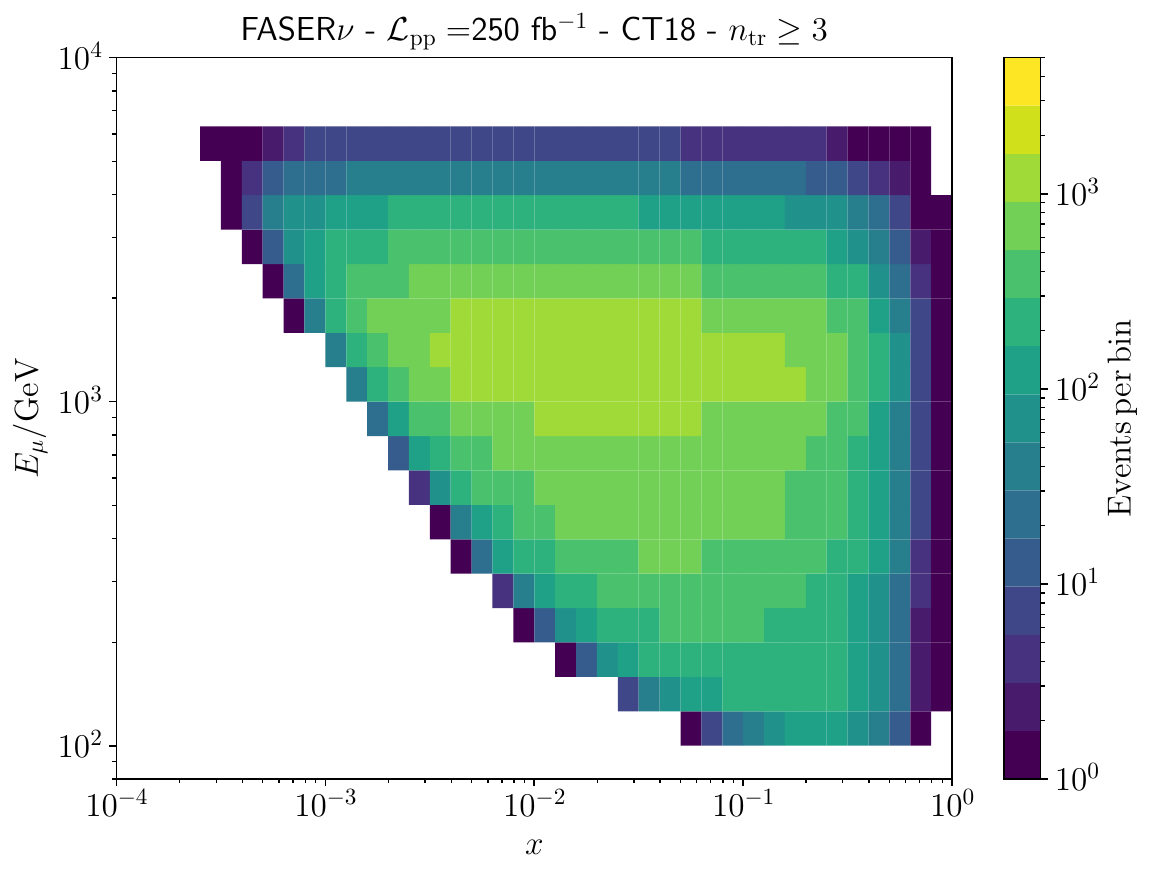} 
    \caption{Same as Fig.~\ref{fig:DIStotal_Q_x} with the selected events binned in $(x,E_\mu)$. }
    \label{fig:DIStotal_correlation_E_x}
\end{figure}

Figs.~\ref{fig:DIStotal_cuts_E_distrib} and~\ref{fig:DIStotal_cuts_x} display the differential distributions of the event yields as a function of $E_\mu$ and $x$ respectively.
To highlight the correlation between the two kinematic variables, Fig.~\ref{fig:DIStotal_correlation_E_x} presents a similar comparison as that of Fig.~\ref{fig:DIStotal_Q_x} now with selected events binned in the $(x,E_\mu)$ plane.
One finds that in general there is a weak correlation between the two variables, indicating that some features of the underlying cross-section may be accessible when measuring the $x$ distribution but not when considering the $E_\mu$ one.
In particular, muons with $E_\mu\sim 1~{\rm TeV}$, which dominate the cross-section, probe a wide range of $x$ values peaking around $x\sim 0.05$.
Events with large-$x$, which are particularly useful to probe the poorly known large-$x$ PDFs, arise from muons with energies spanning the whole kinematically allowed range.
Fig.~\ref{fig:DIStotal_correlation_E_x} highlights the importance of reconstructing the DIS variables $(x,Q^2)$ from the event kinematics, since more directly measurable variables such as $E_\mu$ are less sensitive to the underlying proton structure. 

\paragraph{Predictions for the HL-LHC.}
Table~\ref{table:NeventsFASERnuHL-LHC} provides the same information as Table~\ref{table:Nevents} for FASER$\nu$ and FASER$\nu$2 at the HL-LHC, assuming a total integrated luminosity of $\mathcal{L}_{\rm pp}=3$ ab$^{-1}$~\cite{Azzi:2019yne}.
Results correspond to selected events which satisfy both DIS and fiducial cuts.
For FASER$\nu$2 the length of the tungsten detector is set
to $L_T=6.20$ m.
The event yields grow proportionally to the integrated luminosity and the active volume.
For FASER$\nu$2 at the FPF, our projections indicate up to $3\times 10^7$ inclusive and $2\times 10^6$ charm production events.
These large event yields enable precise muon DIS measurements within a broad kinematic region including, multi-differential measurements. 

\begin{table}[t]
\centering
\renewcommand{\arraystretch}{1.5}
\begin{tabularx}{\textwidth}{XXlcc}
\toprule
\multicolumn{5}{c}{Muon DIS neutral-current event yields at FASER$\nu$ and FASER$\nu$2 for $\mathcal{L}_{\rm pp}=3$ ab$^{-1}$}\\
\midrule
PDF set                     & Charm PDF  & Process			                 & FASER$\nu$             & FASER$\nu$2 \\	
\toprule
\multirow{2}{*}{NNPDF4.0}   & \multirow{2}{*}{Fitted charm}        &$\mu^\pm + W \rightarrow \mu^\pm + X$             & 2.0$\times10^{6}$  
&
2.5$\times10^{7}$ 
\\
& & $\mu^\pm + W \rightarrow \mu^\pm + X + c(\bar{c})$              & 7.0$\times10^{4}$ 
&
8.6$\times10^{5}$
\\
\midrule
\multirow{2}{*}{NNPDF4.0PC} & \multirow{2}{*}{Pert. charm}         & $\mu^\pm + W \rightarrow \mu^\pm + X$             & 2.1$\times10^{6}$  
&
2.6$\times10^{7}$
\\
& &$\mu^\pm + W \rightarrow \mu^\pm + X + c(\bar{c})$              & 7.6$\times10^{4}$ 
&
9.4$\times10^{5}$
\\      
\midrule
\multirow{2}{*}{CT18 FC}    & \multirow{2}{*}{Fitted charm (BHPS3)} & $\mu^\pm + W \rightarrow \mu^\pm + X$      & 2.1$\times10^{6}$ 
&
2.6$\times10^{7}$ 
\\
& &$\mu^\pm + W \rightarrow \mu^\pm + X + c(\bar{c})$              &  1.4$\times10^{5}$
&
1.7$\times10^{6}$
\\
\midrule
\multirow{2}{*}{CT18}       & \multirow{2}{*}{Pert. charm}& $\mu^\pm + W \rightarrow \mu^\pm + X$              & 2.0$\times10^{6}$ & 
2.5$\times10^{7}$ 
\\
&& $\mu^\pm + W \rightarrow \mu^\pm + X + c(\bar{c})$              & 1.1$\times10^{5}$
&
1.4$\times10^{6}$
\\
\bottomrule
\end{tabularx}
\vspace{0.3cm}
\caption{Same as Table~\ref{table:Nevents} for FASER$\nu$ and FASER$\nu$2 at the HL-LHC.
Results correspond to selected events which satisfy both DIS and fiducial cuts.
For FASER$\nu$2 the length of the tungsten detector is set
to $L_T=6.20$ m.
}
\label{table:NeventsFASERnuHL-LHC}
\end{table}

\paragraph{Impact of systematic uncertainties.} 
The study of muon DIS at FASER$\nu$ relies on the measured values of the incoming muon momentum $E_\mu$, the outgoing muon momentum $E'_\mu$, and scattering angle of the latter $\theta_\mu$. 
Muon momenta are measured via the multiple Coulomb scattering method~\cite{Kodama:2007mw}. 
This method uses that charged particles passing through matter are scattered, with the displacement due to scattering following a normal distribution of mean zero and width proportional to the inverse of the particle's momentum. 
Benefiting form the precise position resolution of about 300~nm~\cite{FASER:2025qaf}, the first FASER$\nu$ analysis reported a muon momentum resolution of about 30\% at 200~GeV and
50\% at higher energies~\cite{FASER:2024hoe}.
Dedicated efforts are expected to lead to improvements in future analyses. 
The angular resolution of tracks in the FASER$\nu$ detector is about 0.05~mrad~\cite{FASER:2025qaf}, and can be considered negligible for the purpose of this study. 
We will estimate in Sect.~\ref{sec:charmDIS} for the case of charm production the impact that systematic errors in the measurement of muon momenta may have on the interpretation of muon DIS measurements at FASER$\nu$.
In this work we consider both a conservative muon momentum resolution of 30\% and a more optimistic resolution of 10\%, showing that the latter  strongly helps in isolating the intrinsic charm signal.

\paragraph{Integration into global (n)PDF fits.}
The results presented in this section indicate that a large sample of muon DIS inclusive events with a broad coverage in $x$ and $Q^2$ will be available at FASER$\nu$ as well as at its future upgrades. 
As compared to previous muon DIS measurements in fixed-target experiments, FASER$\nu$ extends the coverage both at small-$x$ and at large-$x$, the latter region specially relevant for BSM searches at the LHC.
Quantifying the impact of muon DIS at FASER$\nu$ in global fits of proton and nuclear PDFs is possible by following the procedure of~\cite{Cruz-Martinez:2023sdv} to estimate the foreseen precision of structure function determinations and then use the {\sc\small EKO}~\cite{Candido:2022tld}, {\sc\small YADISM}~\cite{Candido:2024rkr}, and {\sc\small PineAPPL}~\cite{Carrazza:2020gss} codes to generate fast interpolating grids enabling the inclusion of these projections in (n)PDF fits, as done for the EIC in~\cite{AbdulKhalek:2021gbh, Khalek:2021ulf}.
Such a more detailed study is left for future work and here we restrict ourselves to demonstrate the impact of muon DIS at FASER$\nu$ to constrain the charm content of the proton in the large-$x$ region. 

%% file: sec-charmDIS.tex
\section{Charm production in muon DIS}
\label{sec:charmDIS}

The discussion in Sect.~\ref{sec:inclusiveDIS} has focused on inclusive DIS, which integrates over all the particles of the hadronic final state.
A unique advantage of emulsion detectors such as FASER$\nu$ is the availability to tag charm-hadrons through their secondary decay vertex, which enables the measurement of charm production cross-sections.
This is different as compared to e.g. charm production measurements in inclusive DIS at HERA~\cite{H1:2018flt} where charm-hadron production had to be reconstructed from exclusive decay models.
Furthermore, HERA measurements of $F_2^c$~\cite{H1:2018flt} had excellent coverage of the  small-$x$ region but could not access the large-$x$ one.
The only measurement of $F_2^c$ in the large-$x$ region available to date is the EMC analysis of~\cite{Aubert:1982km}, which claimed evidence for intrinsic charm but that has been contested ever since (for a review see~\cite{Brodsky:2015fna}).

\subsection{Charm PDF overview}
\label{sec:charm_pdf_treatment}

To facilitate the interpretation of the charm production in muon DIS, we first review the treatment of the charm PDF for the four sets used in Sect.~\ref{sec:inclusiveDIS} together with additional variants. 
From the NNPDF4.0 family, we consider three NNLO fit variants: with perturbative charm, the baseline with symmetric fitted charm~\cite{NNPDF:2021njg,Ball:2022qks}, and the variant enabling an asymmetry between the fitted charm and anticharm PDFs and presented in~\cite{NNPDF:2023tyk}.
In all cases, PDF uncertainties are computed from the variance over the $N_{\rm rep}=100$ replicas of the PDF set.
From the CT18 family, we consider the baseline NNLO fit~\cite{Hou:2019efy} based on a perturbative charm PDF as well as the CT18 FC variants of~\cite{Guzzi:2022rca} with a fitted charm PDF based on different model assumptions~\cite{Brodsky:1980pb, Hobbs:2017fom, Hobbs:2013bia}. 
In particular, the realizations of the meson-baryon model based on confining (MBMC) and effective-mass (MBME) quark models allow for a difference between the fitted charm and anticharm PDF, in the same manner as their NNPDF4.0 counterpart of~\cite{NNPDF:2023tyk}.
In Ref.~\cite{Guzzi:2022rca}, CT18 FC variants for different values of the $\Delta\chi^2$ threshold are provided, and here we consider the $\Delta\chi^2=30$ results to match with the standard tolerance criterion of the CT18 analysis at the 68\% CL corresponding to the upper limit in the fitted charm content. 
Note that for the CT18 FC variants of~\cite{Guzzi:2022rca} no PDF error sets are made available. 

\begin{figure}[t]
    \centering
\includegraphics[width=0.99\linewidth]{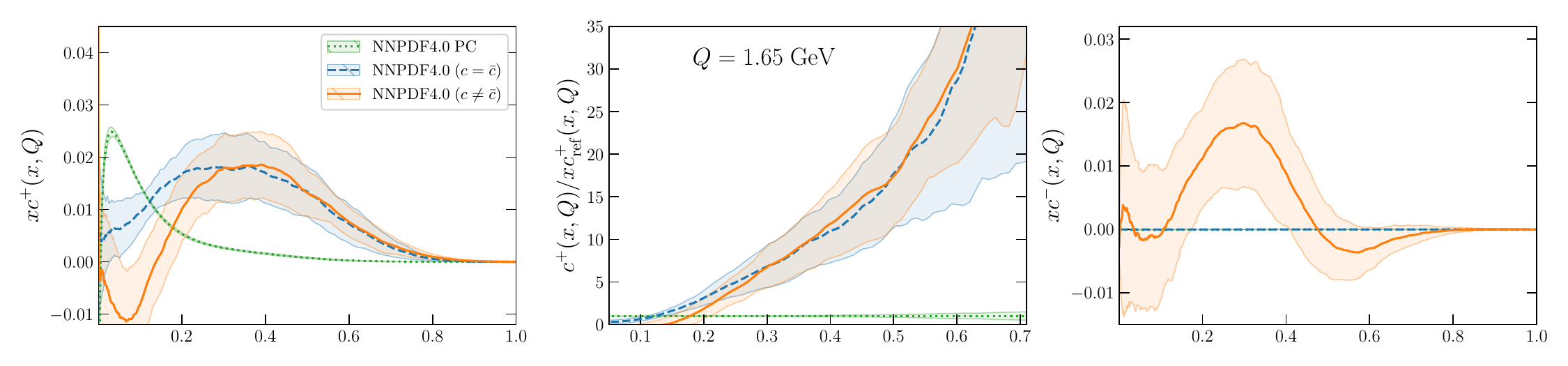}
\includegraphics[width=0.99\linewidth]{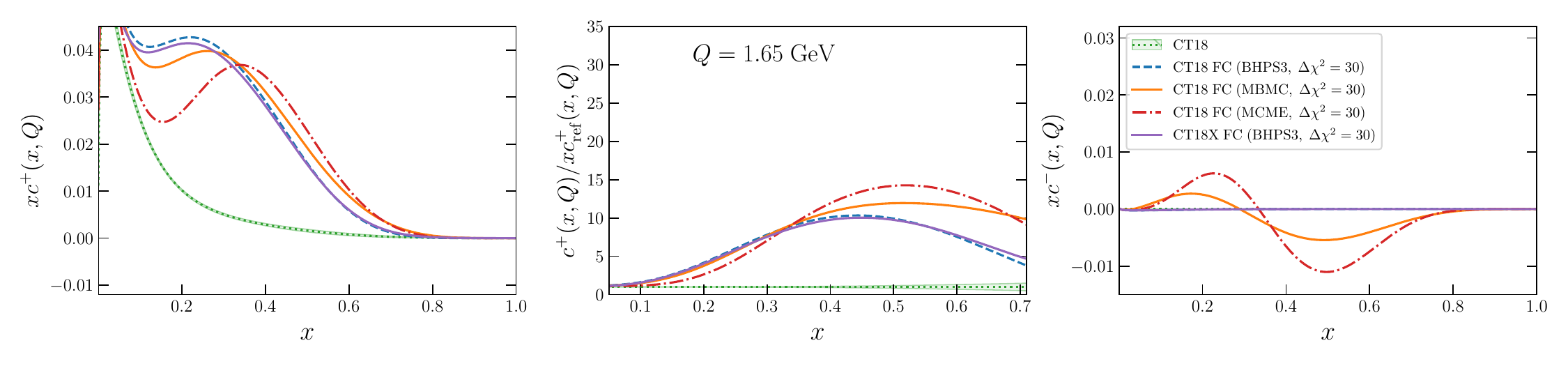}
    \caption{Top: comparison of three NNPDF4.0 NNLO fit variants: with perturbative charm, the baseline with (symmetric) fitted charm~\cite{NNPDF:2021njg}, and the variant enabling an asymmetry between the fitted charm and anticharm PDFs~\cite{NNPDF:2023tyk}. From left to right we show the total charm PDF $xc^+$ in absolute value and normalised to the perturbative charm fit, and then the valence charm PDF $xc^-$.
    Bottom: same for the CT18 NNLO baseline set (with perturbative charm PDF) and the four CT18 FC variants presented in~\cite{Guzzi:2022rca} for $\Delta\chi^2=30$ (matching the standard CT18 tolerance).
    PDF uncertainties are displayed as 68\% CL intervals, and for the CT18 FC variants no error PDF sets are provided. 
    }
    \label{fig:charm-PDFs}
\end{figure}

The top panel of Fig.~\ref{fig:charm-PDFs} displays a comparison of the three NNPDF4.0 sets for the total and valence charm PDFs, $xc^\pm(x,Q^2)= x\lp c\pm \bar{c}\rp(x,Q^2)$.
The bottom panel shows the same results for the CT18 (FC) sets.
The qualitative behaviour of the NNPDF4.0 and CT18 sets with fitted charm is similar for $xc^+$, though in general the CT18 determinations prefers a somewhat larger charm PDF.
For both groups, the fitted charm PDF peaks between $x=0.3$ and $x=0.4$ and then decreases for larger $x$ values. 
Concerning the charm valence asymmetry $xc^-$, the CT18 FC MBMC and MCME variants exhibit a similar trend as the NNPDF4.0 fit (positive asymmetry for $x\sim 0.2$, negative one for $x\sim 0.5$) though the magnitude of the deviations from the symmetric fitted charm scenario is quite different in both PDF sets (nevertheless in agreement within the PDF uncertainties of NNPDF4.0).

The central panels of Fig.~\ref{fig:charm-PDFs} display the ratio between the total charm PDF $xc^+$ in the various fits considered and the perturbative charm baseline.
This ratio highlights the enhancement of the charm PDF if a non-perturbative component is allowed and constrained from the data, with respect to the scenario where the charm PDF is entirely generated by QCD radiative corrections. 
As well known, while differences between perturbative and fitted charm PDFs are small for $x\lsim 0.1$, they grow rapidly with $x$, with an enhancement by a factor 10 for $x\sim 0.4$ both for NNPDF4.0 and for CT18, and by a factor 30 for $x\sim 0.6$ in NNPDF4.0 although PDF uncertainties are significant.
For CT18, the fitted charm enhancement peaks between $x\sim 0.4$ and 0.5 depending on the model and then decreases for larger $x$ values.
Given that charm production cross-sections in NC muon DIS are at leading order directly proportional to the charm PDF, one expects to find for FASER$\nu$ a qualitatively similar enhancement as those found in Fig.~\ref{fig:charm-PDFs} when comparing predictions binned in $x$ based on PDFs with fitted and with perturbative charm.

\subsection{Charm production event yields}
\label{sec:total_charm_production}

Next we revisit the analysis of Sect.~\ref{sec:inclusiveDIS} for charm production cross-sections.
As discussed in Sect.~\ref{sec:muonDIS}, muon DIS neutral current events are tagged as charm production if they contain at least one charmed hadron in the final state passing acceptance cuts, and a value for charm tagging efficiency of $\epsilon_c=0.7$ is adopted.
As indicated in Table~\ref{table:Nevents}, the charm tagging requirement leads to a reduction of the total event yields as compared to the inclusive rate by more than a factor 20.
This suppression is partially explained by the fact that the initial state has zero net charm quark number and hence one expects at least one $c\bar{c}$ pair in the final state.
This implies that most events will include two charm hadrons, and since muon DIS at FASER$\nu$ is dominated at low $W$ values (cfr. Fig.~\ref{fig:DIStotal_Q_x}) their production is kinematically disfavoured. 

\begin{figure}[t]
    \centering
\includegraphics[width=0.49\linewidth]{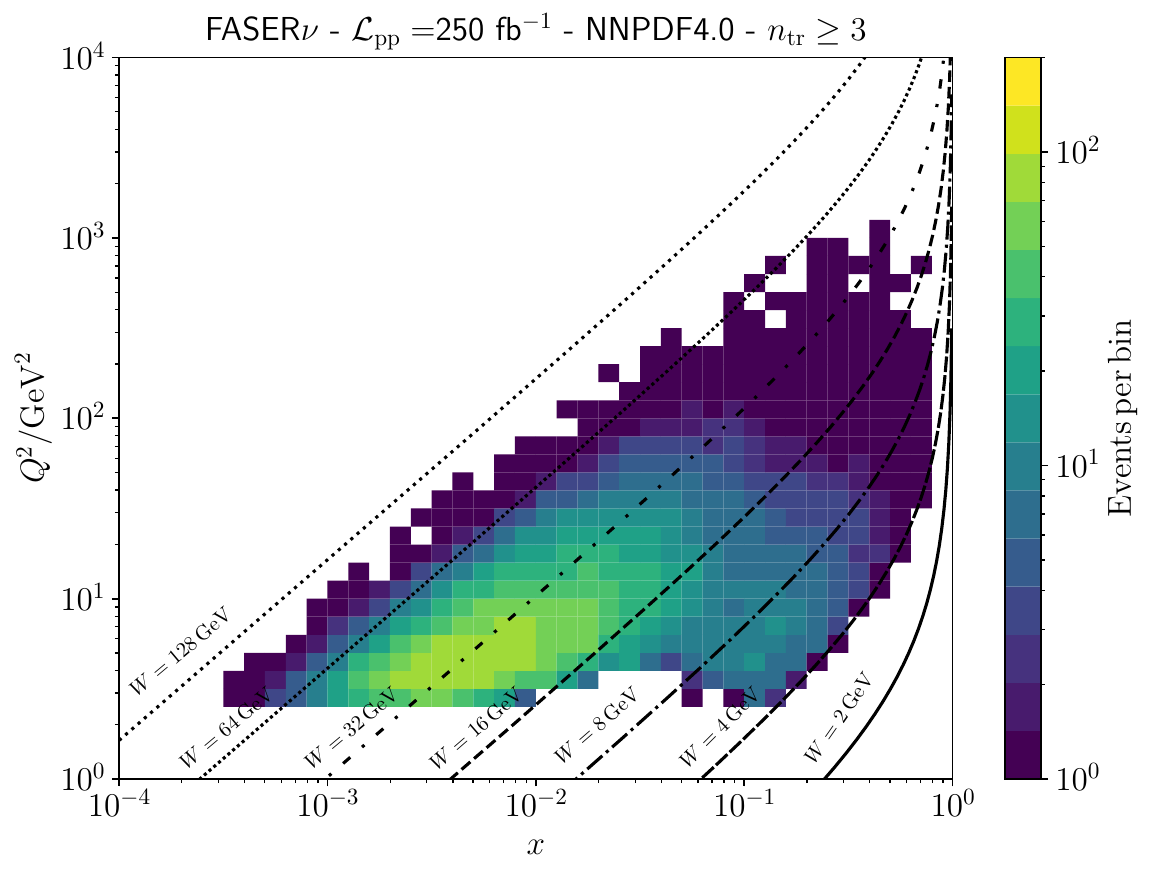}
\includegraphics[width=0.49\linewidth]{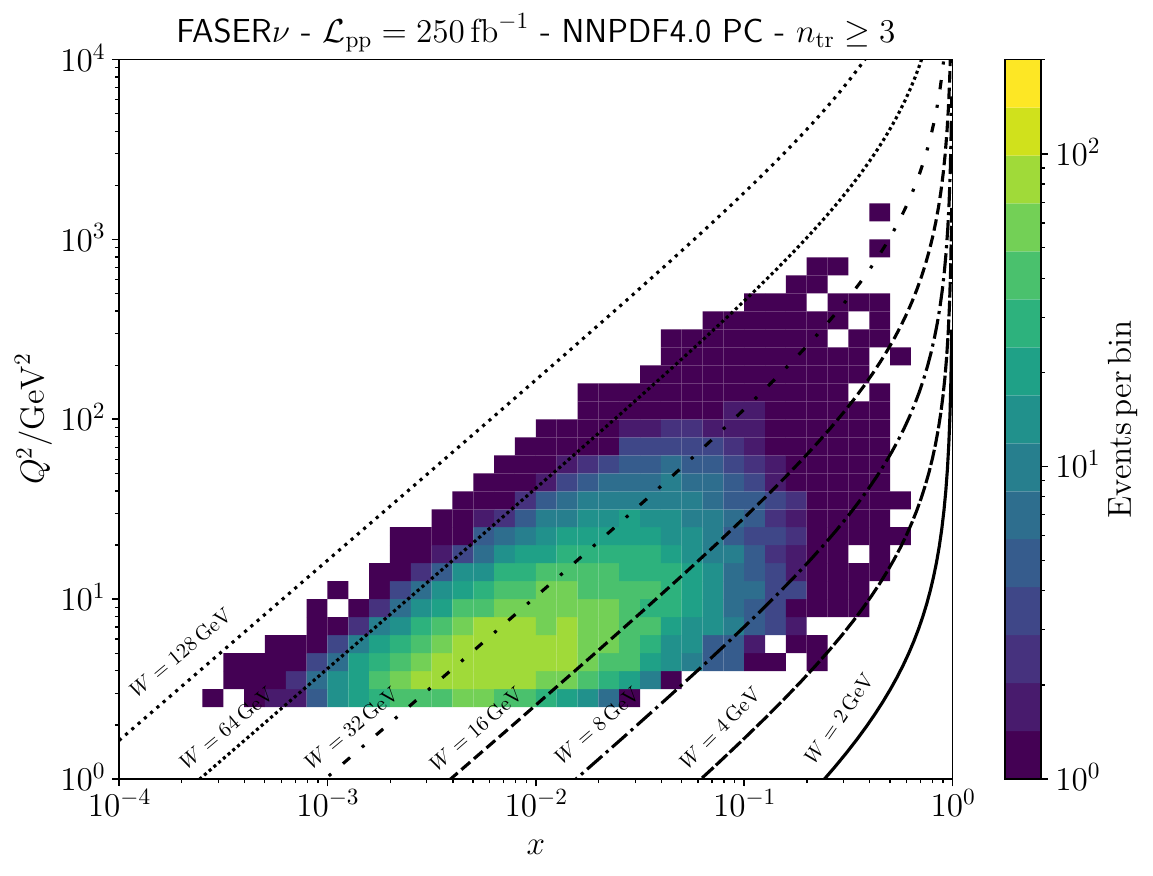}
    \caption{Same as Fig.~\ref{fig:DIStotal_Q_x} for charm production events, $\mu^\pm + W \rightarrow \mu^\pm + X + c(\bar{c})$.
    The left (right) panel display results based on NNPDF4.0 NNLO with fitted (perturbative) charm PDFs.  }
    \label{fig:DIStotal_charm_Q_x}
\end{figure}

Fig.~\ref{fig:DIStotal_charm_Q_x} displays the same predictions as Fig.~\ref{fig:DIStotal_Q_x} now for charm production events, $\mu^\pm + W \rightarrow \mu^\pm + X + c(\bar{c})$, comparing NNPDF4.0 (which has fitted charm PDF as default) with its counterpart based on perturbative charm.
As mentioned above, a single charm hadron tag is required, and we sum over the quark charge. 
As compared to inclusive production, we observe a strong suppression of events with $W\lsim 4$ GeV, a kinematic consequence of the  presence of two charmed hadrons in the final state. 
This suppression of the event yields at low $W^2$ is less marked for the fitted charm PDFs, and nevertheless sensitivity to the large-$x$ region is retained through events with larger $Q^2$ values.
By comparing the left and the right panels of Fig.~\ref{fig:DIStotal_charm_Q_x}, we observe for the baseline NNPDF4.0 predictions an increase of events in the large-$x$ region, which is the sought-for signal for intrinsic charm.

\begin{figure}[t]
    \centering
\includegraphics[width=0.49\linewidth]{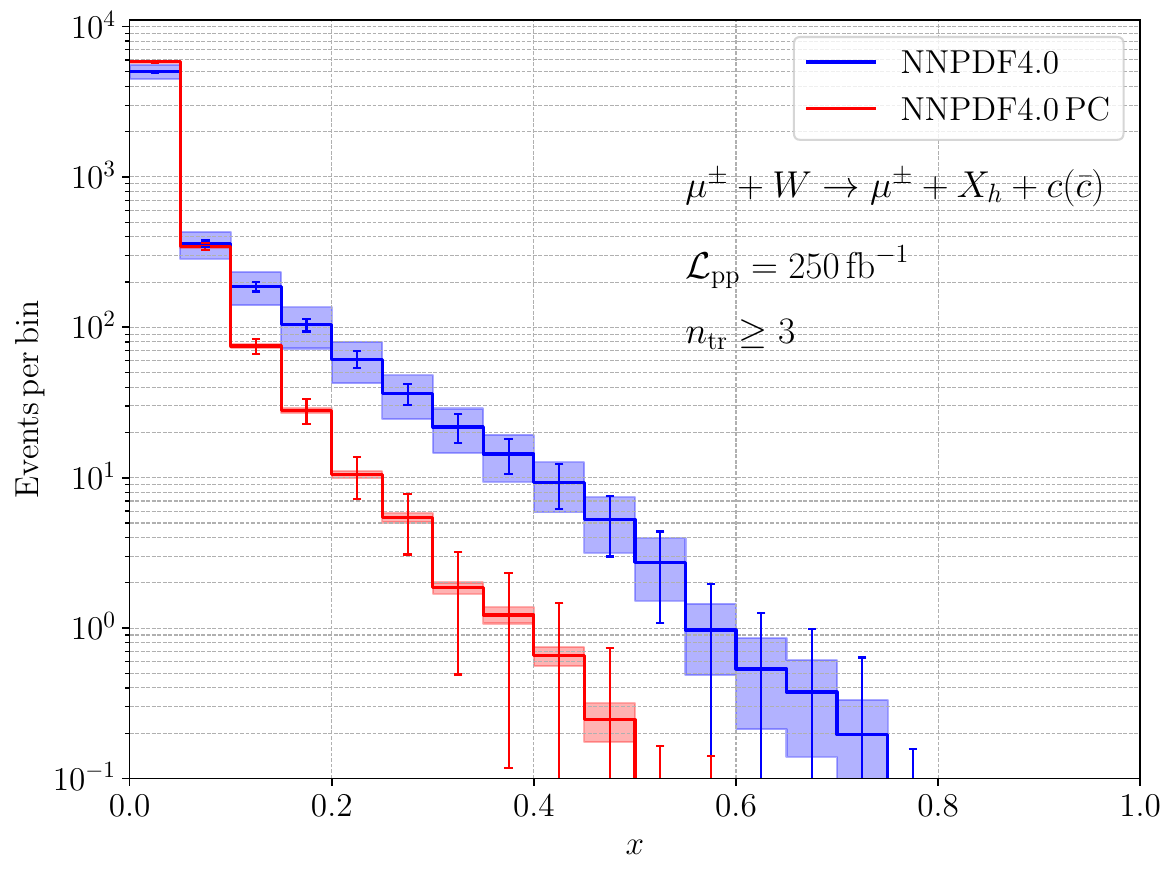}
\includegraphics[width=0.49\linewidth]{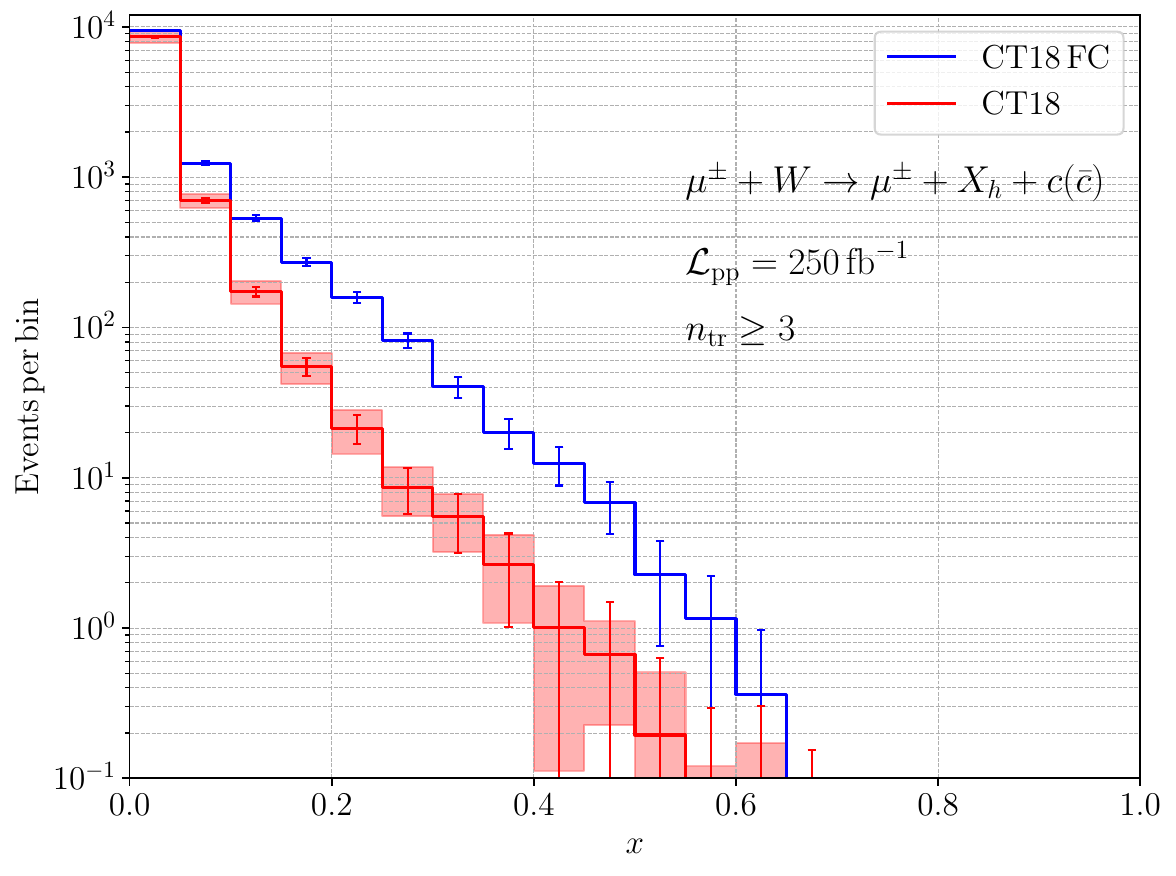}
    \caption{Same as Fig.~\ref{fig:DIStotal_cuts_x}  for charm.
   The error bands indicate the 68\% CL PDF uncertainties, which are not available for the CT18 FC set (BHPS3) displayed here.
   The increase in event yields for the PDF sets with fitted charm as compared to their variants with perturbative charm is consistent with the associated PDF comparison of Fig.~ \ref{fig:charm-PDFs}.
    }
\label{fig:DIStotal_cuts_E_statBands}
\end{figure}

Fig.~\ref{fig:DIStotal_cuts_E_statBands}
shows the same comparison as in Fig.~\ref{fig:DIStotal_cuts_x} in the case of charm production, and is obtained from Fig.~\ref{fig:DIStotal_charm_Q_x} after integrating over all kinematically allowed $Q^2$ values.
The error bands correspond to the PDF uncertainties while the error bars indicate the expected statistical uncertainties.
For the CT18 FC sets no error estimate is provided.
Related information is contained in
Table~\ref{table:Nevents_charm}, with the charm production event yields for two kinematic cuts in $x$, namely $x\ge 0.2$, and $x\ge 0.4$. 
The numbers in parenthesis indicate the ratio between the events yields in the fitted and perturbative charm PDF calculations.

From Fig.~\ref{fig:DIStotal_cuts_E_statBands} and Table~\ref{table:Nevents_charm} one observes for both the NNPDF4.0 and CT18 PDF sets a marked excess in the event yields at large-$x$ in the predictions based on fitted charm PDFs as compared to the perturbative charm counterparts.
In the region $x\gsim 0.2$, where the intrinsic charm component already dominates over the perturbative one, NNPDF4.0~(CT18 FC) predicts that FASER$\nu$ will observe around 150~(320) charm production events, to be compared with the expectations based on the perturbative charm variant of around 20~(40) events.
This increase by a factor 8 is consistent with the corresponding PDF-level comparison shown in the middle panels of Fig.~\ref{fig:charm-PDFs}. 
For $x\gsim 0.4$,  NNPDF4.0 (CT18 FC) predicts an increase in the charm production rates by factor 20~(12) as compared to the perturbative charm baseline.

\begin{table}[t]
\centering
\renewcommand{\arraystretch}{1.5}
\begin{tabularx}{\textwidth}{XXlll}
\toprule
\multicolumn{5}{c}{Charm-tagged muon DIS neutral-current at FASER$\nu$ for $\mathcal{L}_{\rm pp}=250$ fb$^{-1}$}\\
\midrule
PDF set    & Charm PDF           & $x>0$                    & $x\ge 0.2$  & $x\ge 0.4$  \\	
\toprule
NNPDF4.0   & Fitted charm        & 5.8$\times10^{3}$ (0.9) & 153.2 (7.7) & 19.3 (19) \\
\midrule
NNPDF4.0~PC & Pert. charm         & 6.3$\times10^{3}$ (1.0)    & 20.0 (1.0)    & 1.0 (1.0)     \\ 
\midrule
CT18 FC    & Fitted charm (BHPS3) & 1.2$\times10^{4}$ (1.2)  & 324.3 (8.1) & 23.0 (12) \\
\midrule
CT18       & Pert. charm         & 9.5$\times10^{3}$ (1.0)    & 40.2 (1.0)    & 2.0 (1.0)     \\
\bottomrule
\end{tabularx}
\vspace{0.3cm}
\caption{Same as Table~\ref{table:Nevents} for charm production with additional kinematic cuts in $x$ to highlight the differences between theoretical predictions based on fitted and perturbative charm PDFs. 
We consider $x>0$ (last column of Table~\ref{table:Nevents}), $x\ge 0.2$, and $x\ge 0.4$.
The numbers in parenthesis indicate the ratio between the number of events in the fitted and perturbative charm PDF calculations.
}
\label{table:Nevents_charm}
\end{table}

\begin{figure}[t]
    \centering
\includegraphics[width=0.49\linewidth]{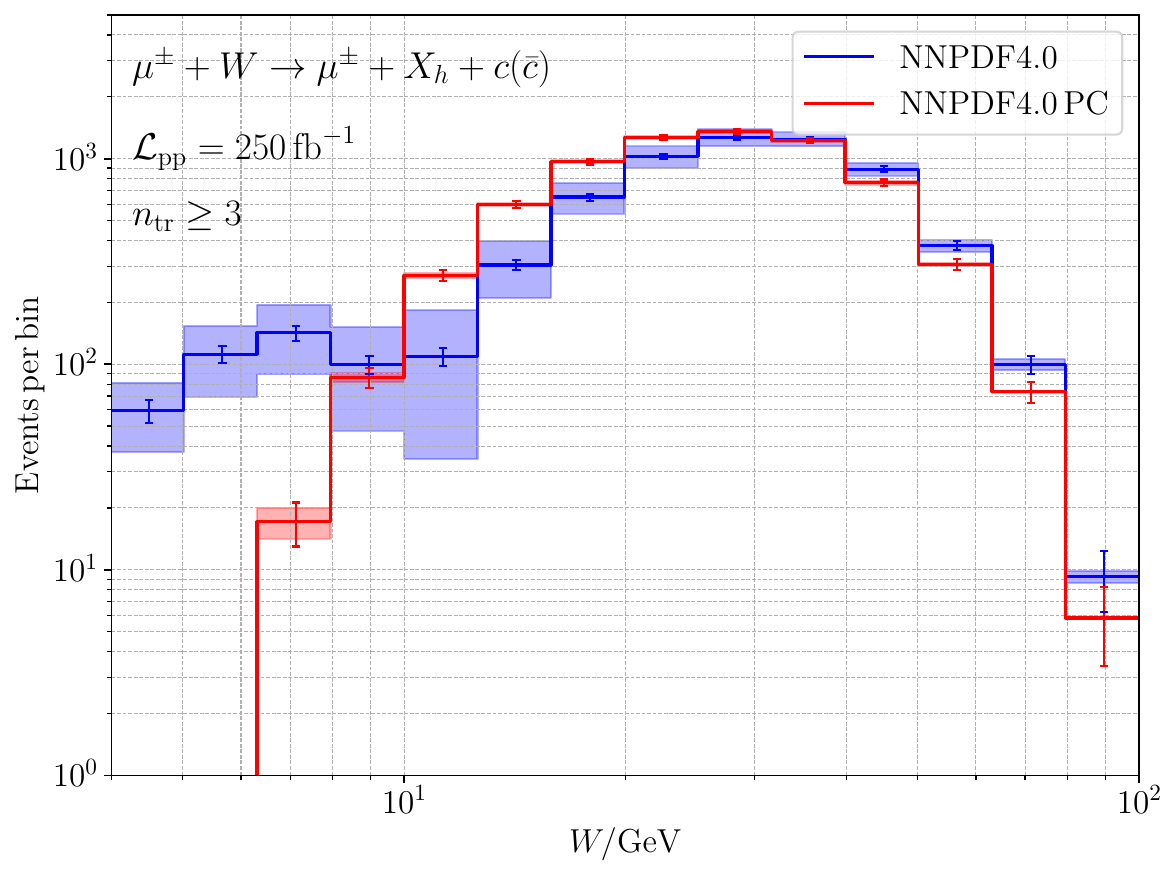}
\includegraphics[width=0.49\linewidth]{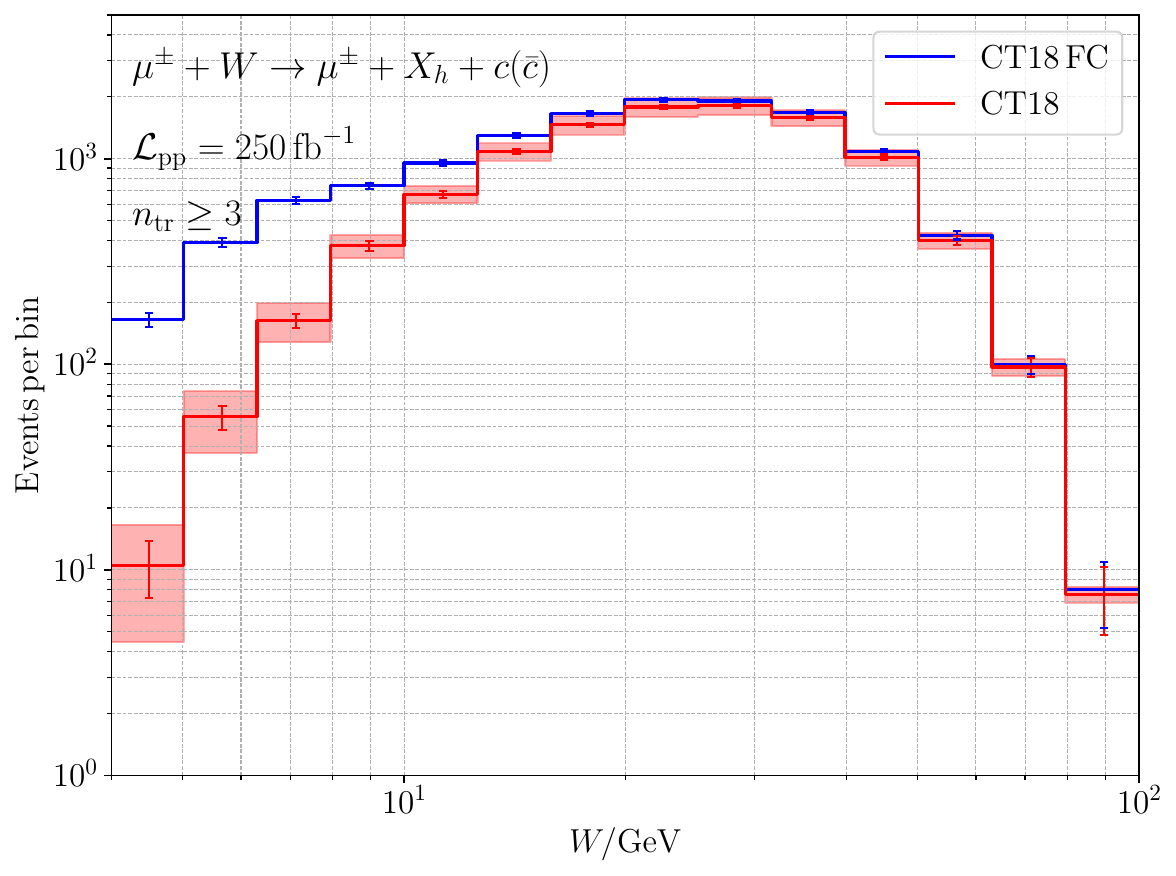}
    \caption{Same as Fig.~\ref{fig:DIStotal_cuts_E_statBands} now with events binned in terms of $W$, the invariant mass of the hadronic final state.
    }
\label{fig:DIScharm_cuts_W}
\end{figure}

In addition to the $x$ distribution, differences between theoretical predictions based on fitted and perturbative charm are also visible at the level of event yields binned in $W$, the invariant mass of the hadronic final state, as shown in Fig.~\ref{fig:DIScharm_cuts_W}.
Both for the NNPDF4.0 and CT18 predictions, the signal of a fitted charm PDF is an increase of the event yields in the low $W$ region with $W\lsim 10$ GeV where the perturbative charm predictions are suppressed. 
Hence this distribution is also sensitive to the charm content of the proton and provides a complementary signature to the $x$ dependence.
For NNPDF4.0, differences between results based on fitted and perturbative charm remain for all $W$ values since as shown by the upper left panel of Fig.~\ref{fig:charm-PDFs} (see also~\cite{Ball:2022qks}) the two charm PDFs only coincide at small-$x$ and large-$Q^2$.
Instead, in the CT18 FC approach, the intrinsic component is added on top of the perturbative contribution and by construction the fitted charm PDF coincides with its perturbative counterpart for $x\lsim 0.1$.

\begin{figure}[t]
    \centering
\includegraphics[width=0.49\linewidth]{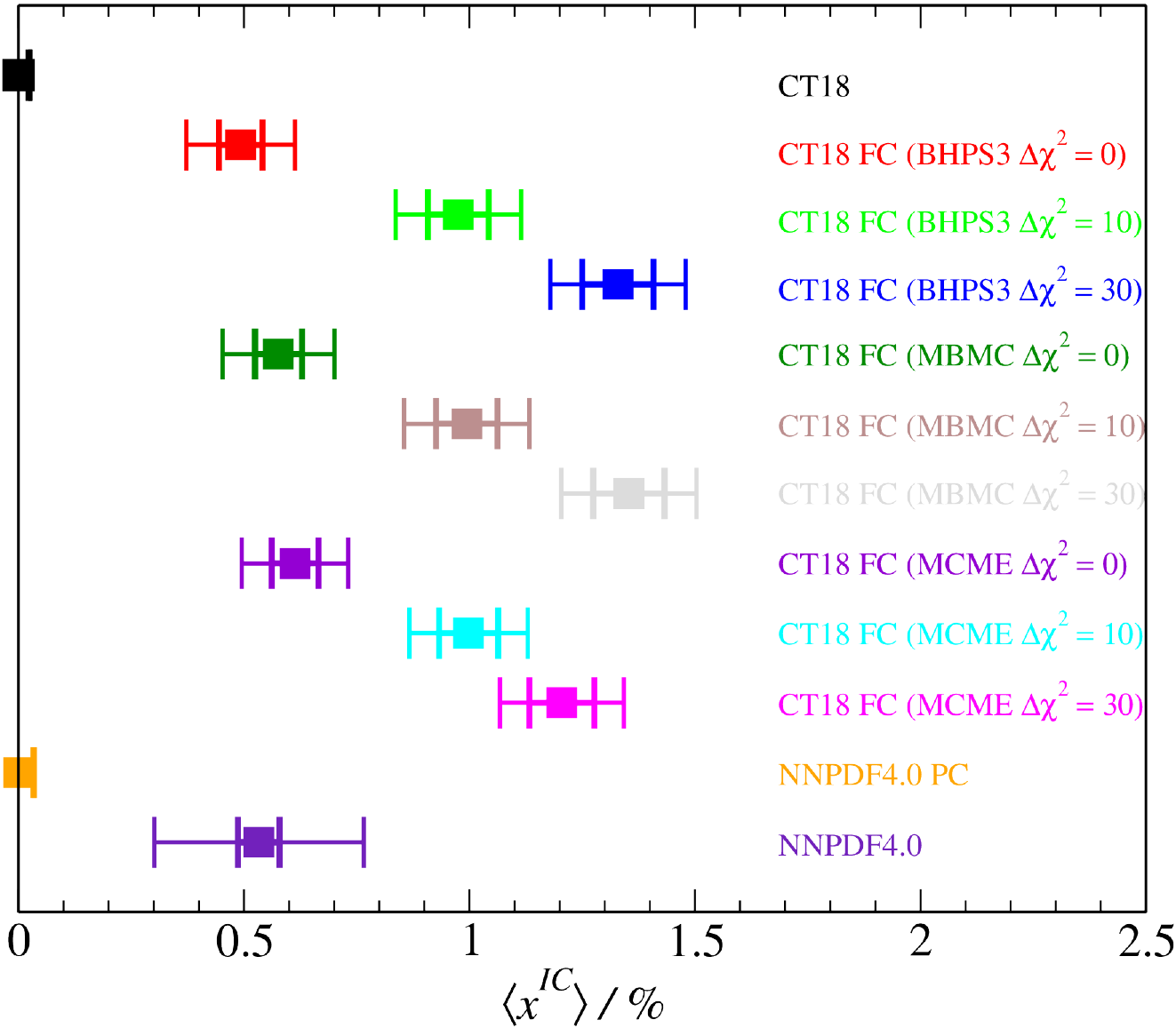}
\includegraphics[width=0.49\linewidth]{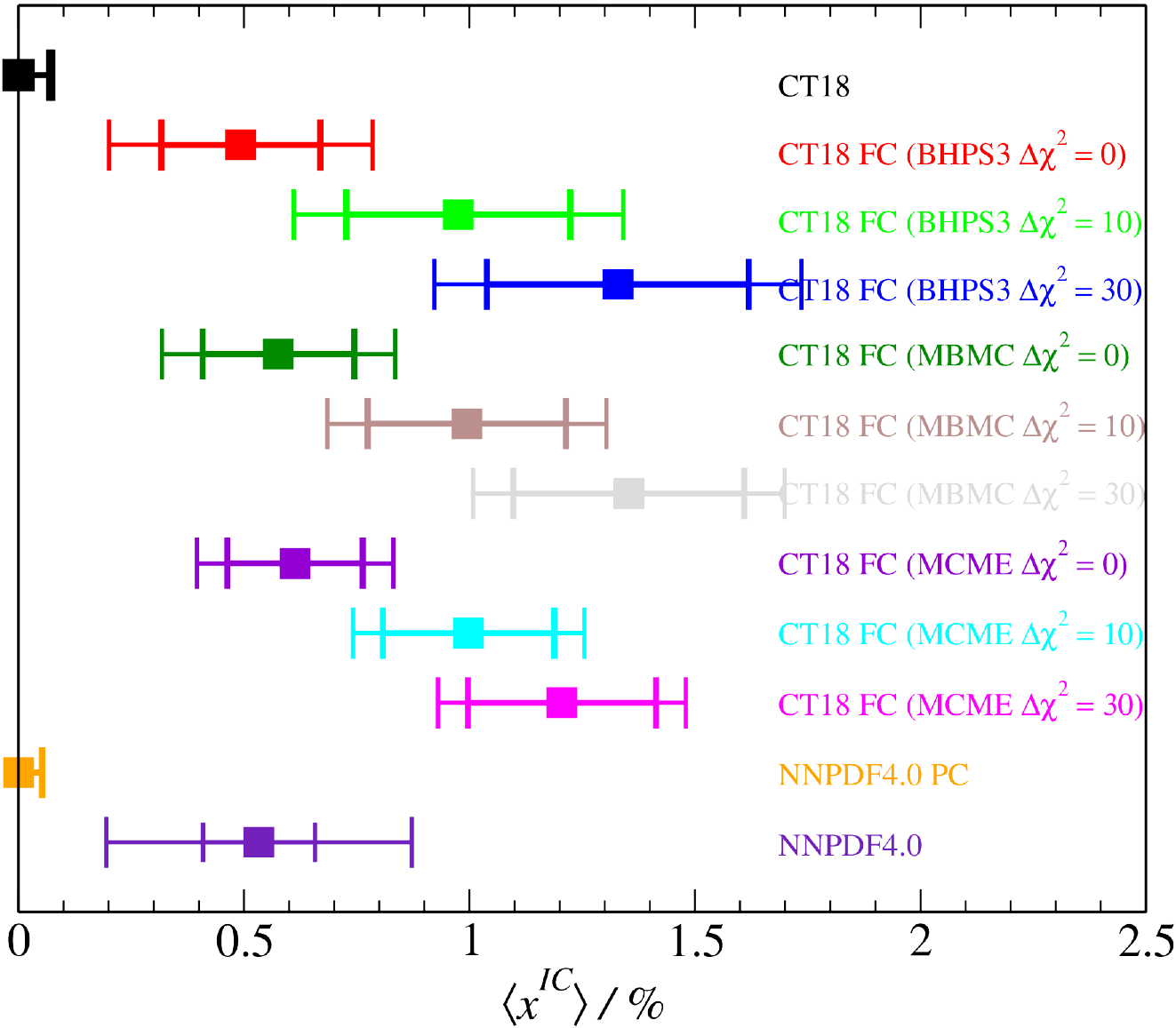}
    \caption{Sensitivity to the intrinsic charm momentum fraction (in percentage), defined in Eq.~(\ref{eq:ICmomfrac}), at FASER$\nu$ considering the acceptance and selection cuts described in Sect.~\ref{sec:settings} for an integrated luminosity of $\mathcal{L_\mathrm{pp}} = 250\,\mathrm{fb}^{-1}$ for events with $x\geq 0.2$ (left) and $x \geq 0.4$ (right). 
    The inner error bars represent the statistical uncertainty, while the outer error bars include both statistical and the uncertainties at the $1\sigma$ level.
    }
\label{fig:xcIC_FASERnu}
\end{figure}

Traditionally, the sensitivity of a given analysis to intrinsic charm has been estimated in terms of the non-perturbative charm momentum fraction.
Since available PDF sets with fitted charm include both a perturbative and an intrinsic component, one can define the intrinsic charm momentum fraction as
\be
\label{eq:ICmomfrac}
\langle x^{\mathrm{IC}}\rangle \equiv \int_{0}^{1}\mathrm{d}x\, x  \lc \lp c+\bar{c}\rp_{\rm FC}-\lp c+\bar{c}\rp_{\rm PC}\rc (x , Q = 1.65\,\mathrm{GeV})  \, ,
\ee
where $Q_0=1.65$ GeV being a scale slightly above the charm threshold $\mu_c=m_c$, and where FC and PC stand for the fitted and perturbative charm variants of a given PDF set.
Subtracting the perturbatively generated component to the fitted charm PDF resolves then the intrinsic charm momentum fraction. 
In Fig.~\ref{fig:xcIC_FASERnu} we present the expected sensitivity of FASER$\nu$ to the IC momentum fraction as defined in Eq.~(\ref{eq:ICmomfrac}), considering events for $x\geq 0.2$ (left) and $x\geq 0.4$ (right) for CT18 and NNPDF4.0. 
The inner error bar corresponds to  statistical uncertainties, while the outer one takes also into account PDF uncertainties. 
These results provide a complementary estimator highlighting the  FASER$\nu$ potential to study the intrinsic charm PDF.

\paragraph{Dependence on the $n_{\rm tr}$ cut.}
As discussed in Sect.~\ref{sec:muonDIS}, in the current experimental selection strategy used in FASER$\nu$, muon DIS events are requested to have at least a certain number of charged tracks $n_{\rm tr}$ to be selected by the interaction vertex finding algorithm (the values of $n_{\rm tr}$ are highly correlated with both $E_h$ and $W^2$).
In the case of charm production, most events are located in the small-$Q^2$ and large-$x$ region and hence they are sensitive to the choice of the value of the $n_{\rm tr}$ cut: a too stringent value will remove from the data sample the kinematic region where predictions based on fitted and on perturbative charm differ the most, washing out a possible signal for intrinsic charm.

\begin{figure}[t]
\centering
\hspace{-12mm}
\includegraphics[width=0.49\linewidth]{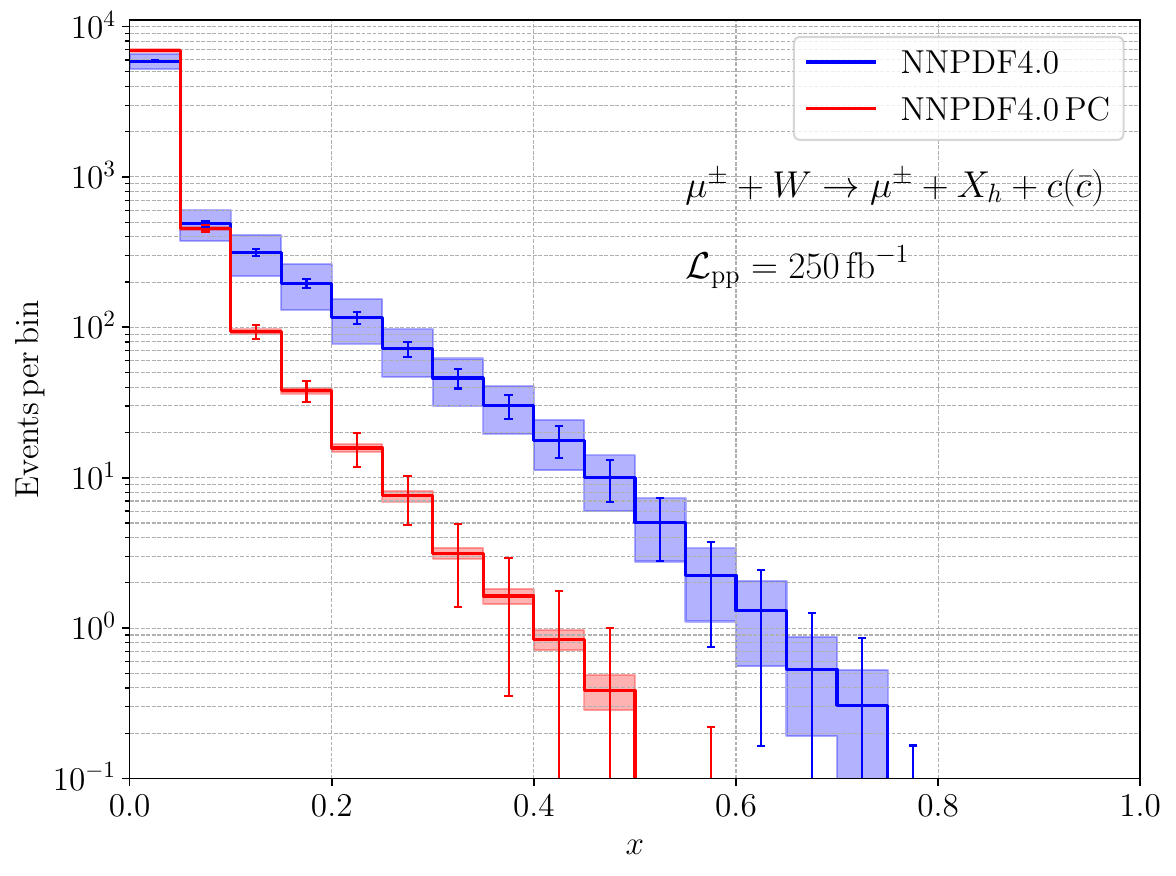}
\includegraphics[width=0.49\linewidth]{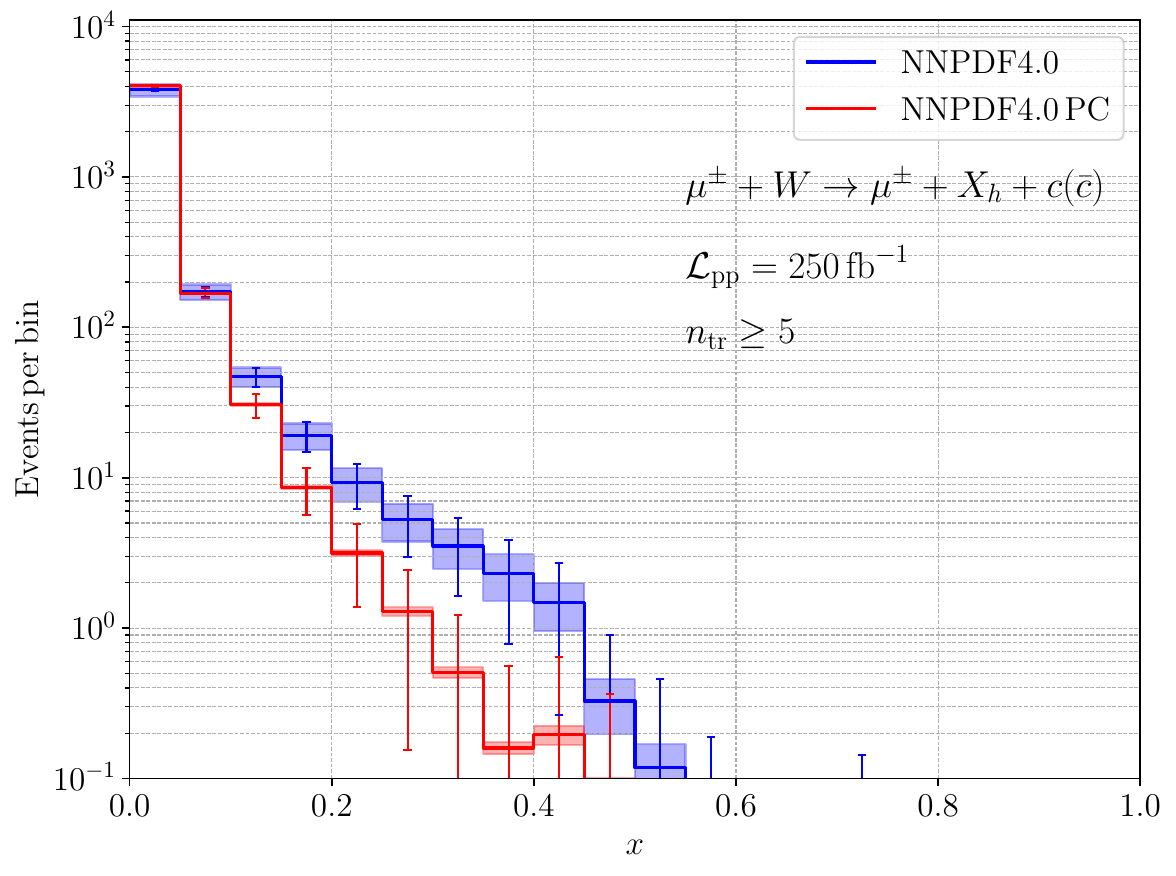}
\includegraphics[width=0.49\linewidth]{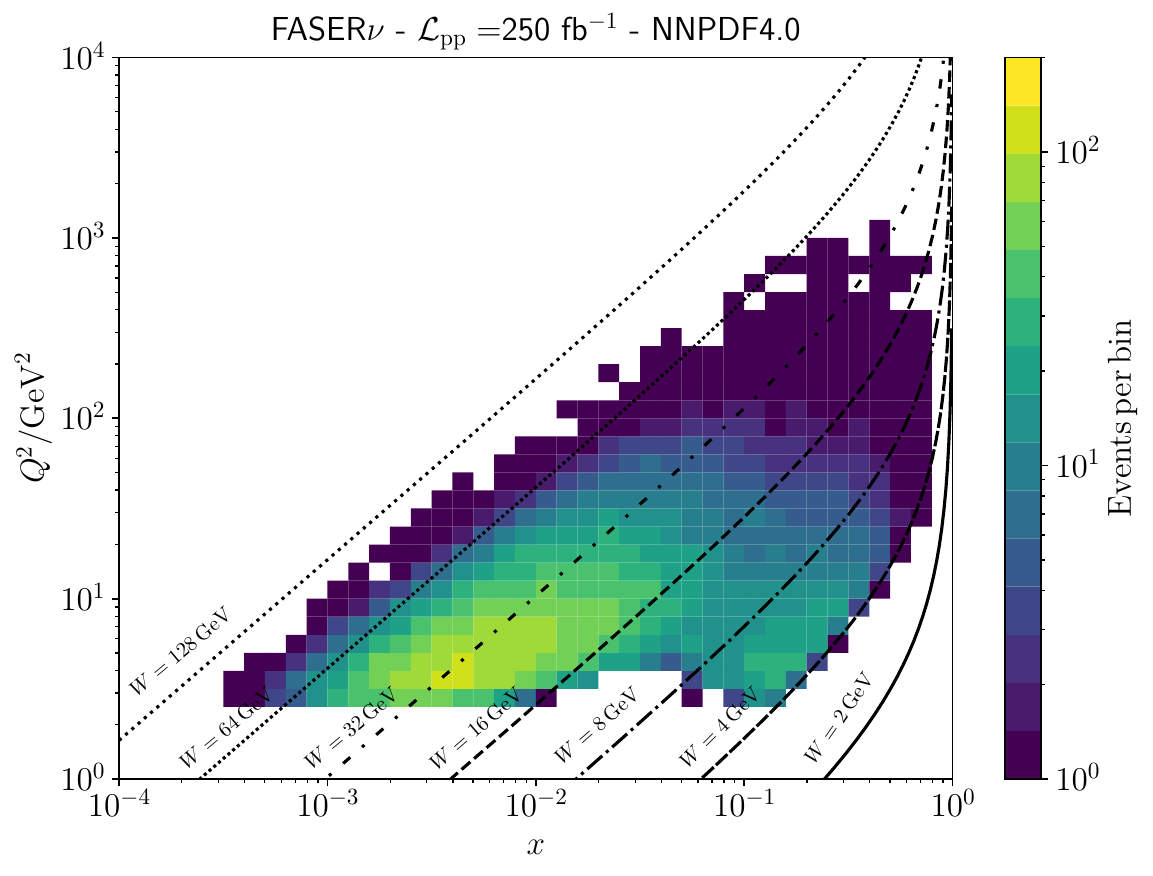}
\includegraphics[width=0.49\linewidth]{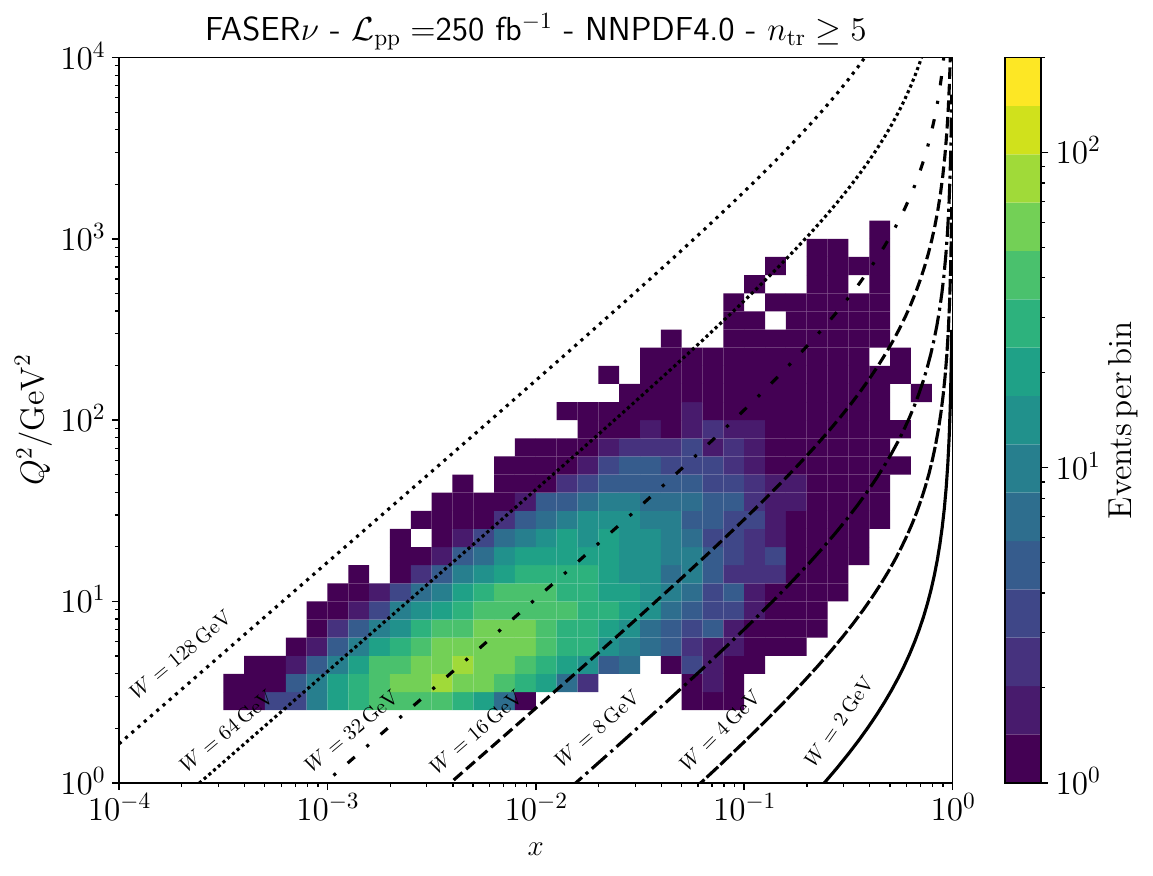}     
\caption{Top: same as Fig.~\ref{fig:DIStotal_cuts_E_statBands} in the case of the NNPDF4.0 predictions comparing results based on requesting $n_{\rm tr}\ge 1$ (left) and and $n_{\rm tr}\ge 5$ (right panel) charged tracks in the  final state. 
Bottom: same for Fig.~\ref{fig:DIStotal_charm_Q_x}.
}
\label{fig:DIStotal_cuts_x_nTracks}
\end{figure}

To illustrate the dependence of the charm production event yields with respect to the value of $n_{\rm tr}$, the top panels of Fig.~\ref{fig:DIStotal_cuts_x_nTracks} display the same comparison as Fig.~\ref{fig:DIStotal_cuts_E_statBands} (event yields binned in $x$) in the case of the NNPDF4.0 predictions comparing results based on requesting $n_{\rm tr}\ge 1$ and $n_{\rm tr}\ge 5$ charged tracks in the hadronic final state, to be compared with  the baseline predictions based on $n_{\rm tr}\ge 3$.
The bottom panels show the same comparison as in Fig.~\ref{fig:DIStotal_charm_Q_x} for the distribution of selected events in $(x,Q^2)$ also for different values of the $n_{\rm tr}$ cut.
The results of Fig.~\ref{fig:DIStotal_cuts_x_nTracks} highlight how the sensitivity to the large-$x$ charm PDF is worsened as $n_{\rm tr}$ is increased.
For $n_{\rm tr}\ge 5$, the predictions for the charm-tagged events at FASER$\nu$ between fitted and perturbative charm PDFs are compatible within statistical uncertainties, due to this cut removing most of the relevant events with $W<8$ GeV.
This finding indicates that optimising the cut in $n_{\rm tr}$, or more in general the vertex finding algorithm of FASER$\nu$, is essential to ensure that sufficient sensitivity to intrinsic charm is retained with the Run~3 dataset.

\paragraph{Impact of systematic uncertainties.}
While a complete simulation of event reconstruction in the case of a realistic detector-level analysis is beyond the scope of this work, we can estimate the impact of systematic uncertainties for large-$x$ charm production as follows.
In muon DIS, one measures both the initial and final state muon energies, $E_\mu$ and $E'_{\mu}$ respectively, as well as the hadronic final state energy $E_h$. 
As mentioned in Sect.~\ref{sec:inclusiveDIS}, initial FASER$\nu$ studies of neutrino identification report a resolution in the muon momentum $E_\mu$ of about 30\% at 200~GeV and
50\% at higher energies~\cite{FASER:2024hoe}, while the resolution in the muon scattering angle $\theta_\mu$ can be considered as negligible.
To model the finite detector resolution, here we apply Gaussian smearing to the initial and final-state muon momenta produced by the {\sc\small POWHEG+Pythia8} simulation by a fixed factor $\sigma_p$, which for simplicity is taken to be the same for $E_h$.
We then impose that the true energies satisfy the kinematic constraint $E_\mu = E'_{\mu}+E_h$ (neglecting proton mass effects) to extract the reconstructed energies using a $\chi^2$ fit. 
From them we evaluate the reconstructed DIS kinematical variables $x_{\rm reco}$ and $Q^2_{\rm reco}$.
We consider for illustration two values of this energy smearing, namely $\sigma_p=10\%$ and $\sigma_p=30\%$.
This simple approach is likely to represent a worse case scenario as compared to the real analysis and hence overestimate the impact of finite detector resolution.

\begin{figure}[t]
    \centering
\includegraphics[width=0.49\linewidth]{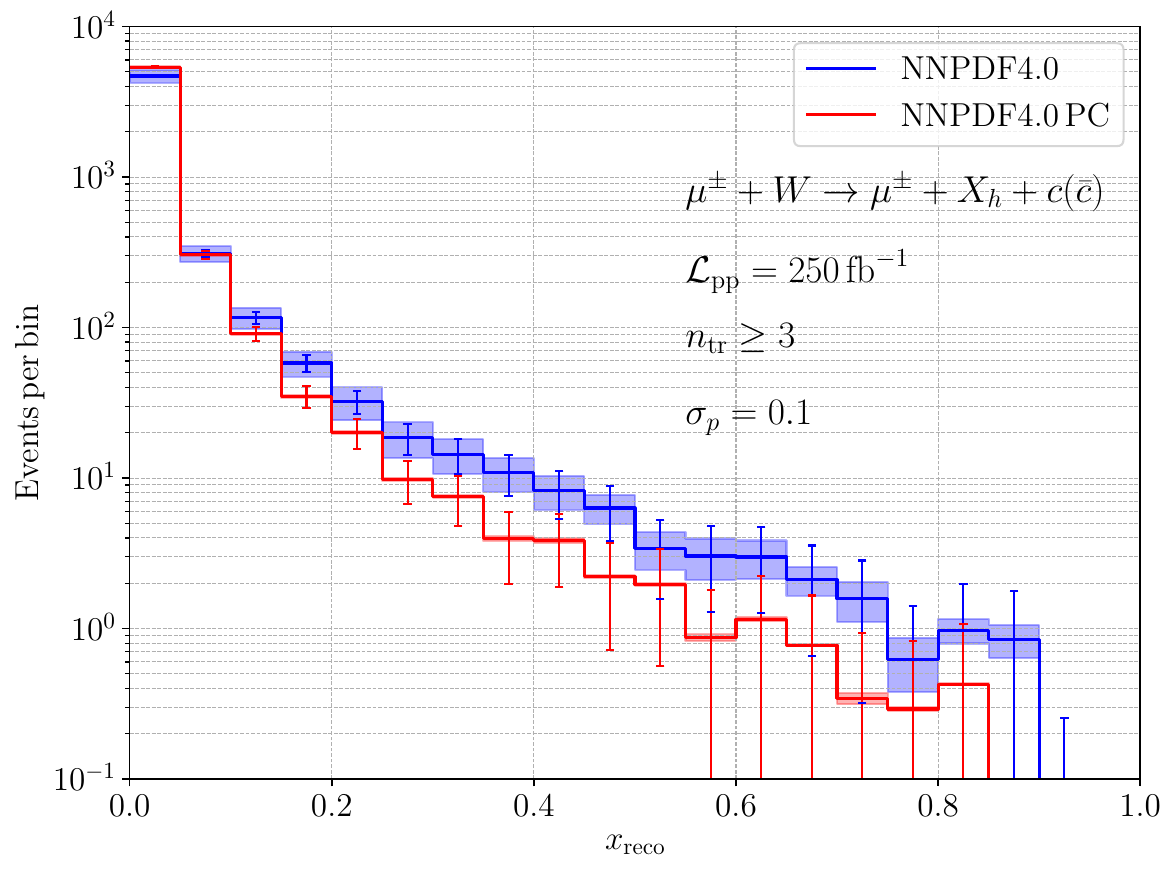}
\includegraphics[width=0.49\linewidth]{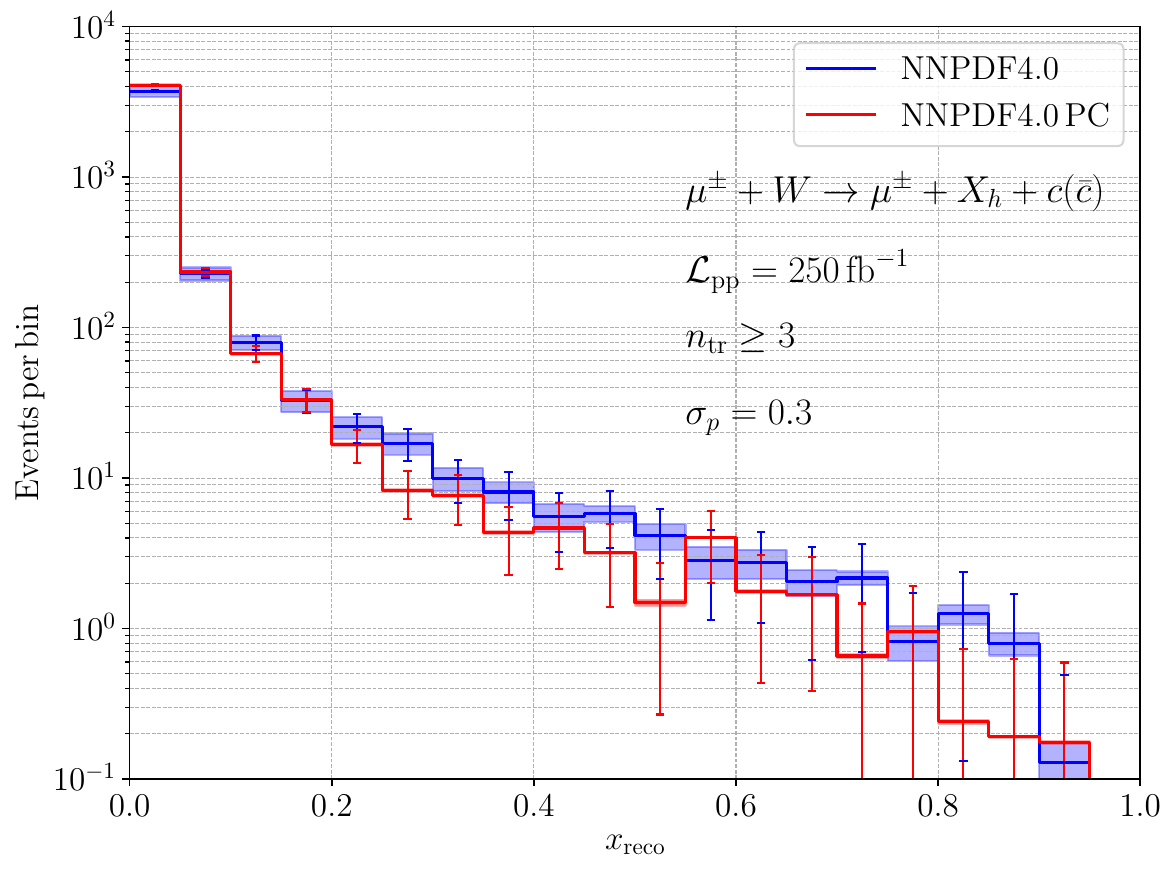}
  \caption{Same as Fig.~\ref{fig:DIStotal_cuts_E_statBands} in the case of the NNPDF4.0 predictions comparing the predictions with an smearing of the muon momentum by of $\sigma_p=10\%$ (left) and $
  \sigma_p=30\%$ (right). 
    }
    \label{fig:DIStotal_cuts_x_sigma}
\end{figure}

\begin{figure}[t]
    \centering
\includegraphics[width=0.49\linewidth]{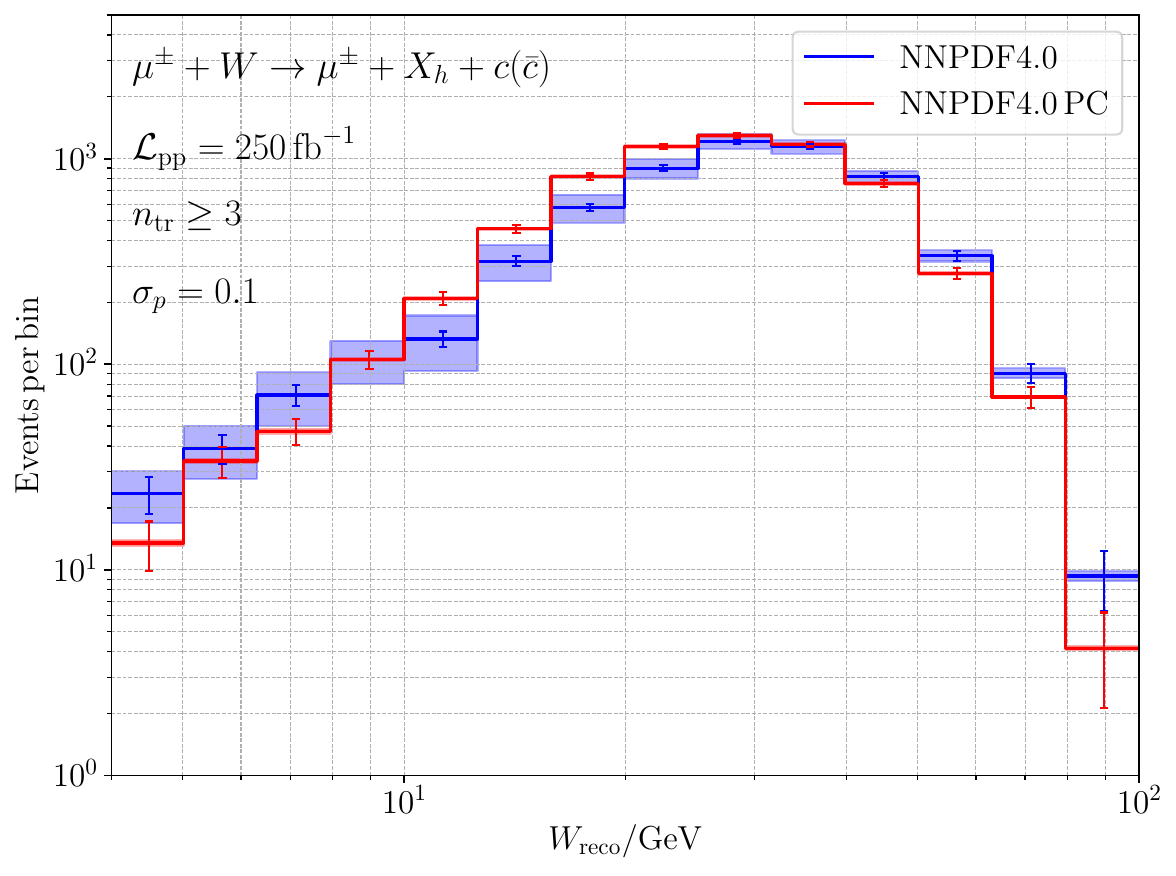}
  \caption{Same as the left panel in Fig.~\ref{fig:DIStotal_cuts_x_sigma} for the event yields binned in $W_{\rm reco}$.
    }
    \label{fig:DIStotal_cuts_W_sigma}
\end{figure}

\begin{table}[t]
\centering
\renewcommand{\arraystretch}{1.5}
\begin{tabularx}{\textwidth}{XXXlll}
\toprule
\multicolumn{6}{c}{Charm-tagged muon DIS neutral-current at FASER$\nu$ for $\mathcal{L}_{\rm pp}=250$ fb$^{-1}$}\\
\midrule
PDF set    & Charm PDF    & $\sigma_p$ & $x_{\rm reco}>0$             & $x_{\rm reco}\ge 0.2$ & $x_{\rm reco}\ge 0.4$  \\	
\toprule
NNPDF4.0   & Fitted charm & 0.1      & 5.3$\times10^{3}$ (0.9) & 104.6 (2.1)       & 27.6 (1.9) \\
\midrule
NNPDF4.0~PC & Pert. charm  & 0.1      & 5.8$\times10^{3}$ (1.0)    & 50.6 (1.0)          & 14.5 (1.0)     \\ 
\toprule
NNPDF4.0   & Fitted charm & 0.3      & 4.2$\times10^{3}$ (0.9) & 81.4 (1.3)        & 22.7 (1.2) \\
\midrule
NNPDF4.0~PC & Pert. charm  & 0.3      & 4.5$\times10^{3}$ (1.0)    & 60.6 (1.0)          & 18.4 (1.0)     \\ 
\bottomrule
\end{tabularx}
\vspace{0.3cm}
\caption{Same as Table~\ref{table:Nevents_charm} now applying a muon momentum smearing of $\sigma_p=10\%$ and $\sigma_p=30\%$ in the reconstruction of the Bjorken-$x$ variable $x_{\rm reco}$ in the case of the NNPDF4.0 NNLO predictions, see text.
}
\label{table:Nevents_charm_smeared}
\end{table}

Fig.~\ref{fig:DIStotal_cuts_x_sigma} presents the same as  Fig.~\ref{fig:DIStotal_cuts_E_statBands} for NNPDF4.0 with a Gaussian smearing of energies by a factor $\sigma_p=10\%$ and $30\%$. 
The tabulated results with two different $x_{\rm reco}$ cuts are provided in Table~\ref{table:Nevents_charm_smeared} and can be compared with the particle-level predictions of Table~\ref{table:Nevents_charm}.
It is found that the applied smearing washes out most of the differences between the predictions based on fitted and perturbative charm PDFs.
For $\sigma_p=10\%$ the increase in event yields for the fitted charm case is reduced to a factor 2 for both $x_{\rm reco}\ge 0.2$ and $\ge 0.4$, to be compared with factors 8 and 19 for the particle-level results.
For $\sigma_p=30\%$ it is not possible to distinguish between the two scenarios. 
Similar conclusions are derived from other distributions, such as the $W_{\rm reco}$ distribution shown in Fig.~\ref{fig:DIStotal_cuts_W_sigma}: for $\sigma_p=10\%$, the significance of the differences between fitted and perturbative charm, while still there, is markedly reduced. 
Here we used the measured muon momenta and scattering angle to evaluate $W_{\rm reco}$. 
An independent measurement of $W_{\rm reco}$ can also be performed using the final-state hadron momenta, offering different systematic uncertainties that could help improve the overall measurement precision.

\begin{figure}[t]
    \centering
\includegraphics[width=0.49\linewidth]{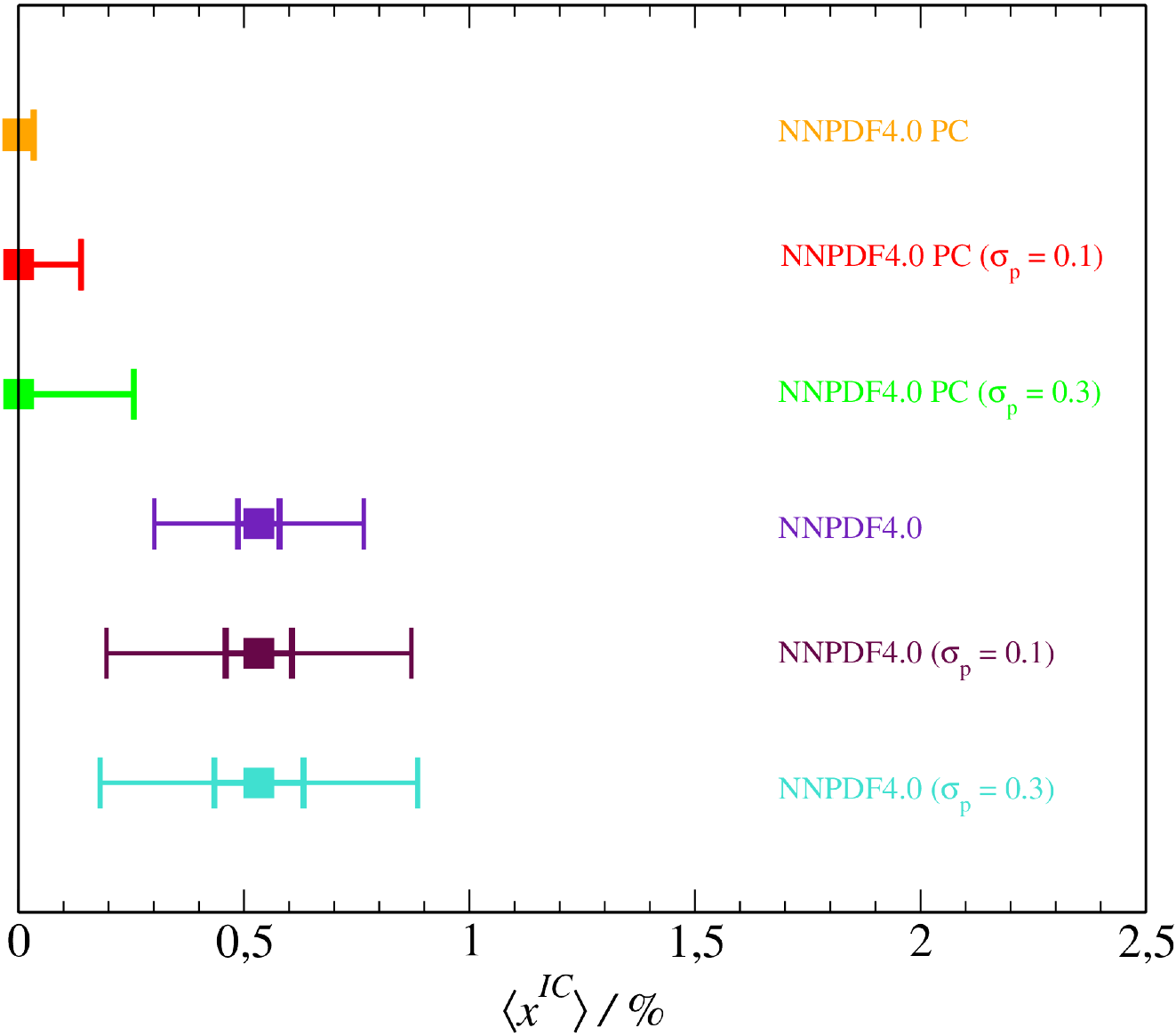}
\includegraphics[width=0.49\linewidth]{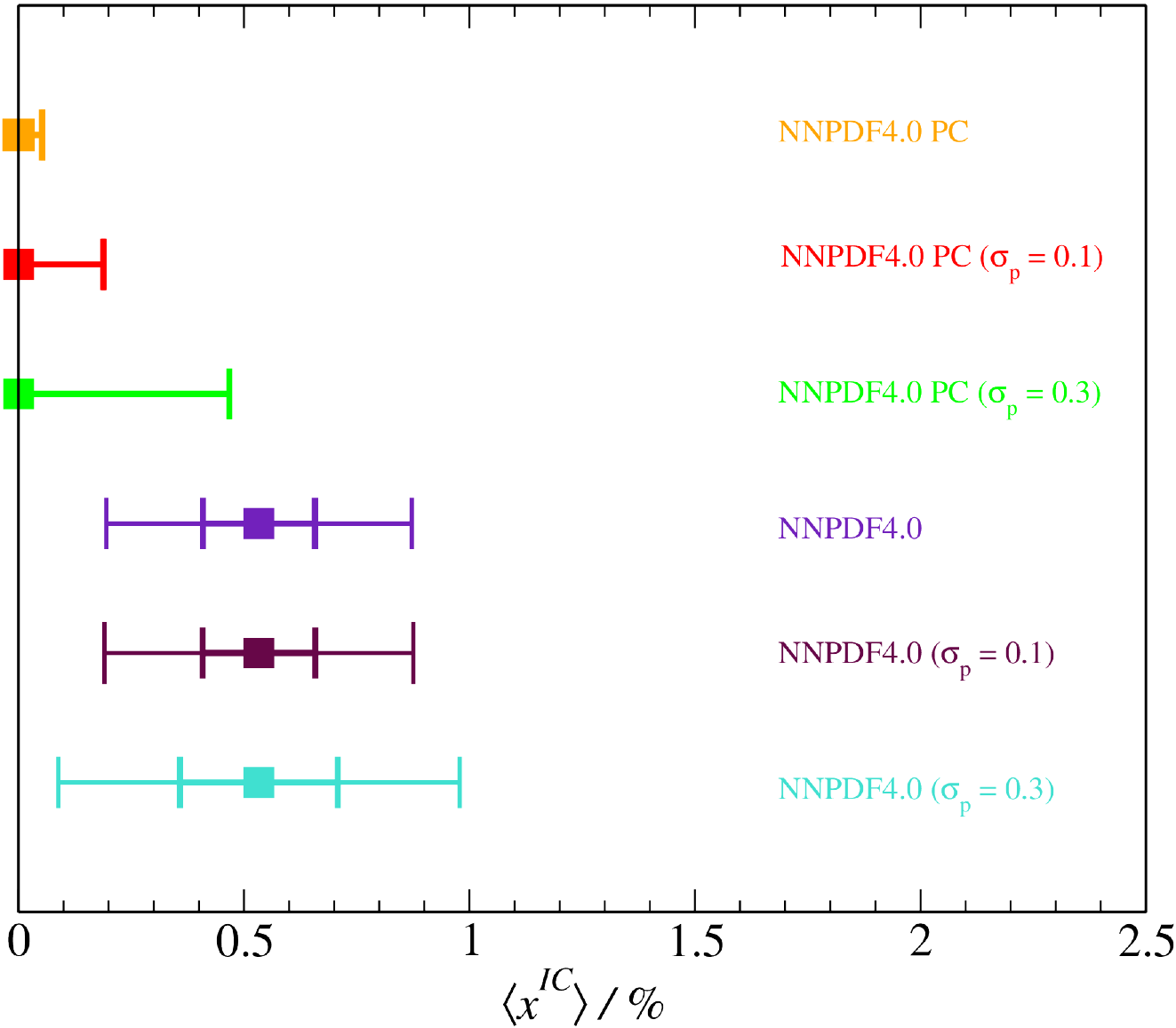}
    \caption{ Same as Fig.~\ref{fig:xcIC_FASERnu} for NNPDF4.0, comparing unsmeared results with those assuming an smearing of either $\sigma_p=10\%$ or $\sigma_p=30\%$ in the final-state muon and hadronic momenta.
    }
\label{fig:xcIC_smearing_FASERnu}
\end{figure}

Fig.~\ref{fig:xcIC_smearing_FASERnu} presents the same analysis as in Fig.~\ref{fig:xcIC_FASERnu} for the intrinsic charm momentum fraction evaluated for NNPDF4.0, now taking into account the experimental smearing in the final-state muon and hadronic energies. 
This smearing increases the uncertainties in the determination of the IC momentum fraction, making it challenging to separate it from the perturbative charm predictions even for the optimistic case of $\sigma_p = 10\%$.
This result indicates that, at least for FASER$\nu$ with the Run 3 dataset, the  traditionally adopted charm momentum fraction Eq.~(\ref{eq:ICmomfrac})  is not the most sensitive observable to intrinsic charm scenarios.

\begin{figure}[t]
    \centering
\includegraphics[width=0.49\linewidth]{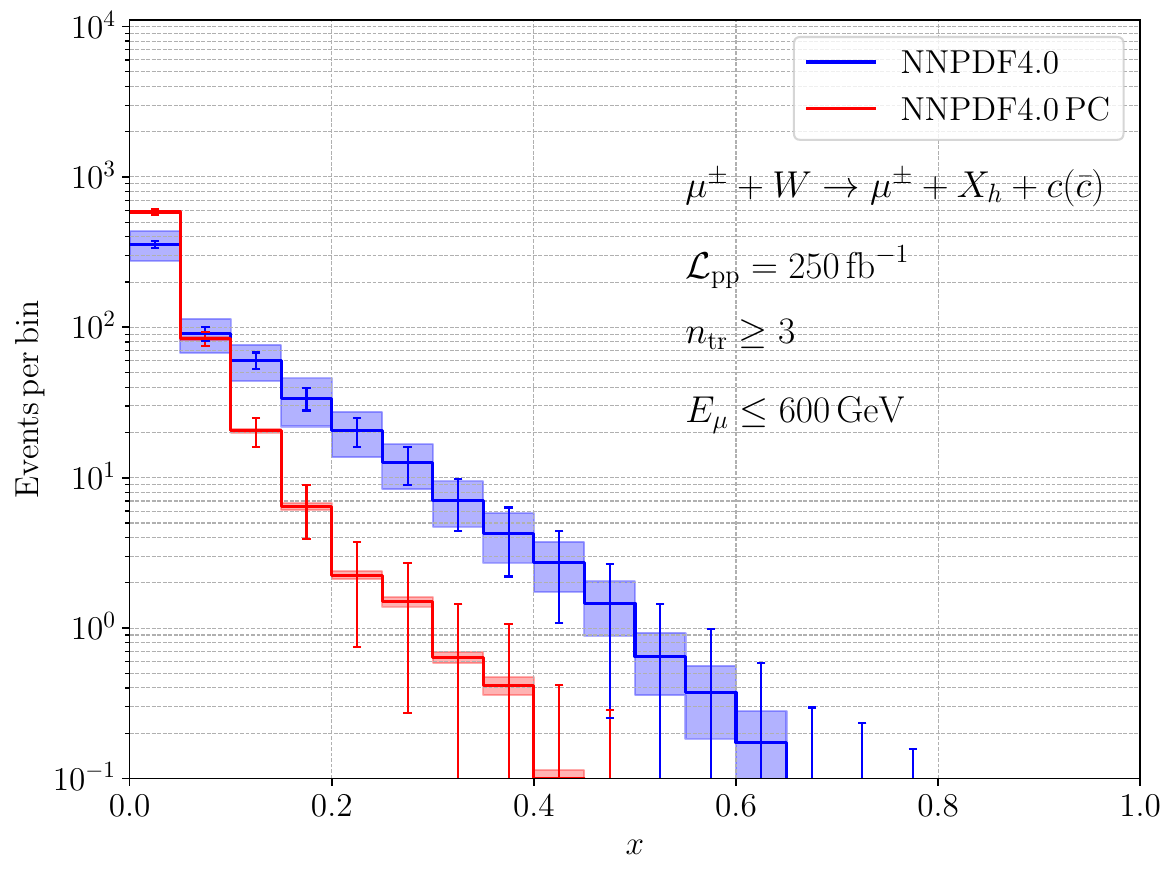}
\includegraphics[width=0.49\linewidth]{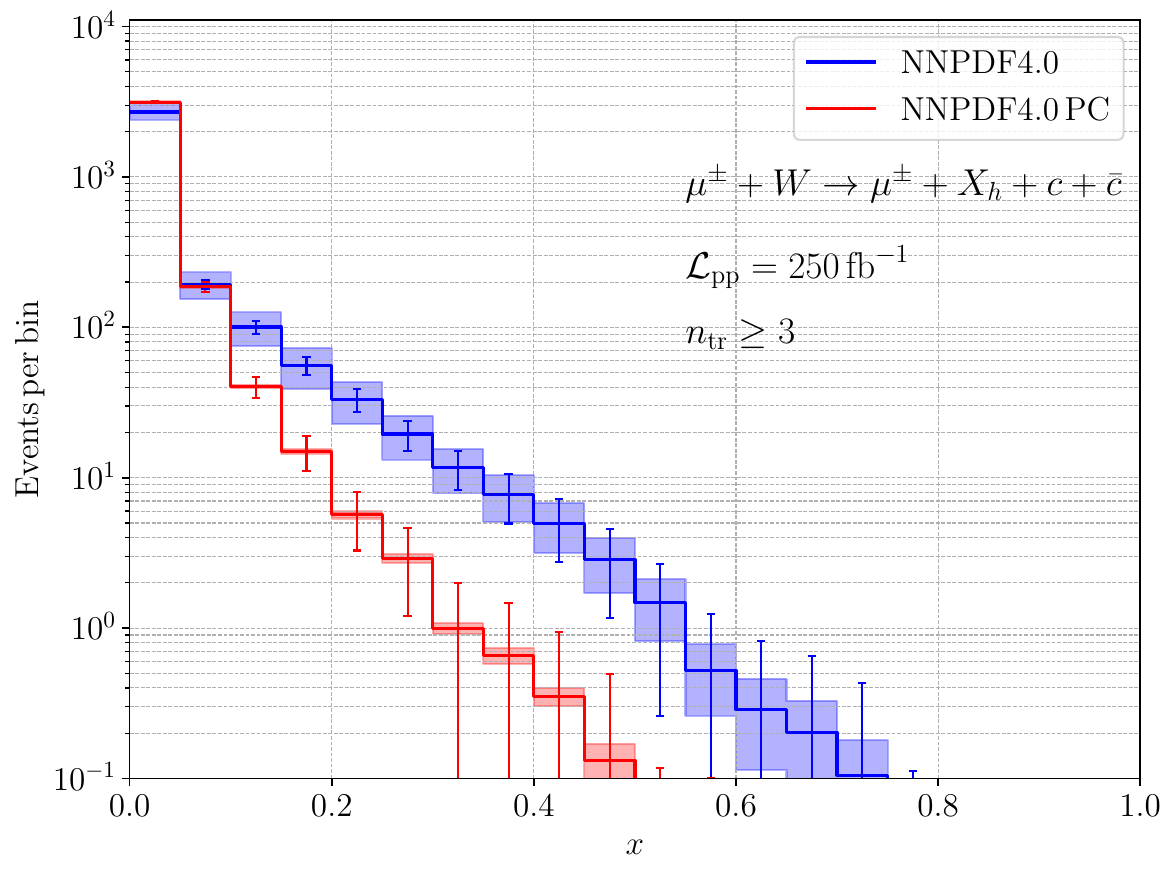}
    \caption{Same as Fig.~\ref{fig:DIStotal_cuts_E_statBands} with additional cuts: initial muon energy less than 600 GeV (left) and requiring a double charm tagging (right).
    }
\label{fig:DIScharm_cuts_x_Ei_600GeV}
\end{figure}

Another strategy to decrease the impact of systematic uncertainties entering charm production is to select only events with initial muon energies of $E_\mu \le 600$ GeV, for which final-state momenta can be reconstructed better as compared to events initiated by higher energy muons.
Fig.~\ref{fig:DIScharm_cuts_x_Ei_600GeV} (left) presents the events yields binned in $x$ after applying this additional cut: one observes a reduction of the number of events in the large-$x$ region, by approximately a factor of three, compared to the case without the upper initial muon energy cut, but the ratio between predictions based on intrinsic and perturbative charm PDFs remains stable.
In the right panel of Fig.~\ref{fig:DIScharm_cuts_x_Ei_600GeV}, we consider instead only events with double charm tagging. 
In practice, requiring double charm significantly suppresses the background from secondary hadron interactions and yields a cleaner sample dominated by genuine charm production.

Beyond a traditional binned analysis, significant improvements in event reconstruction and interpretation can be achieved using machine learning methods using full event information at the detector level such as the neural simulation based inference~\cite{ATLAS:2024jry} or transformer architectures~\cite{ATLAS:2025dkv} deployed in ATLAS.
The benefits of unbinned measurements in DIS for proton structure and QCD studies has been demonstrated e.g. in~\cite{Aggarwal:2022cki, H1:2021wkz}. 

These findings, together with the $n_{\tr}$-dependence study of Fig.~\ref{fig:DIStotal_cuts_x_nTracks}, indicate that dedicated improvements in event selection and kinematic reconstruction in muon DIS would significantly enhance the prospects of impactful measurements for QCD analyses.

\subsection{The intrinsic charm asymmetry at FASER$\nu$}
\label{subsec:asy}

The analysis of Sect.~\ref{sec:total_charm_production} shows that charm production in muon DIS at large-$x$ provides a sensitive discriminator for the total charm content of the proton. 
There we sum over the electric charges of the final state charm-hadrons and hence cannot disentangle the charm from the anticharm PDF.
However, the ultimate smoking gun for non-perturbative (intrinsic) charm would be provided by evidence of a non-zero asymmetry in the charm PDF~\cite{NNPDF:2023tyk}, which cannot be generated by any known QCD perturbative mechanism.\footnote{With the exception of a very small asymmetry induced by DGLAP evolution starting at NNLO~\cite{Catani:2004nc}.}

While FASER$\nu$ cannot access the electric charge of the final-state particles due to the lack of a magnetic field, the combination of FASER$\nu$ signals with information from the FASER electronic detector, in particular the spectrometer, provides a handle on the separation between charmed and anticharmed hadrons produced in muon DIS.
Indeed, as opposed to FASER$\nu$, the electronic detector is embedded into a magnetic field enabling electric charge separation. 
In this case the targeted signature would be a muon DIS charm production event tagged in the emulsion detector correlated to a dimuon event in the electronic detector, where the second muon arises from the charm-hadron decay taking place in the FASER$\nu$ fiducial volume.
The branching ratio for the latter according to the latest PDG average~\cite{ParticleDataGroup:2022pth} is 
\be
\label{eq:charm_BR}
{\rm BR}\lp c \to \ell^+ +X\rp= 
{\rm BR}\lp \bar{c} \to \ell^- +X\rp = 
0.096 \pm 0.004 \, ,
\ee
a suppression by a factor 10 as compared to tagging directly the final-state charm quark. 
The correlation of muons observed in FASER$\nu$ and the FASER spectrometer proceeds through the interface tracker, a dedicated tracking station placed right behind FASER$\nu$. 
Using that one can individually track the muons from the primary interaction and from the charm decay vertex, we can identify the charm quark charge in the latter.
Here we assume that, by combining information from the electronic detector, FASER$\nu$ can resolve the tagged-charm electric charge, and correspondingly reduce the projected event yields by a factor 10 to account for the semi-leptonic branching fraction Eq.~(\ref{eq:charm_BR}).

In this scenario we can construct dedicated observables representing a proxy for an asymmetry between the charm and anticharm content of the proton.
Specifically, in muon DIS at FASER$\nu$ and in analogy with the corresponding analysis at the EIC proposed in~\cite{NNPDF:2023tyk}, one can probe the intrinsic charm asymmetry by defining an asymmetry between events whose final states contain charm quarks with opposite charge:
\begin{equation}
\label{eq:charm_asymmetry}
    \mathcal{A}_c \equiv \frac{ N(\mu^\pm + W \to \mu^\pm + \widetilde{X}_h + c)-N(\mu^\pm + W \to \mu^\pm + \widetilde{X}_h + \bar{c})}{N(\mu^\pm + W \to \mu^\pm + \widetilde{X}_h + c) + N(\mu^\pm + W \to \mu^\pm + \widetilde{X}_h + \bar{c})} \, ,
\end{equation}
where $N$ stands for the number of muon DIS events satisfying the acceptance and selection cuts and account for the charm tagging efficiency $\epsilon_c$ and for the semileptonic branching fraction Eq.~(\ref{eq:charm_BR}).
That is, Eq.~(\ref{eq:charm_asymmetry}) measures the difference between events with a charm quark in the final state and those with a charm antiquark.
At leading order in the QCD expansion, Eq.~(\ref{eq:charm_asymmetry}) is directly proportional to $xc^-=x(c-\bar{c})$ which is fully dominated by its non-perturbative component.
In constructing Eq.~(\ref{eq:charm_asymmetry}), we take the hardest charm hadron of the event, which is associated with the charm quark participating in the hard scattering. 
Conservation of charm number implies the presence of another charm (anti)quark, which will be less energetic and likely part of the target nucleus remnant and often remains undetected.

\begin{figure}[t]
    \centering
\includegraphics[width=0.45\linewidth]{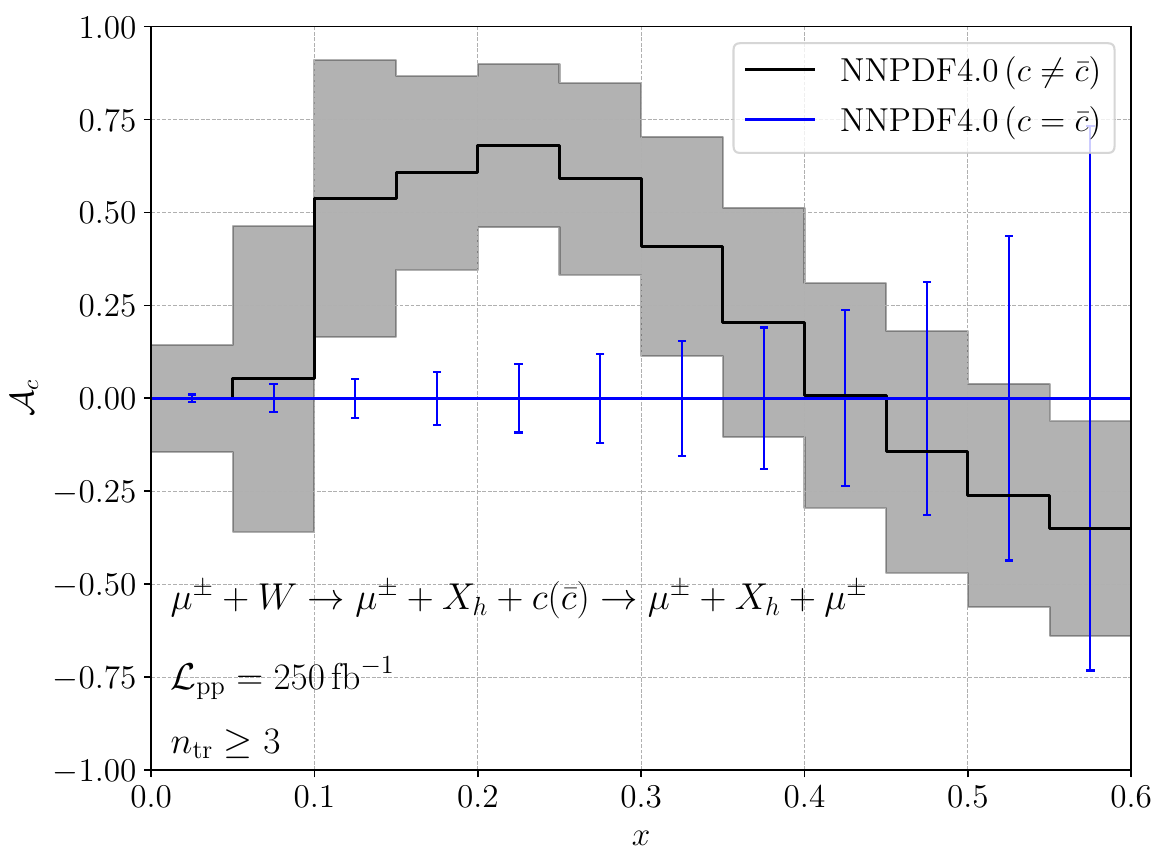} 
\includegraphics[width=0.45\linewidth]{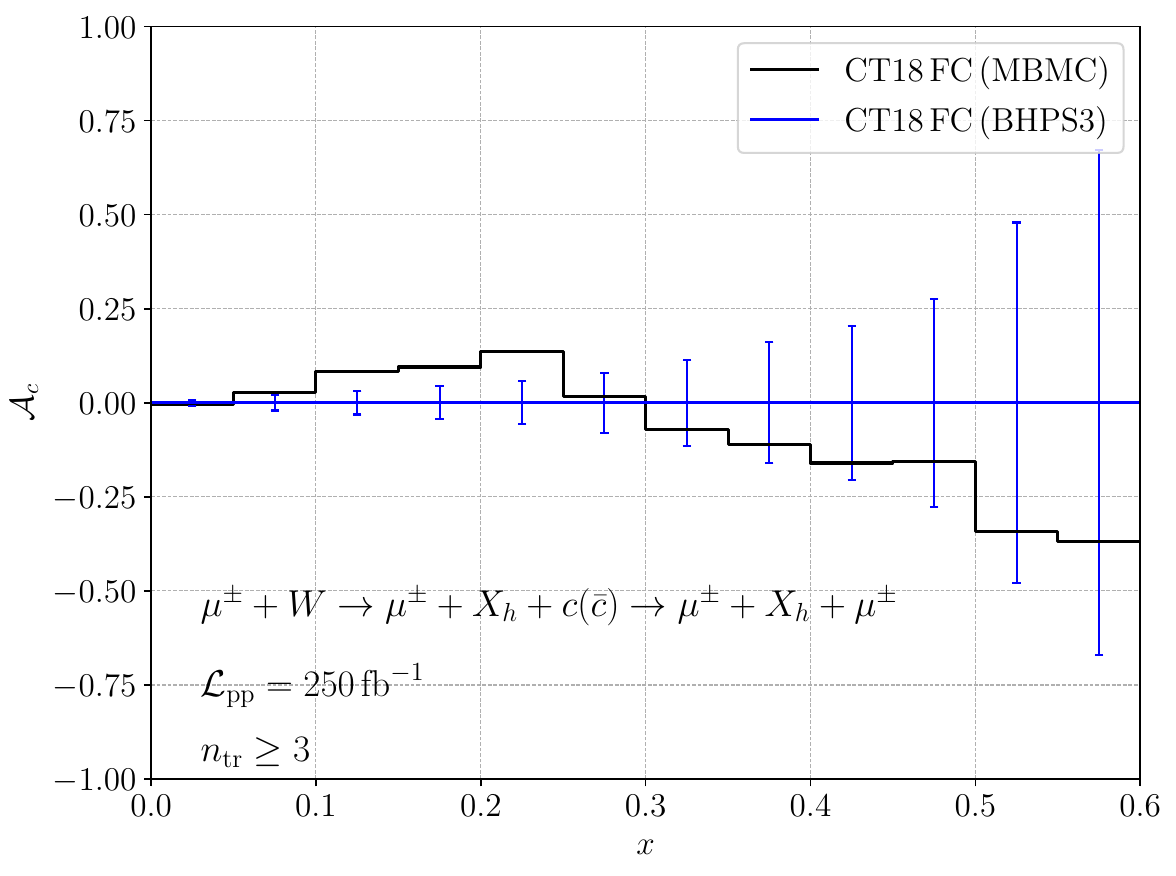}
    \caption{ 
    Left: the charm asymmetry $\mathcal{A}_c$ defined in Eq.~(\ref{eq:charm_asymmetry}) 
    for the NNPDF4.0 set with a fitted charm asymmetry~\cite{NNPDF:2023tyk}, together with the predictions using the baseline NNPDF4.0 set (with symmetric fitted charm), see also Fig.~\ref{fig:charm-PDFs}.
    The error bands correspond to the PDF uncertainties, while the error bars indicate the expected statistical uncertainty in the measurement. 
    Right: same for the CT18 FC sets in the MBMC model where an intrinsic charm asymmetry is allowed, compared with the predictions based on BHPS3 where the charm PDF is symmetric.
     }
    \label{fig:asy_x_charm}
\end{figure}

Fig.~\ref{fig:asy_x_charm} displays the asymmetry $\mathcal{A}_c$, Eq.~(\ref{eq:charm_asymmetry}), for the NNPDF4.0 and CT18 sets with (fitted) charm, both $c=\bar{c}$ and $c\ne \bar{c}$  shown in the right panels of Fig.~\ref{fig:charm-PDFs}.
The error bands correspond to the PDF uncertainties (for the NNPDF4.0 fit), while the error bars indicate the expected statistical uncertainty in the measurement for the case of a symmetric charm PDF. 
As in the rest of the paper, the calculation is carried out with the {\sc\small POWHEG+Pythia8} simulation framework.
As expected, the {\sc\small POWHEG+Pythia8} predictions for $\mathcal{A}_c$ reproduce the qualitative behaviour of underlying charm PDFs.
For simplicity, the present analysis does consider systematic uncertainties. 
A more realistic estimate of the prospects for measuring the intrinsic charm asymmetry at FASER could be made by same smearing approach used in the previous section. 

Both NNPDF4.0 ($c\ne\bar{c}$) and CT18 FC (MBMC) predict a non-zero, positive asymmetry peaking around $x\sim 0.2$ and then a negative asymmetry for $x\gsim 0.4$ (for NNPDF4.0) and $x\gsim 0.3$ for CT18 FC.
In the case of NNPDF4.0 (CT18 FC), the expected value of this positive asymmetry is 75\% (15\%) at $x\sim 0.2$ and $-30\%$ ($-35\%$) at $x\sim 0.6$, albeit in the NNPDF4.0 large variations remain possible within the estimated PDF uncertainties.
For the PDF sets with a fitted symmetric charm PDF, the asymmetry satisfies $\mathcal{A}_c\sim 0$ providing a solid control baseline.
When comparing the {\sc\small POWHEG+Pythia8} predictions with the expected statistical uncertainties, we see that the FASER$\nu$ precision should be sufficient to identify positive values of $\mathcal{A}_c$ in the region around $x\sim 0.2$, specially in the case of NNPDF4.0.
 
The results of Fig.~\ref{fig:asy_x_charm} stablish the measurement of $\mathcal{A}_c$ from the Run~3 dataset as an important and accessible target for the neutrino program of FASER$\nu$, realizing a prominent QCD analysis which otherwise would be delayed until the realisation and commissioning of the EIC.

\subsection{Predictions for the HL-LHC}
\label{sec:charmprod_projections}

\begin{figure}[t]
    \centering
\includegraphics[width=0.45\linewidth]{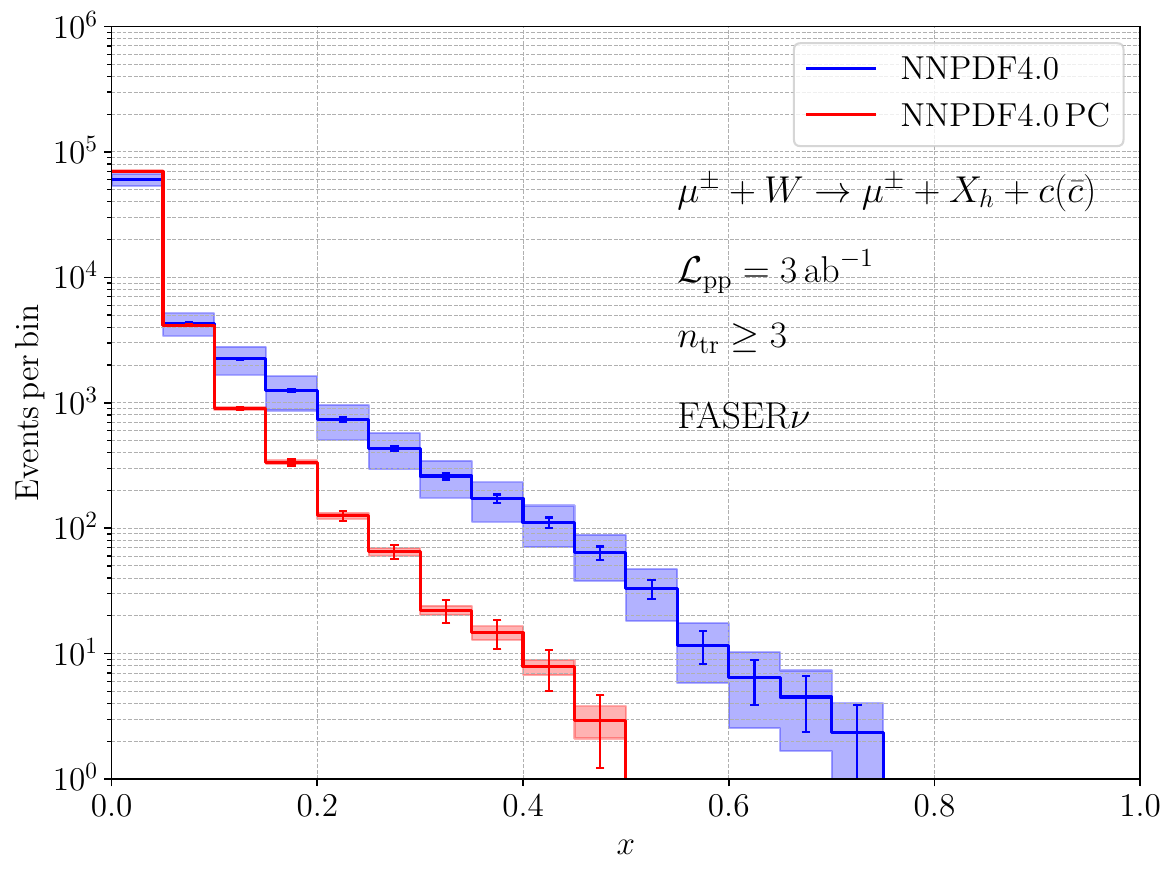} 
\includegraphics[width=0.45\linewidth]{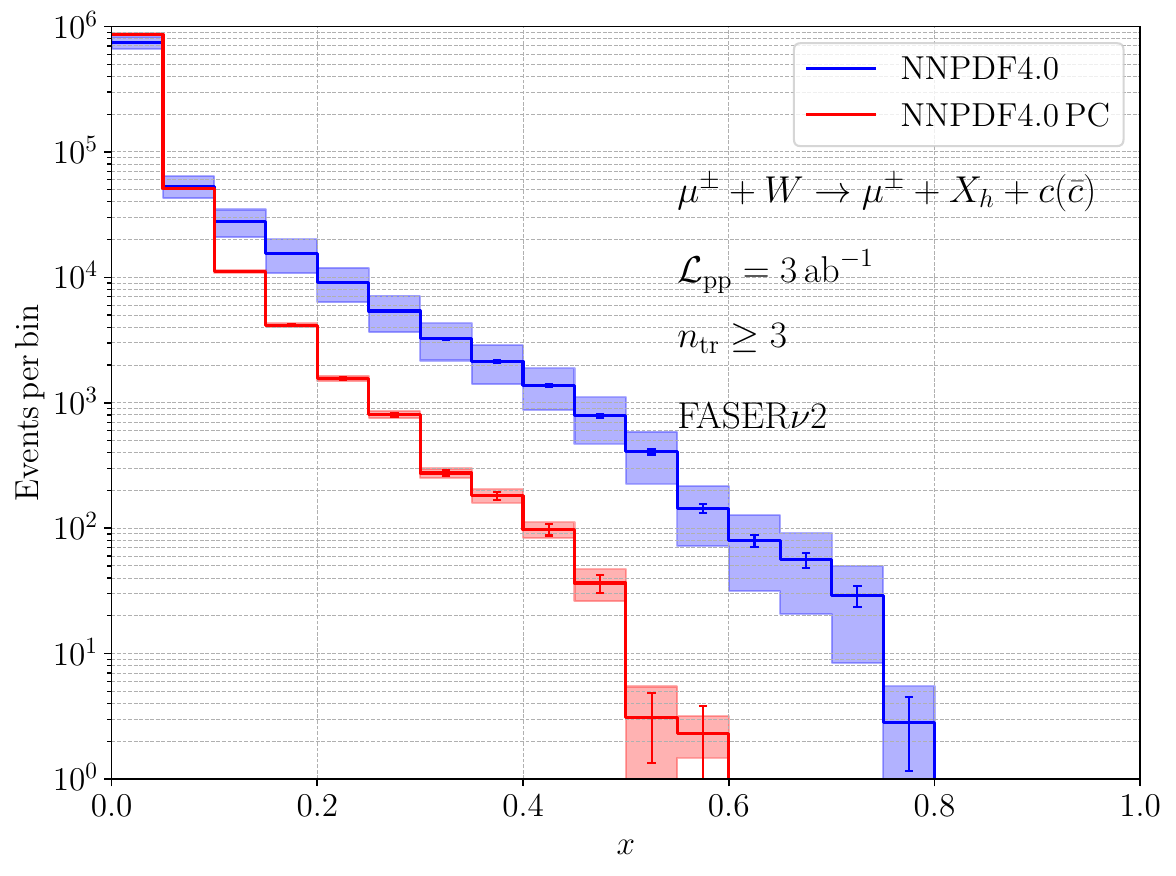}
\includegraphics[width=0.45\linewidth]{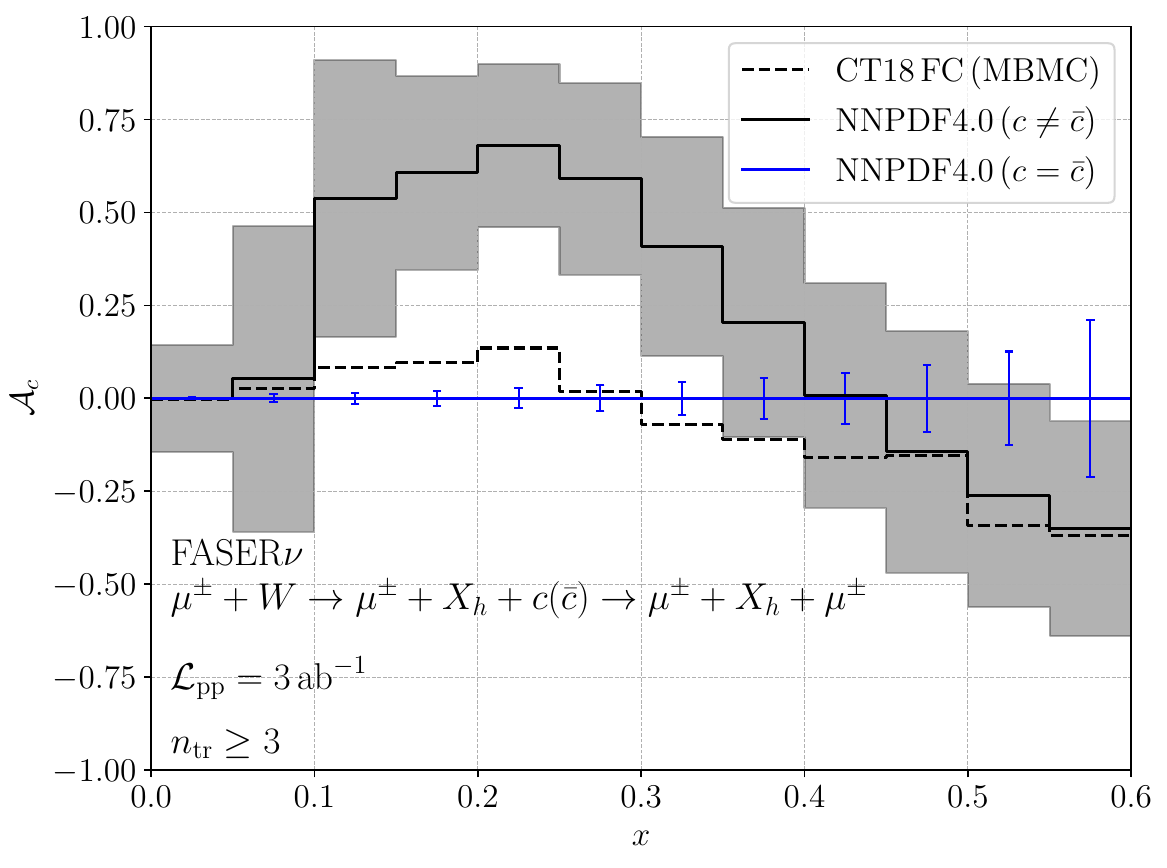} 
\includegraphics[width=0.45\linewidth]{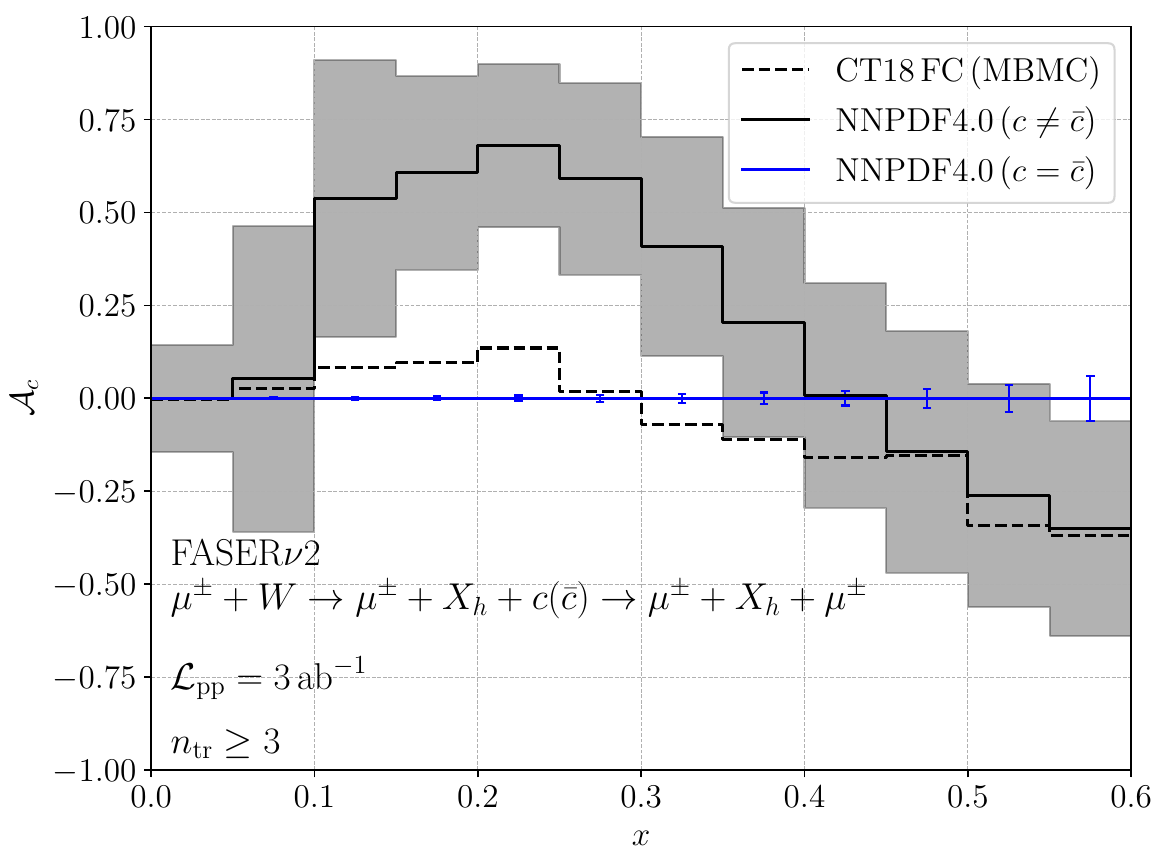}
    \caption{ 
    Top: same as Fig.~\ref{fig:DIStotal_cuts_E_statBands} for the FASER$\nu$ (left) and FASER$\nu$2 (right panel) detectors at the HL-LHC for $\mathcal{L}_{\rm pp}=3$ ab$^{-1}$.
    Bottom: same as  Fig.~\ref{fig:asy_x_charm} for the charm production asymmetry $\mathcal{A}_c$.
    In both cases, predictions are based on the NNPDF4.0 sets, and in the bottom panel also the predictions based on CT18 FC (MBMC) are shown.
     }
\label{fig:Events_x_charm_HL-LHC}
\end{figure}

Finally, following Sect.~\ref{sec:inclusiveDIS}, we provide predictions for both inclusive charm production in muon DIS and for the charm production asymmetry $\mathcal{A}_c$ for the HL-LHC data taking period.
We consider both the current FASER$\nu$ detector and for larger FASER$\nu$2 hosted at the FPF, in the two cases for an integrated luminosity of $\mathcal{L}_{\rm pp}=3$ ab$^{-1}$.
Fig.~\ref{fig:Events_x_charm_HL-LHC} displays these projections, in the same format as in Fig.~\ref{fig:DIStotal_cuts_E_statBands} for the event yields for total charm production binned in $x$ (top) and for the charm production asymmetry $\mathcal{A}_c$ (bottom).

Fig.~\ref{fig:Events_x_charm_HL-LHC} illustrates the benefits of the higher statistics expected for the HL-LHC data taking period. 
For instance, while for NNPDF4.0 at FASER$\nu$ (Run~3) one expects only 20 charm production events for $x\gsim 0.4$, where fitted charm is most enhanced, more than 2000 events would be recorded by FASER$\nu$2 in the same kinematic region, offering unparalleled sensitivity to the charm PDF.
These higher statistics also offer direct access to the ``background-free'' large-$x$ region, e.g. at FASER$\nu$2 no events are expected at $x\gsim 0.6$ in the perturbative charm scenario, while more than 100 events would be recorded for the NNPDF4.0 fitted charm case, proving an unambiguous and striking signal. 

Similar considerations apply to the charm production asymmetry projections.
While the theoretical predictions based on NNPDF4.0 are essentially unchanged, given that the size of the predicted asymmetry does not depend on the magnitude of the total event yields, the statistical uncertainties estimated for the measurement of $\mathcal{A}_c$ become much smaller at future LHC runs. 
For the specific case of FASER$\nu$2, statistical errors at the few percent level in relevant region of $x$ are expected, and even for $x\sim 0.6$ one can measure $\mathcal{A}_c$ with a 5\% uncertainties. 
These results imply that FASER$\nu$(2) at the HL-LHC will precisely fingerprint a possible charm asymmetry in the proton with excellent sensitivity to the signals predicted by the wide majority of IC models that allow $c\ne \bar{c}$, including the MBMC model adopted in the CT18 FC study.

\begin{figure}[t]
    \centering
\includegraphics[width=0.49\linewidth]{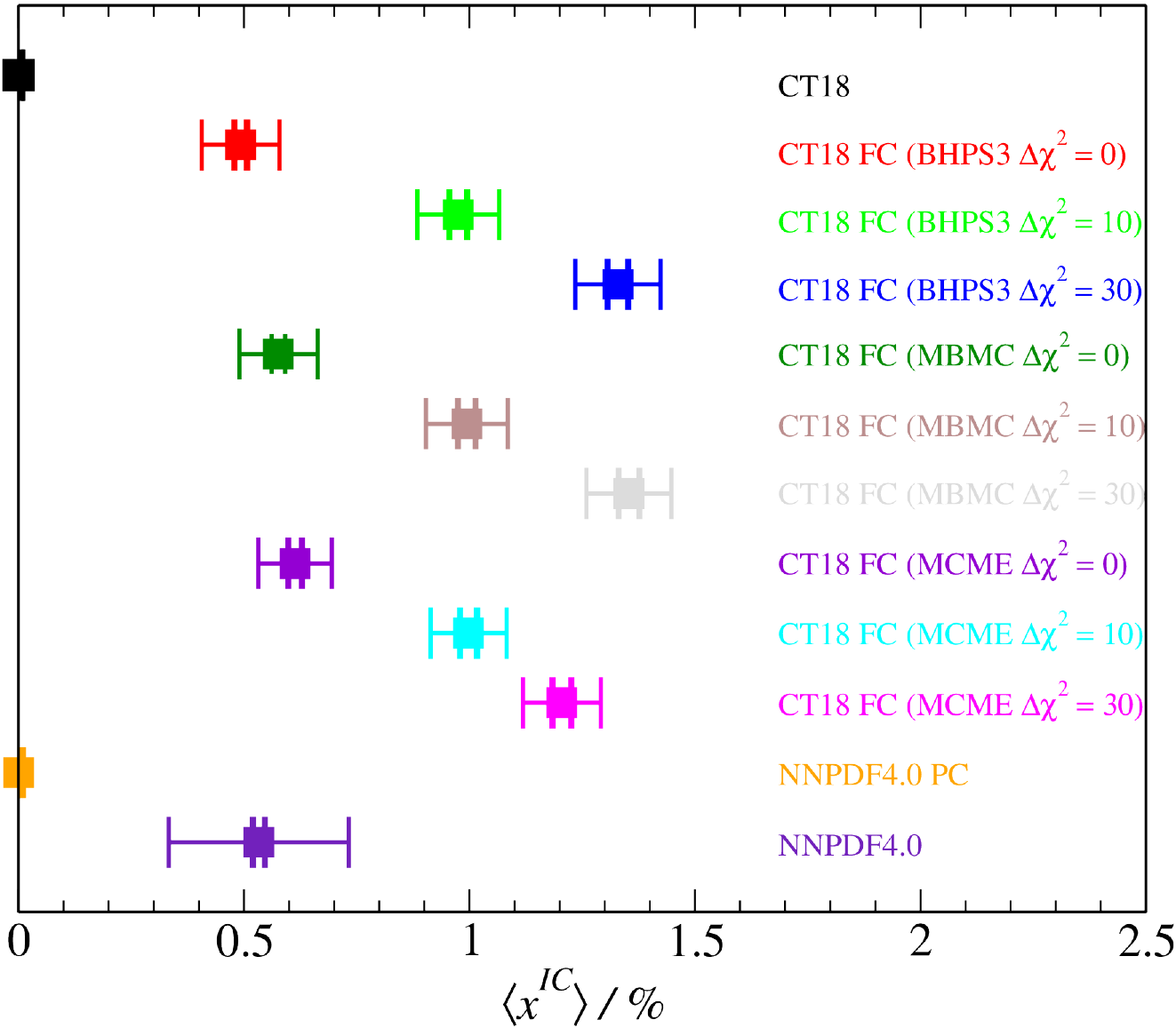}
\includegraphics[width=0.49\linewidth]{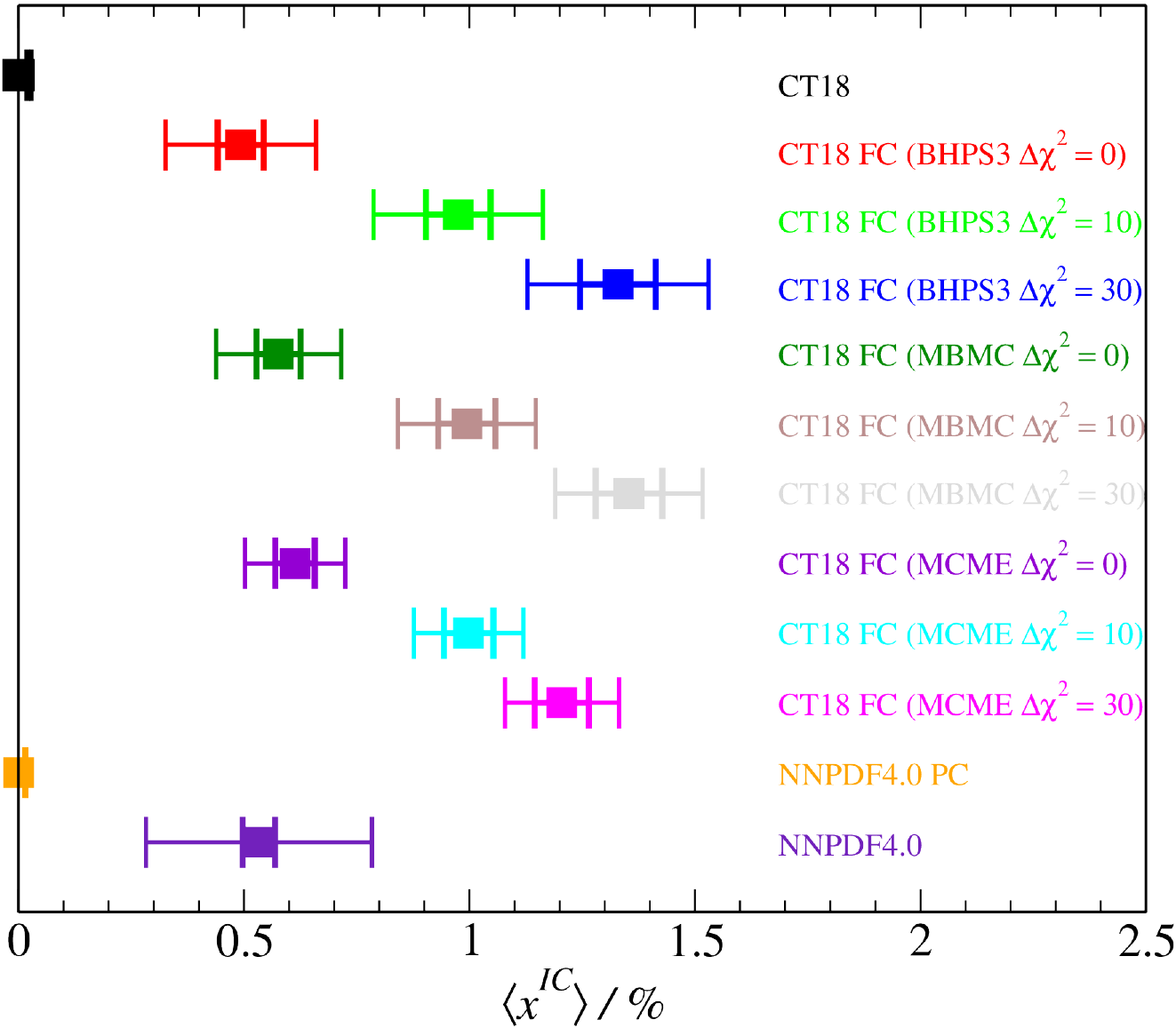} \\
\includegraphics[width=0.49\linewidth]{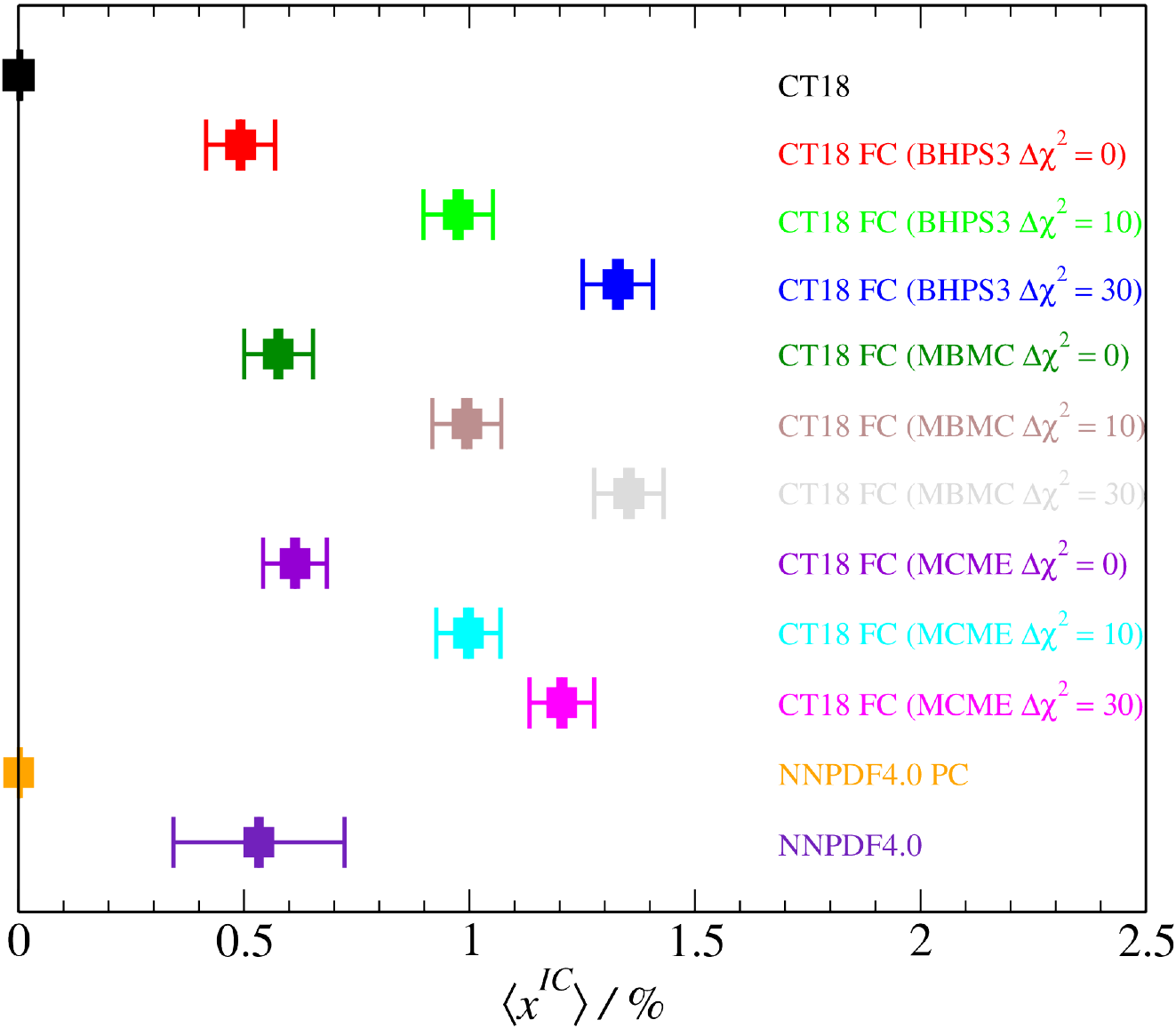}
\includegraphics[width=0.49\linewidth]{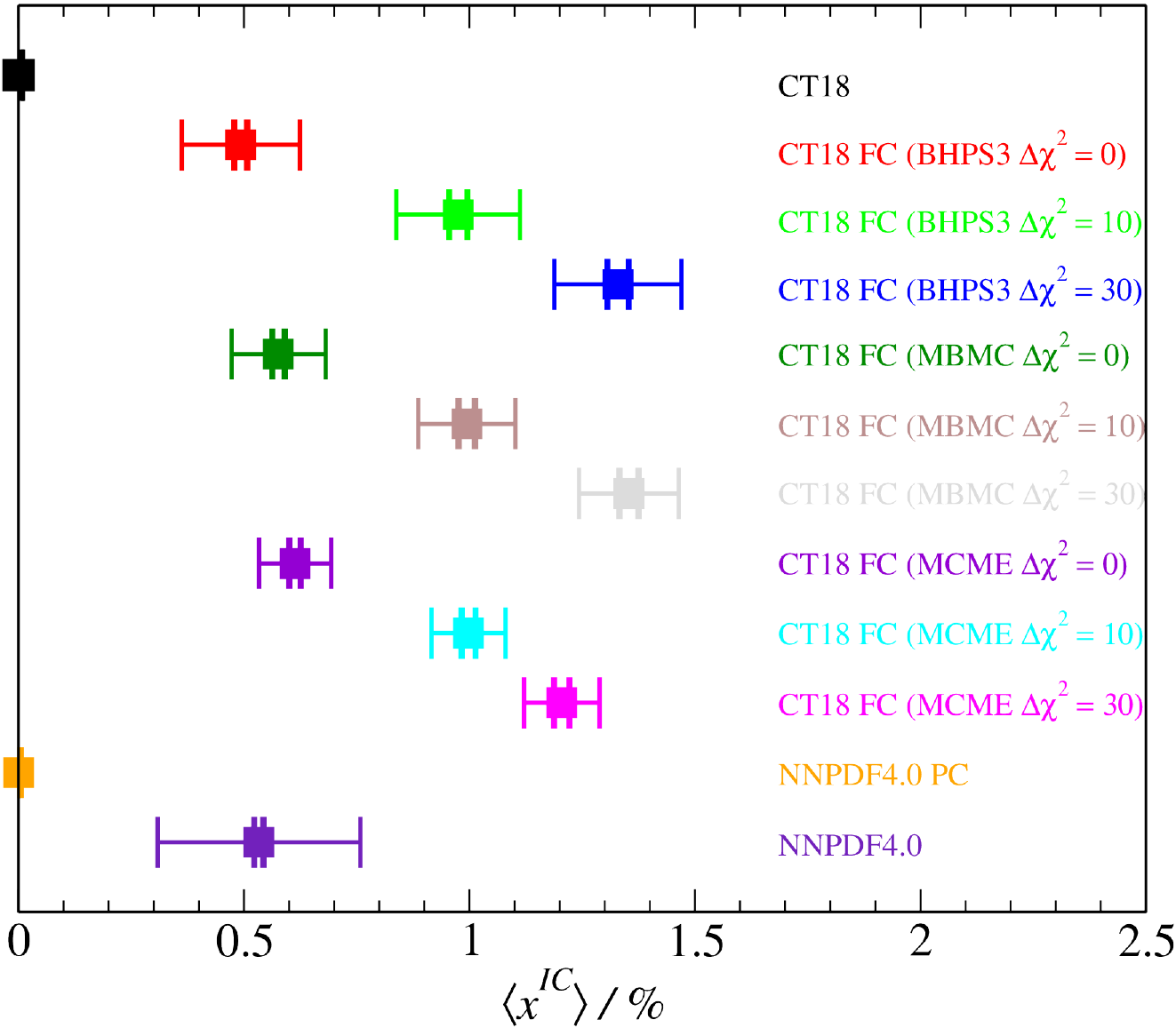}
    \caption{ Same as Fig.~\ref{fig:xcIC_FASERnu} for FASER$\nu$ (top) and FASER$\nu2$ (bottom) detectors at the HL-LHC for $\mathcal{L_{\mathrm{pp}}} = 3\,\mathrm{ab}^{-1}$.
    }
\label{fig:xcIC_FASERnu_HL}
\end{figure}

Fig.~\ref{fig:xcIC_FASERnu_HL} presents the same as Fig~\ref{fig:xcIC_FASERnu} now for FASER$\nu$ (top) and FASER$\nu 2$ (bottom) at the HL-LHC. 
In this case, total uncertainties are dominated by PDFs uncertainties, mainly when one considers events with $x\geq 0.2$. 
The high-statistics available at the HL-LHC therefore will result in strong constraints on the intrinsic charm momentum fraction, even considering only events with $x \geq 0.4$.

%% file: sec-summary.tex
\section{Summary and outlook}
\label{sec:summary}

Since the original proposals of the LHC far-forward detectors,  the science potential of LHC neutrino physics has attracted a significant amount of attention.
This interest has only been strengthened by the recent measurements of neutrino fluxes and TeV cross-sections by FASER, which impose the first constraints on the mechanisms driving forward neutrino production in pp collisions.
It is now recognised that, in addition to a rich program of light BSM physics searches, these far-forward detectors can operate as a high-intensity neutrino DIS experiment at TeV energies, offering unique opportunities to pin down strong interaction dynamics and hadron structure in poorly understood regimes by means of the most energetic neutrino beam ever achieved in a laboratory.

In these LHC neutrino physics studies as well as in most BSM searches, the abundant flux of high-energy muons reaching the far-forward detectors has been  considered as an inconvenient background, whose reduction is essential to achieve the physics reach of the analyses. 
In this work we provide a different viewpoint and demonstrate that this intense flux of high energy muons reaching FASER enables novel physics opportunities with value on its own and complementing those enabled by the neutrino program.
We find that the event yields for muon DIS at FASER are significantly larger than those of neutrino DIS, reaching $\mathcal{O}(10^{5})$ inclusive events satisfying acceptance cuts for $\mathcal{L}_{\rm pp}=250$ fb$^{-1}$ of which up to $\mathcal{O}(10^{4})$ being accompanied by the production of charm quarks. 

As a representative application of the physics potential of muon DIS at FASER we have quantified its sensitivity to the charm content of the proton.
We have highlighted its discrimination potential to separate between the intrinsic and perturbative charm scenarios, and in the former case constructed a dedicated observable that enables the detection of the ultimate signal of intrinsic charm, namely an asymmetry between the charm and anticharm PDFs in the initial state of the scattering.
This unique capability arises both from the excellent coverage of the large-$x$ region and from the combination of information provided by the FASER electronic and emulsion detectors. 

The results of this work can be extended in multiple directions.
First, by quantifying the precision with which NC inclusive and charm structure functions can be extracted at FASER and studying their impact on proton and nuclear PDFs by following the procedure of~\cite{Cruz-Martinez:2023sdv} for neutrino DIS based on Hessian profiling and NNPDF fits.
Second, the analysis of~\cite{Hammou:2024xuj} demonstrates that NC DIS data at the EIC enables breaking degeneracies between PDFs and BSM signals arising at the high-$p_T$ tails at the LHC, and hence one would like to assess whether this feature also holds for muon DIS measurements at FASER$\nu$.
Third, recent studies have shown~\cite{Boughezal:2020uwq, Boughezal:2022pmb,Duarte:2025zrg,Bissolotti:2023pjh} how high-energy DIS at the EIC and LHeC~\cite{LHeC:2020van} can  constrain Wilson coefficients in the Standard Model Effective Field Theory (SMEFT) that cannot be accessed with LEP or LHC data alone.
Further studies found this also applied to neutrino DIS measurement at FASER~\cite{Falkowski:2021bkq, Kling:2023tgr}.
Muon DIS at FASER$\nu$ should therefore be able to provide comparable constraints to the EIC and LHeC.
Our study emphasizes that to achieve this potential reducing systematic uncertainties is essential, especially regarding the muon energy resolution and the minimum $n_{\mathrm{tr}}$ value in the event reconstruction algorithm.

From the theoretical modelling point of view, the NLO {\sc\small POWHEG+Pythia8} simulations used in this work could be improved benefitting from recent progress in event generators for deep-inelastic scattering, such as accounting for charm mass effects in the matrix element~\cite{Buonocore:2024pdv}, matching matrix elements with different multiplicities~\cite{Meinzinger:2025pam}, increasing the perturbative accuracy of the parton shower to next-to-leading logarithmic order~\cite{vanBeekveld:2023chs} and beyond, and considering NNLO and N$^3$LO corrections to both the matrix elements and to the input PDFs~\cite{NNPDF:2024nan,McGowan:2022nag}.
These potential improvements should be combined with {\sc\small PineAPPL}-like fast interpolation interfaces, to streamline the interpretation of the FASER measurements and enable their inclusion of global analyses of proton and nuclear PDFs. 

Beyond Run 3, the much higher statistics available for FASER$\nu$ during Run 4 and beyond can only make the physics potential of muon DIS stronger, while the eventual realisation of the FPF with a dedicated muon DIS program would then enable multi-differential measurements giving access to more complex QCD objects beyond collinear PDFs such as TMD-PDFs~\cite{Angeles-Martinez:2015sea}. 
Our analysis also motivates the explicit consideration of muon DIS measurement in the optimization and design of the far-forward experiments and their upgrades~\cite{FASER:2025myb}. 
All in all, the muon DIS program at FASER will anticipate some of the key scientific goals of the EIC, providing a complementary probe of its investigations of the strong nuclear force, with the combination of neutrino and charged-lepton DIS measurements providing one of the backbones of global fits of proton and nuclear PDFs for the coming decade. 

\vspace{0.2cm}
\begin{center}
\rule{0.6\linewidth}{0.4pt}
\end{center}
\vspace{0.2cm}

\noindent
The results of this work can be reproduced by means of the variant of the {\sc\small POWHEG} DIS code~\cite{Banfi:2023mhz,FerrarioRavasio:2024kem,vanBeekveld:2024ziz} adapted to muon scattering at FASER and available in:
\begin{center}
\url{https://github.com/LHCfitNikhef/DIS_with_LHC_muons}
\end{center}
which also includes run cards and example analysis files illustrating its main functionalities.

\subsection*{Acknowledgements.}
We are grateful to Akitaka Ariga, Tomoko Ariga, Jamie Boyd and Ken Ohashi for discussions on this topic. 
We further thank the CERN STI group for performing detailed FLUKA simulations of the muon fluence along the LoS and the FASER collaboration for sharing the simulated muon flux passing through the FASER detector.
The work of R. F. is partially supported by Conselho Nacional de Desenvolvimento Cient\'{\i}fico e Tecnol\'ogico (CNPq, Brazil), Grant No. 161770/2022-3, and by DERI/Santander mobility program.
The work of V.P.G. is partially supported by CNPq, FAPERGS and INCT-FNA (Process No. 464898/2014-5).
The work of J.~R. is partially supported by the Dutch Science Council (NWO).
The work of F.~K. is supported in part by Heising-Simons Foundation Grant 2020-1840 and in part by U.S. National Science Foundation Grant PHY-2210283.